\documentclass[useAMs,a4paper]{mn2e}
\usepackage{savesym}
\usepackage{graphicx}
\expandafter\let\csname equation*\endcsname\relax
  \expandafter\let\csname endequation*\endcsname\relax 
\usepackage{subfig}
\usepackage{amsmath}
\usepackage{amssymb}
\usepackage{verbatim}
\usepackage[yyyymmdd,hhmmss]{datetime}
\usepackage{array}
\usepackage{times}
\usepackage[total={17.8cm,24.0cm},centering]{geometry} 
\usepackage{color}

\newcommand{\be}{\begin{equation}}
\newcommand{\beq}{\begin{equation}}
\newcommand{\ee}{\end{equation}}
\newcommand{\eeq}{\end{equation}}
\newcommand{\eea}{\end{eqnarray}}
\newcommand{\bea}{\begin{eqnarray}}

\newcommand\W {{W^r_{\ \phi}}}

\title[A unified model of disc-dominated TDEs]{A unified model of tidal destruction events in the  disc-dominated  phase}
\author [Andrew Mummery]{Andrew Mummery\thanks{E-mail:
andrew.mummery@physics.ox.ac.uk}
\\
Oxford Astrophysics, Denys Wilkinson Building, Keble Road, Oxford, OX1 3RH, United Kingdom}
\begin{document}

\date{}

\pagerange{\pageref{firstpage}--\pageref{lastpage}} \pubyear{2021}

\maketitle

\label{firstpage}

\begin{abstract} 
We develop a unification scheme which explains the varied observed properties of TDEs in terms of simple disc physics. The unification scheme postulates that the different observed properties of TDEs are controlled by the peak Eddington ratio of the accretion discs which form following a stellar disruption. Our primary result is that the TDE population can be split into four subpopulations, which are (in order of decreasing peak Eddington ratio): ``obscured'' UV-bright and X-ray dim TDEs; X-ray bright soft-state TDEs; UV-bright and X-ray dim ``cool'' TDEs; and X-ray bright hard-state TDEs. These 4 subpopulations of TDEs will occur around black holes of well defined masses, and our unification scheme is therefore directly testable with observations. As an initial test, we model the X-ray and UV light curves of six TDEs taken from three of the four subpopulations: ASASSN-14ae, ASASSN-15oi, ASASSN-18pg, AT2019dsg, XMMSL1 J0740 \& XMMSL2 J1446. We show that all six TDEs, spanning a wide range of observed properties, are well modelled by evolving relativistic thin discs. The peak Eddington ratio's of the six best-fitting disc solutions lie exactly as predicted by the unified model. The mean stellar mass of the six sources is $\left\langle M_\star \right\rangle \sim 0.24 M_\odot$. The so-called `missing energy problem' is resolved by demonstrating that only $\sim 1\%$ of the radiated accretion disc energy is observed at X-ray and UV frequencies. Finally, we present an empirical, approximately linear, relationship between the total radiated energy of the accretion disc and the total radiated energy of an early-time, rapidly-decaying, UV component, seen in all TDEs. 
\end{abstract}

\begin{keywords}
accretion, accretion discs --- black hole physics --- transients, tidal disruption events
\end{keywords}
\noindent

\section{introduction}
The complete destruction and subsequent accretion of a star by a super-massive black hole at the center of a galaxy, a so-called tidal destruction event (TDE), offers a novel probe of the physics of black hole accretion. {The theoretical modelling of TDEs has a long history, with} basic  models of TDEs evolution first put forward in the 1970s (e.g. Hills 1975; Rees 1988). In these models the stellar debris returns to the pericenter of the disrupted stars orbit at a rate approximately given by $\sim t^{-5/3}$. This returning material then powers  accretion onto the black hole, resulting in luminous flares that last for $\sim$ years. 

In the last two decades TDEs have been discovered and observed at almost every observing frequency. This includes hard X-rays (e.g. Cenko {\it et al}. 2012), soft X-rays (e.g. Greiner {\it et al}. 2000), optical/UV (e.g. Gezari {\it et al}. 2008, van Velzen {\it et al}. 2020), infared (e.g. van Velzen {\it et al}. 2016b) and radio frequencies (e.g. Alexander {\it et al}. 2016). TDEs with powerful radio and X-ray bright jets have also been discovered (e.g. Burrows {\it et al}. 2011). 

Before this wealth of observational information can be used to learn in detail about super massive black holes and the discs which surround them, a concrete theoretical framework through which these observations can be interpreted must be developed. The ultimate goal here is to develop a unification scheme in which the varied observed properties of TDEs can be consistently explained in terms of the different properties of, for example, the star or black hole involved in a particular TDE. 

Previously, a unification scheme of this type has been put forward by Dai {\it et al}. (2018). Dai {\it et al}. argued (based on a GRMHD simulation) that each TDE results in a geometrically thick accretion flow accreting at super-Eddington rates. Outflows from this super-Eddington flow can, for certain viewing angles, obscure the inner-most X-ray producing region, leading to detections of only optical/UV radiation. Conversely, for different viewing angles the innermost regions will not be obscured, and bright X-ray emission will be detected. The unification scheme of Dai {\it et al}.(2018) therefore suggests that the disc-observer orientation angle is the key parameter determining the observed properties of an individual TDE.

However, the GRMHD simulation which Dai {\it et al}. (2018) based their argument upon involved a relatively low mass $M = 5\times 10^6M_\odot$, and relatively rapidly rotating $a/r_g = 0.8$ black hole. In a previous paper (Mummery 2021) we have demonstrated that black holes with high spins and low masses are exactly those TDE systems which will produce initially super-Eddington accretion states.  We also demonstrated, however, that TDEs around black holes of larger masses (closer to $10^7 M_\odot$) are extremely unlikely to form discs with initially super-Eddington luminosities. It  therefore seems unlikely that the conditions examined by Dai {\it et al}. (2018) will be truly universal. In this paper we shall put forward an alternative unification scheme. While for low black hole masses our new unification scheme does in effect agree with the Dai {\it et al}. (2018) model,  our predictions are very different at large black hole masses. 

By now there is strong evidence, both theoretical and observational, that the dominant emission components of many TDE light curves result from evolving thin discs. The X-ray (Mummery \& Balbus 2020a), late-time UV (van Velzen {\it et al}. 2019, Mummery \& Balbus 2020a, b) and late-time optical (Mummery \& Balbus 2020b) light curves of a number of TDEs are  all well described by evolving thin-disc models. Furthermore, recent evidence of TDE disc accretion state changes (Wevers 2020, Jonker {\it et al}. 2020, Wevers {\it et al}. 2021), {and TDE QPO measurements (Pasham {\it et al}. (2019)}, strongly suggests that the accretion discs that form in the aftermath of a TDE have similar accretion properties as the more well-studied galactic X-ray binaries (e.g. Fender {\it et al}. 2004). 

While it is clear that a sub-set of all TDEs have disc-dominated light curves, the current TDE population is of course notable for its varied observed properties. Excluding jetted TDEs, which are distinct category of  events all together, TDEs have been observed in broadly three distinct X-ray states. Some are X-ray bright with quasi-thermal (blackbody) X-ray spectra (e.g. ASASSN-14li, Holoien {\it et al}. 2016), others are UV bright with no detected X-ray emission (e.g. ASASSN-14ae, Holoien {\it et al}. 2014), and finally some sources are X-ray bright but with a nonthermal power-law-with-energy X-ray spectrum (e.g. XMMSL2 J1446, Saxton {\it et al}. 2019). All sources are observed to be bright at optical and UV energies. How much of this variation in observed properties can be explained by the physics of time-dependent accretion discs is clearly an extremely important question. 

It is the purpose of this paper, in addition to our previous work, to examine the answer to this question. We will argue that the different types of observed TDEs can be unified into a single framework. In this framework the  amplitude of the peak Eddington ratio of the discs which form in the aftermath of a TDE {is the key parameter which controls the} observed properties of TDEs. 

In a series of three papers (Mummery \& Balbus 2021a, Mummery \& Balbus 2021b, Mummery 2021; hereafter Papers I, II and III) we studied the properties of  time-dependent accretion disc systems across the typical range of black hole masses expected to host TDEs, $10^6 < M/M_\odot < 10^8$. This work resulted in a number of key predictions about the observed properties of the TDE population as a whole, which will be summarised and extended in this paper. Principally, we showed that standard time-dependent accretion disc theory predicts a peak Eddington ratio which is strongly dependent on black hole mass.  Over the factor $\sim 100$ change in black hole mass of anticipated TDE hosts, there is a huge potential variance in peak Eddington ratios of TDE discs, of order $\sim 10^7$ {(Mummery 2021; a derivation of this potentially surprising result will be outlined in section \ref{UM})}.  By making analogy with X-ray binary systems we can make predictions about the different observed properties of TDE systems which form at different Eddington ratios. As these different Eddington ratios will be principally controlled by the central black hole mass of the TDE (equation \ref{edrat}), we can make quantitative predictions of the observed black hole mass distributions of the different TDE spectral sub-populations. 

In section \ref{UM} of this paper we summarise the key findings of Papers I, II and III, bringing them together into a single unified framework. We will then demonstrate that the current population of observed TDEs -- comprising of approximately 40 sources -- have observed black hole mass and X-ray luminosity distributions which are in strong agreement with the predictions of the unified model. 

While this unification scheme makes accurate predictions about the observed distribution of TDEs on the population level, it is extremely important that it reproduces the observed light curves of individual TDE sources. To that end we model the X-ray and UV light curves of six TDEs from across the different TDE sub-populations. In section \ref{method} we summarise the methodology used to model the six sources. The techniques used are similar to our previous work (Mummery \& Balbus 2020a, b), but with a more physical treatment of the initial disc condition and radiative transfer effects. 

In sections 4, 5 \& 6 we model two TDEs from each of three different spectral sub-populations. In section 4 we model the X-ray bright TDEs AT2019dsg and ASASSN-15oi, both of which have quasi-thermal X-ray spectra. In section 5 we model the X-ray dim TDEs ASASSN-14ae and ASASSN-18pg. In section 6 we model two TDEs with bright nonthermal X-ray spectra, XMMSL1 J0740 and XMMSL2 J1446.  

In section 7 we demonstrate that each of the six TDEs have peak Eddington ratios as predicted within the unified framework. We show that the so-called missing energy problem is resolved by only $\sim 1\%$ of the total disc energy being radiated into observable frequencies, and present empirical scaling relationships between properties of the early-time UV luminosity and physical parameters of the best-fitting disc solutions. These empirical scaling relationships will serve both to elucidate the nature of this early-time UV component, but also to infer physical parameters of TDE systems from early-time observations of TDEs. 

We discuss how the properties of jetted TDEs may be brought into the unification scheme in section 8, where we make predictions about the black hole mass distribution of bright jetted TDEs. We conclude in section 9.

\section{The Unified  model of disc-dominated TDEs  }\label{UM}
The unified model of disc-dominated TDEs makes a number of key predictions about the observed properties of the TDE population based on a relatively small number of assumptions about the physical properties of TDE discs. Perhaps the most important  result of this unification scheme is that, of all the parameters which can be {\it directly inferred from observations}, it is the central black hole mass which is the key parameter in determining the observed properties of a particular TDE. {Note that of course, as we stressed in the introduction, it is the discs Eddington ratio which physically sets the  properties of an individual TDEs light curves. However, as we shall shortly demonstrate,  the Eddington ratio is strongly constrained -- although not uniquely determined -- by the TDE hosts black hole mass. In the following section we shall cast our results in terms of the black hole mass of the TDE host for the following simple reason: the central black hole mass of a TDE can be inferred directly from observations of the TDEs host galaxy.  This is in contrast with the discs bolometric luminosity, which can never truly be measured.  By casting our results in this manner our predictions become much more simple to test through direct observations.  }

In this section we summarise the key model assumptions, and key predictions, of the unified TDE disc model. We then demonstrate that the properties of the observed TDE population are in good accord with the model predictions.

\subsection{Key model assumptions}
The first key assumption of our unified model is the following:
\begin{itemize}
\item{A TDE results in an amount of matter $M_{\rm deb}$ which is gravitationally bound to the central black hole. All of this material can potentially form into an accretion disc around the central black hole. This accretion disc then dominates the TDE emission at X-ray energies, and at late times in the UV and optical.  }
\end{itemize}
It follows that the observed TDE X-ray and late-time UV light curves can be understood from the properties of evolving relativistic accretion discs. {Note that TDE emission observed at early-times at UV and optical frequencies is not well described by disc emission, and additional components are required. The physical origin of this early emission will be discussed firther in section \ref{cor_sec}. } Two further assumptions are then required to make progress in describing evolving TDE discs:
\begin{itemize}
\item{Once a disc initial condition has been specified, the evolution of the TDE accretion disc is well described by the relativistic thin disc evolution equation (Balbus 2017, Balbus \& Mummery 2018)} 
\item{The turbulent angular momentum transport within the disc is well approximated by local $\alpha$-type models (Shakura \& Sunyaev 1973). }
\end{itemize}
To leading order our TDE accretion model  is therefore described by three free parameters, the disc mass $M_d$ (which will be related to the initial debris mass), an $\alpha$-type parameter (or viscous timescale), and a central black hole mass $M$. While additional parameters (e.g., the black hole spin, the disc-observer inclination angle, and the properties of the ISCO stress) of course modify the properties of individual disc systems, it is these three parameters which set the overarching scales of emission and therefore which dominate the observed distribution of the TDE population as a whole.  

There is at least a factor $100$ range of central black hole masses that TDEs are likely to occur around, with expected black hole masses ranging roughly between $M \sim 10^6 - 10^8 M_\odot$. A result of this central mass variance is that TDE discs can form with widely varying initial Eddington ratios. To model TDE discs across the whole  black hole mass range, we make a fourth assumption about the accretion properties of TDEs as a function of Eddington ratio:
\begin{itemize}
\item{TDE discs have similar accretion state properties, as a function of Eddington ratio, as X-ray binaries (e.g., Fender {\it et al}. 2004). }
\end{itemize}
Specifically, this fourth assumption essentially splits the TDE populations into three regimes, based on the discs peak Eddington ratio. Discs which form with Eddington ratios ($l \equiv L_{\rm bol, peak}/L_{\rm edd}$) in the range  $1 \gtrsim l \gtrsim 0.01$, will form in the so-called `soft' state, with intrinsic emission well described by superpositions of thermal blackbody emission from each disc radii. 

In addition, those TDE discs which form with smaller Eddington ratios, $l \lesssim 0.01$, will likely form in the `hard' state (Maccarone 2003). Hard state TDEs can be observationally distinguished by the presence of a non-thermal power-law X-ray spectral component. Theoretically, we assume that this nonthermal X-ray component results from the Compton up-scattering of thermal disc photons by a compact electron scattering corona located in the innermost disc regions (see Paper II for full details). 

Finally, TDE discs may form with $l > 1$, an initially super-Eddington accretion state. The modelling of these sources requires some care, as a super-Eddington luminosity appears to contradict our second key assumption. Specifically, it is unlikely that the governing assumptions required to derive the thin disc evolution equation (Balbus 2017) will be valid in a supper-Eddington  accretion state. In Paper III we developed in detail the additional assumptions required to model TDEs in this regime, which are recapped below.  

\subsection{From assumptions to predictions}
A useful physical insight governing the predictions of the unified model is that the peak temperature reached in an evolving accretion disc is the key parameter which governs the properties of a TDE when observed at X-ray energies. By further demonstrating that this peak temperature is a sensitive function of the system parameters, and is particularly sensitive to the central black hole mass, the dependence of observed TDE X-ray properties on system parameters can be determined. 

In Paper I, we demonstrated that the peak temperature in a time-dependent relativistic accretion disc depends sensitively upon the black hole mass.  Expressed in terms of the disc mass $M_d$, $\alpha$-parameter $\alpha$, and black hole mass $M$, the temperature of the hottest disc regions scales as
\beq\label{temp}
T_p \propto \alpha^{1/3} M_d^{5/12} M^{-7/6} .
\eeq
As the bolometric luminosity of a TDE disc scales like $T_p^4$ multiplied by the disc area  (proportional to the square of the black hole mass), the disc Eddington ratio scales as  (Paper I):
\beq\label{edrat}
l \equiv {L_{\rm bol, peak} \over L_{\rm edd}} \propto {\alpha^{4/3} M_d^{5/3} \over M^{11/3}} .
\eeq
Note the strong black hole mass dependence of the Eddington ratio $l \propto M^{-11/3}$.   Equation \ref{edrat} implies that discs with identical disc masses and $\alpha$ parameters around black holes which differ by a factor 100 will have Eddington ratios which differ by a factor $\sim 10^7$. It is therefore no surprise that TDE discs show such varied observational properties, nor that the black hole mass plays a key role in governing the observed properties of an individual TDE. 

{Equation \ref{edrat} is a critical component of the unified TDE disc model put forward in this paper.  It is important, therefore, to understand how this new result differs from the `Eddington ratio' that is often inferred from a `fall-back rate' calculation.  The fall-back rate is defined as the rate at which the stellar debris return to the pericentre of the disrupted stars orbit. It can be shown that this rate varies like $\dot M_{\rm fb} \sim M^{-1/2}$ (e.g. Metzger \& Stone 2016).  It is common to assume that the disc bolometric luminosity will behave as $L_{\rm disc} \propto \dot M_{\rm fb}$.   However, this simple $L-\dot M_{\rm fb}$ relationship is a blending of {\it steady-state} scaling relationships with a {\it time-dependent} regime, and this is not generally valid.  Only in the limit of near steady-state behaviour (i.e., many viscous timescales into the evolution $t/t_{\rm visc} \gg 1$)  does the disc luminosity scale proportionately to the rate at which the disc is fed material.  At early times (e.g., $t/t_{\rm visc} \simeq 1-10$), as is always the case for TDE observations, the disc accretion rate is a dynamic quantity, which {\it varies at every radius and time within the disc}.  Knowing the rate at which the disc is fed mass is not by itself sufficient to determine the disc luminosity; a more detailed treatment is required.    The full time-dependent calculation (equation \ref{edrat}; full derivation in Paper I) leads to a much more sensitive black hole mass dependence of the disc luminosity than the fall-back rate estimate. This stronger sensitivity naturally leads to a wider variance in the observed properties of the TDE population.  }

{Now that the physical parameter dependence of the temperature and luminosity of TDE accretion discs has been determined, we can more carefully define the different subpopulations of TDEs that comprise our unification scheme.} Consider a disc with mass $M_d$, a prescribed $\alpha$-parameter, and a central black hole mass of $M_{\rm edd}$ for which the bolometric luminosity peaks at the Eddington luminosity, $l = 1$.   The mass value $M_{\rm edd}$ represents a key scale in the unified TDE model, and for a Schwarzschild black hole takes the approximate value
\beq
M_{\rm edd} \simeq 5\times 10^6 M_\odot \left({M_d\over 0.5M_\odot}\right)^{5/11} \left({\alpha\over 0.1}\right)^{4/11} .
\eeq

In the first instance let us consider those TDEs which occur around black holes with masses such that the disc forms in the soft accretion state ($1 \gtrsim l \gtrsim 0.01$). From equation \ref{edrat} we see that this corresponds to those TDEs around black holes with masses $M_{\rm edd} < M < M_{\rm HS}$, where $M_{\rm HS}$ is the black hole mass  at which the disc transitions into the hard accretion state. This hard state mass will be defined more carefully shortly. 

We demonstrated in Paper I that in this accretion regime the thermal X-ray luminosity emergent from TDE discs is strongly suppressed around large black hole masses. More precisely,   the X-ray luminosity $L_X$ is cut-off according to (assuming a vanishing ISCO stress, Paper I):
 \beq\label{lxsoft}
L_X \propto M^{1/4} \exp\left(-m^{7/6}\right) , 
\eeq
where $m$ is a dimensionless, normalised mass variable, in essence the black hole mass in units of $\sim 10^6$M$_\odot$. This strong suppression results in upper-observable black hole mass limits, which we demonstrate (by assuming canonical disc parameter values) to be of order $M_{\rm lim} \simeq 3 \times 10^7M_\odot$, above which thermal X-ray emission will not be observable. This upper observable black hole mass limit is a function of the remaining disc parameters, and the full dependence can be described analytically (Paper I, eq. 81).

While the thermal X-ray luminosity of TDE discs is strongly suppressed around large mass black holes, at yet higher black hole masses TDE discs will have Eddington ratios of order $\sim 1\%$, and will likely form in (or promptly transition into) the hard accretion state. The typical black hole mass scale at which this transition occurs is given by 
\begin{equation}
M_{\rm HS} = M_{\rm edd} \, {l_{\rm HS}}^{-3/11},  
\end{equation}
where $l_{\rm HS}$ is the Eddington ratio at which the transition into the hard state occurs. More explicitly
\begin{equation}\label{MHS}
M_{\rm HS} \simeq 2\times 10^7 M_\odot \left({M_d\over 0.5M_\odot}\right)^{5/11} \left({\alpha\over 0.1}\right)^{4/11} \left({l_{\rm HS} \over 0.01}\right)^{-3/11}. 
\end{equation}
In Paper II we developed a model of nonthermal hard-state TDE disc emission. We assume that upon transitioning into the hard state, an electron scattering corona forms in the innermost disc regions (parameterised by a radial size $R_{\rm Cor}$), which Compton up-scatters a fraction $f_{\rm SC}$ of the thermal disc photons into a power-law spectrum of index $\Gamma$. The detailed modelling of the resultant disc spectrum is then computed using photon number conservation. This analysis is presented in detail in Paper II. 

A key result developed in Paper II was that this hard state TDE model produced observable levels of nonthermal X-ray emission for black hole masses $M > M_{\rm HS}$.  The typical mass scale at which TDE discs are expected to form in the hard state is comparable to the mass scale at which thermal X-ray TDEs produce unobservable luminosities. The two populations of TDEs (those with thermal and nonthermal X-ray spectra) will therefore be drawn from populations of black holes with systematically different masses. While there may be some overlap between the two populations at intermediate black hole mass ($M \sim 10^7 M_\odot$), the majority of thermal X-ray TDEs will have masses $M \lesssim 10^7 M_\odot$, and the majority of nonthermal TDEs $M \gtrsim 10^7 M_\odot$.

In addition to determining the black hole mass dependence of observed TDE properties, an interesting observational prediction of the unified TDE disc model relates to the maximum observed  X-ray luminosities of both thermal and nonthermal X-ray TDEs (Paper III). By assuming that all photons emitted from the inner disc are Compton up-scattered by the corona ($f_{\rm SC} \rightarrow 1$), we can calculate an upper X-ray luminosity limit for disc-dominated hard state X-ray TDEs. By direct numerical calculation we determine that nonthermal X-ray TDEs will be observed at luminosities below a maximum X-ray luminosity of $L_{X, {\rm max}} \equiv L_M(a) \sim 10^{43} - 10^{44}$ erg/s. The variance in this luminosity scale results primarily from the range of allowed black hole spins. 

Finally, at the lowest black hole masses (for $M < M_{\rm edd}$) TDE discs are expected, in the simplest modelling, to form in super-Eddington accretion states  ($l > 1$). However, as has been noted by a variety of authors (see Jonker {\it et al}. (2020) for more details) there is as-of-yet {very little} observational evidence for this super-Eddington accretion state in {the}  observed TDE X-ray spectra {of disc-dominated TDEs}. {(This is in contrast with bright jetted TDEs, which may have been launched by super-Eddington accretion, Burrows {\it et al}. 2011.)} In fact, early-time soft X-ray detections of TDEs generally find quasi-thermal spectra that are analogous to an XRBs high-luminosity, spectrally soft state (e.g., five recently discovered sources in van Velzen {\it et al}. 2020{; although see Lin et al. (2015) and Kara et al. (2018) for arguments that two outflows inferred from soft X-ray spectra may be linked to super-Eddington accretion flows}). The remaining X-ray bright TDEs have been found in the nonthermal hard accretion state.

In Paper III we put forward a model to explain this lack of observed super-Eddington X-ray TDE discs. We assume (Paper III) that the amount of material which forms into a compact accretion disc in the aftermath of a TDE is set so that the bolometric luminosity of the disc never exceeds its Eddington luminosity. This approximation is motivated, in addition to the observational absence of any super-Eddington TDEs, by the fact that the matter within a TDE disc is gradually deposited, with debris returning to the pericentre of the disrupted stars orbit approximately according to $\dot M_{\rm fb} \sim t^{-5/3}$ (Rees 1988). Much of this returning matter is likely to be only tenuously gravitationally bound (Metzger \& Stone 2016), and so is susceptible to being expelled by large radiation pressures resulting from a particularly bright accretion disc. In fact, it is known that not all of the debris mass forms into a disc: exponentially declining light curve components are observed across optical and UV bands at early times, prior to these light curves transitioning to a disc dominated state (van Velzen et al. 2019, Mummery \& Balbus 2020a, b). We propose that the amount of matter that reaches orbits close to the black holes ISCO is set so that, at peak, $L_{\rm bol} \leq L_{\rm edd}$. 

This additional assumption means that for black hole masses lower than the Eddington mass ($M < M_{\rm edd}$) only a fraction of the debris mass forms into an accretion disc. The surviving disc mass fraction in the $M < M_{\rm edd}$ regime can be calculated explicitly (from eq. \ref{edrat}): $M_d = M_{\rm deb}\, (M/M_{\rm edd})^{11/5}$. 

This final modelling assumption results in two key predictions which can be tested observationally. Firstly, the parameter dependence of the X-ray luminosity emergent from  TDE discs in this Eddington limited regime is qualitatively different from those solutions discussed earlier (i.e., equation \ref{lxsoft}). For a vanishing ISCO stress disc, for example, the X-ray luminosity now scales with black hole mass according to  
\beq\label{lxedd}
L_X \propto M^{13/8} \, \exp\left(-C_1 M^{1/4}\right), \quad M < M_{\rm edd} ,
\eeq
where $C_1$ is a dimensional parameter which depends on the disc properties (Paper III). These two expressions (eq. \ref{lxsoft} \& \ref{lxedd}) have qualitatively different properties. When sub-Eddington (eq. \ref{lxsoft}, $M > M_{\rm edd}$) the X-ray flux is given by the product of a polynomial which is a weak function of mass, and a negative exponential which is strongly mass dependent. However, when in the Eddington limited regime (eq. \ref{lxedd}, $M < M_{\rm edd}$), the X-ray flux is given by the product of a polynomial which is a relatively strongly increasing function of mass, and an exponential which is only weakly mass dependent.

In fact numerical studies (performed in detail in Paper III) show that for $M < M_{\rm edd}$ these two black hole mass dependences effectively cancel out, and the  X-ray luminosity is approximately constant, depending  only weakly on the black hole spin. The amplitude of this peak X-ray emission is demonstrated to vary between  $L_M(a) \sim 10^{43} - 10^{44}$ erg/s. This X-ray luminosity scale is the  same level as for the maximum hard-state X-ray emission derived earlier. 

Finally, if TDE discs are to form in this Eddington limited state, then the vast majority of the stellar debris of TDEs around low black hole masses must be expelled from the system. We assume that this material is removed by radiatively driven winds, and a simple calculation shows that the outflow mass is given explicitly by  
\beq
M_{\rm out} = M_{\rm deb} \, \left[ 1 - \left( {M \over M_{\rm edd}}\right)^{11/5} \right], \quad M \leq M_{\rm edd}.  
\eeq
For TDEs around low mass black holes this outflow mass can be extremely large, representing the majority of the debris mass. As has been demonstrated in previous works (e.g., Metzger \& Stone 2016), those TDEs which form with small disc-to-outflow ratios are likely to be only observed at optical and UV photon energies.  This is because the ejected material will remain sufficiently neutral to block hard EUV and X-ray radiation from the hot inner disk, which becomes trapped in a radiation-dominated nebula. Ionising radiation from this nebula then heats the inner edge of the ejecta to temperatures of $T \sim {\rm few} \times 10^4$ K, converting the emission to optical/near-UV wavelengths. At these frequencies the photons more readily escape the nebula due to the lower opacity and are then observed (see Metzger \& Stone (2016) for full details). The effect of this inner-disc obscuring at low black hole masses is similar to the Dai {\it et al}. (2018) model. 

\subsection{Unified model predictions}
The unified model of disc-dominated TDEs makes a number of predictions about the properties of the observed TDE population. Firstly, and perhaps most importantly, we expect the observed population of TDEs to be split into four distinct sub-populations, which are separated in black hole mass. 

In order of increasing black hole mass these sub-populations are: 
\begin{itemize}
\item{A population of TDEs around  black holes with the lowest masses ($M \lesssim 10^6 M_\odot$)  which are observed at only optical and UV photon frequencies. This results from the inner-disc being obscured by a large mass radiatively driven outflow (Paper III).  }

\item{The bright quasi-thermal X-ray TDE population to contain black holes with masses primarily between $M \sim 10^6 - 10^7 M_\odot$, which sources above $M_{\rm lim} \simeq 3\times 10^7 M_\odot$ being extremely rare (Paper I). }

\item{The existence of some optical and UV bright TDEs with black hole masses around $M \sim 10^7 M_\odot$, which are not observed at X-ray energies. These sources are insufficiently hot to produce observable levels of thermal X-ray radiation, a result of the strong dependence of the disc temperature on black hole mass (Paper I).   }

\item{The bright nonthermal X-ray TDE population to contain blackholes with masses primarily in excess of $M \gtrsim 2 \times 10^7 M_\odot$, with sources with masses below $M \lesssim 5 \times 10^6 M_\odot$ being extremely rare (Paper II). }
\end{itemize}

\begin{figure}
  \includegraphics[width=.5\textwidth]{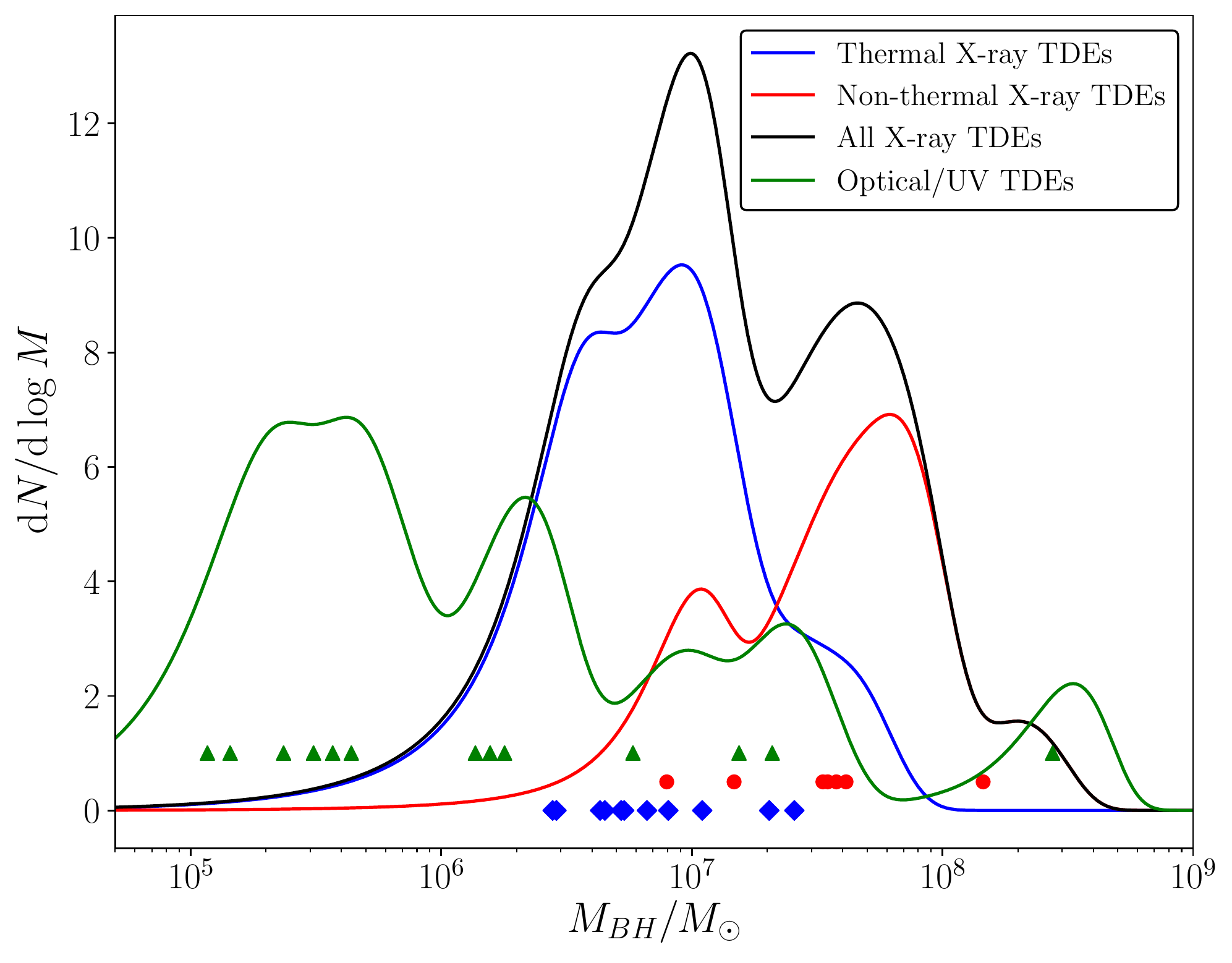} 
 \caption{The black hole masses of thermal X-ray TDEs (blue points \& curve, Paper I), non-thermal X-ray TDEs (red points \& curve,  Paper II) and optical/UV-only TDEs (green points \& curve, Wevers {\it et al}. (2019), their table A1). The inferred distributions (solid curves) are calculated  using kernel density estimation using a kernel width equal to the uncertainty in each TDEs black hole mass. The different populations of TDEs fit exactly as predicted by the unified disc model. All TDEs with inferred masses $M< 2\times 10^6M_\odot$ are only observed in optical and UV bands (corresponding to the 9 TDEs with the lowest inferred masses). The populations of the thermal and nonthermal X-ray TDEs are systematically offset in mass, being distributed either above (nonthermal) or below (thermal) $M \sim 10^7 M_\odot$.  The larger mass $M \sim 10^7 M_\odot$ optical/UV-only TDEs likely lack X-ray radiation due to the suppression of X-ray emission from large mass black holes (Paper I). }  
 \label{UM1}
\end{figure}

These predictions of the unified TDE disc model  can therefore be simply verified, on the population level, by determining the black hole mass distribution of the different spectral types of the observed TDE population. In Figure \ref{UM1} we plot the inferred black hole mass distributions of the three spectral types of observed TDEs. {In Papers I \& II we used well-established galactic scaling relationships between the black hole mass and (i) the galactic bulge mass $M : M_{\rm bulge}$, (ii) the galactic velocity dispersion $M : \sigma$, and (iii) the bulge V-band luminosity $M : L_V$. All of the scaling relationships are taken from McConnell \& Ma (2013).  Where available, values of $ M_{\rm bulge}$, $\sigma$ and $L_V$ were taken from the literature for each TDE, and the mean black hole mass for each TDE was computed (summarised in Appendix \ref{masses}).} The mean black hole mass values of the 11 quasi-thermal X-ray TDEs are shown by blue diamonds, the 7 nonthermal X-ray TDEs by red dots, and the optically-bright X-ray-dim TDEs by green triangles. The black hole masses of the optically bright TDEs are taken from the galactic velocity dispersion measurements of Wevers {\it et al}. (2019), their Table A1. The inferred distributions (solid curves) are calculated  using kernel density estimation using a kernel width equal to the uncertainty in each TDEs black hole mass.

It is clear that the black hole mass distributions of the observed TDE population fit exactly as predicted by the unified disc model of TDEs. The 9 black holes with the lowest inferred masses are all only observed at optical/UV energies ($M < 2\times10^6M_\odot$). The black hole masses of the bright quasi-thermal X-ray TDE population predominantly lie between $M \sim 10^6 - 10^7 M_\odot$, with no sources detected at extremely large black hole masses $M \gtrsim 3 \times 10^7 M_\odot$. The population of bright nonthermal X-ray TDEs are at systematically higher black hole masses than the bright quasi-thermal population, being predominantly observed when the black hole mass $M \gtrsim 2 \times 10^7 M_\odot$. The null-hypothesis of identical black hole mass distributions is rejected with $p$-value = 0.01 for a two-sample K-S test, and $p = 0.002$ for a two-sample Anderson-Darling test (Paper II). 

The main predictions of the unified TDE disc model for the X-ray luminosity distributions of the TDE population can be summarised as:

\begin{itemize}
\item The X-ray luminosity of quasi-thermal disc-dominated TDEs evolving with bolometric luminosities comparable to their Eddington luminosity will lie at a near universal scale $L_M(a) \sim 10^{43} -10^{44}$ erg/s, with the variance primarily stemming from the range of allowed black hole spins (Paper III). 
 
\item The nonthermal X-ray luminosity of disc-dominated hard state sources is also limited by a maximum X-ray luminosity scale $L_M(a) \sim 10^{43}- 10^{44}$ erg/s (although by completely different physics, Paper III).
\end{itemize}
These two findings then lead to two key observational predictions: 
\begin{itemize}
\item The brightest quasi-thermal and nonthermal X-ray TDEs will both peak with X-ray luminosities in the range $L_X \sim 10^{43} - 10^{44}$ erg/s, meaning that the X-ray luminosity of the brightest TDEs will be independent of both black hole mass and X-ray spectral state. 

\item There will be a lack of observations of disc-dominated TDEs with X-ray luminosities $L_X \gtrsim 10^{44}$ erg/s. 
\end{itemize}

 \begin{figure}
  \includegraphics[width=.5\textwidth]{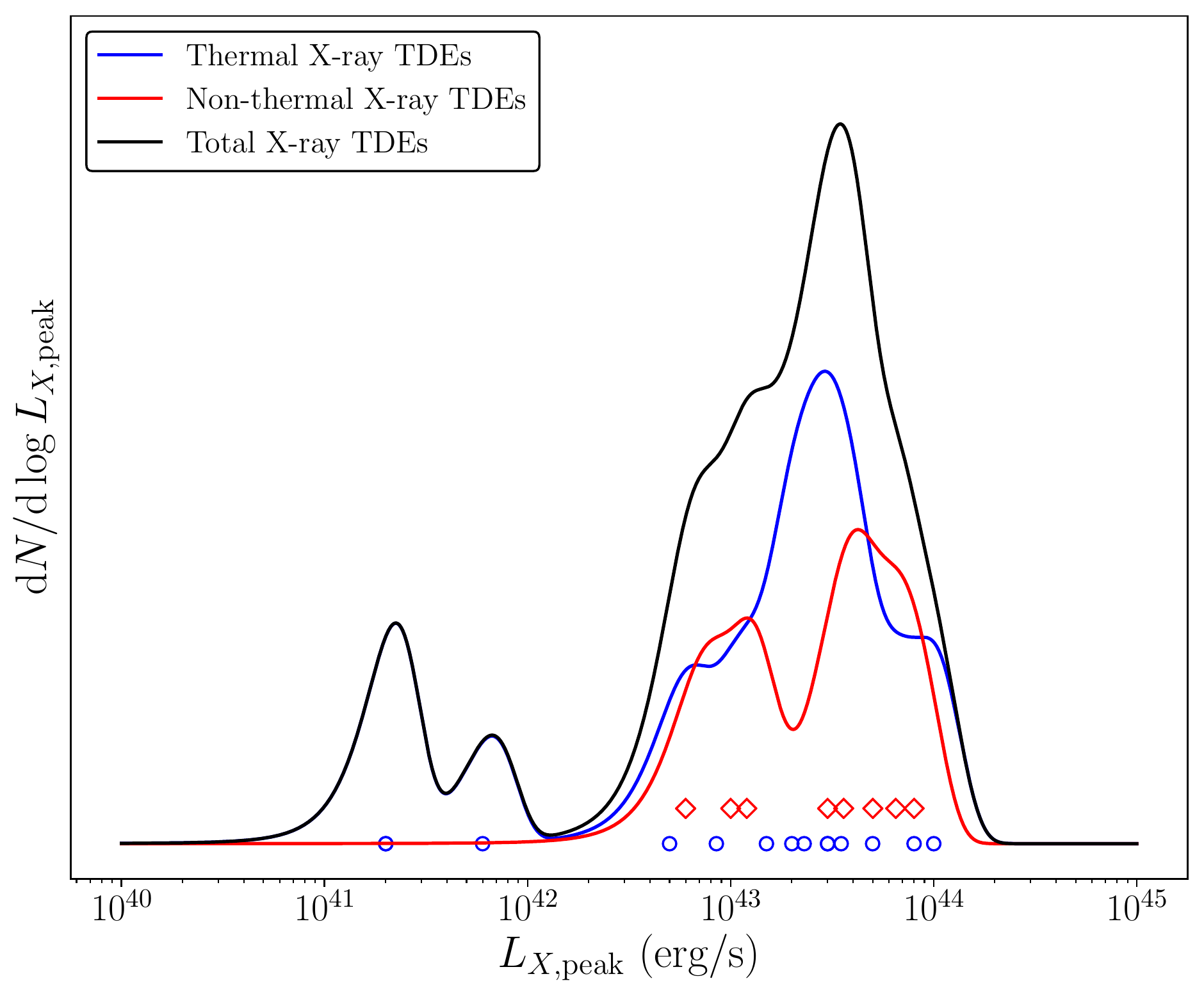} 
 \caption{The peak observed X-ray ($0.3$--$10$ keV) luminosities of the historic TDE population (Paper III). 18 of the 22 disc-dominated X-ray TDEs lie in the black hole spin-dependent characteristic luminosity scale predicted by the unified TDE model.  No disc dominated TDEs have yet been observed in the predicted barren region of the X-ray TDE luminosity diagram, with $L_X > 10^{44}$ erg/s. } 
\label{UM2}
\end{figure}

 \begin{figure}
  \includegraphics[width=.5\textwidth]{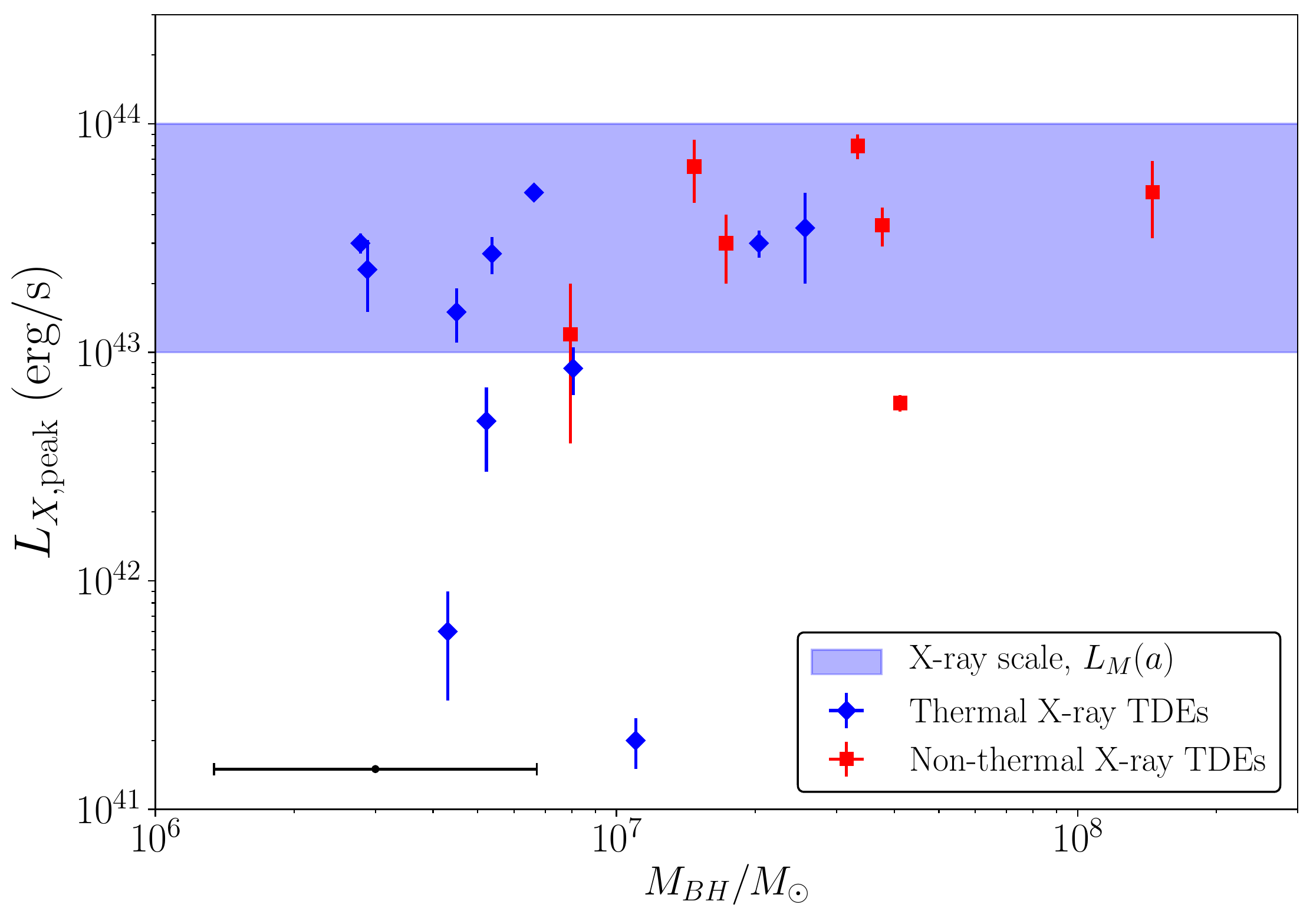} 
 \caption{The peak observed X-ray luminosities of the 18 X-ray TDEs with black hole mass estimates from galactic scaling relationships (Papers I \& II). Different X-ray spectral types are denoted by blue diamonds (quasi-thermal emission) and red squares (non-thermal emission).  The black point shows the typical uncertainty in the black hole masses resulting from intrinsic scatter in galactic scaling relationships. While there is a systematic difference in the black hole masses of the different X-ray spectral types, the peak X-ray luminosity is approximately independent of the central black hole mass.  } 
 \label{UM3}
\end{figure}

Figures \ref{UM2} and \ref{UM3} show that these predictions are borne out by the current TDE population. In Paper III we collated the maximum observed X-ray luminosities of all  the disc-dominated X-ray bright TDEs from the literature. Figure \ref{UM2} demonstrates that the vast majority (18 out of 22) of X-ray bright TDEs have luminosities which lie approximately at the level $L_X \sim 10^{43}-10^{44}$ erg/s.  The fact that this X-ray luminosity scale is independent of both black hole mass and X-ray spectral state is further emphasised in Figure \ref{UM3}, where we plot the peak observed X-ray luminosity against black hole mass for the 18 X-ray bright TDEs for which we have estimates of the black hole mass. Although the masses of thermal (blue diamonds) and nonthermal (red squares) X-ray TDEs are systematically offset, the X-ray luminosity scale at which all the TDEs peak is independent of both black hole mass and spectral state.

\section{Modelling individual TDEs: Methodology}\label{method}
While the unified TDE disc model successfully explains the properties of TDEs on the  population level, it remains an important question as to whether the individual light curves of TDEs can be reproduced across each of the different TDE sub-populations. In previous works we have examined ASASSN-14li, a well-observed bright quasi-thermal X-ray TDE (Mummery \& Balbus 2020a), and ASASSN-15lh, a UV bright X-ray dim TDE with an extremely large central mass (Mummery \& Balbus 2020b). Both of these TDEs were well described by an evolving relativistic thin disc. 

In this paper we will model six different TDEs from 3 different sub-populations: bright quasi-thermal X-ray sources, X-ray dim sources which undergo a clear disc-dominated UV plateau, and bright nonthermal X-ray sources. We model these sources in a very similar, but quantitively different,  manner to our earlier work. The entire methodology is summarised below. By modelling six sources from across the different sub-populations in an identical manner we are able to test both the ability of disc models to reproduce the observed TDE light curves of a wide range of different TDEs, but also whether the best-fit properties of the different systems align with the predictions of the unified model. 

\subsection{Disc evolution and initial condition}
The evolving disc density profiles are calculated using the methods developed in (Mummery \& Balbus 2020a), to which the reader should refer for detailed information.  A methodological difference between the current work and our previous modelling is that, rather than introducing the disc material as a delta-function ring at a single time $t_D$, here we gradually feed matter into a ring using a source term ${\cal S}_M$ in the governing disc equation (Balbus 2017) 
\beq
{\partial \zeta \over \partial t} = {\cal W} {\partial \over \partial r} \left( {U^0 \over U_\phi ' } {\partial \zeta \over \partial r} \right) + {\cal S}_M .
\eeq
Here $\zeta \equiv r \Sigma W^r_\phi /U^0$ allows the evolution equation for the disc surface density $\Sigma(r,t)$, under the influence of a turbulent stress tensor $W^r_\phi$, to be written in a compact form.  We have defined ${\mathcal W} = (1/U^0)^2\left[W^r_\phi + \Sigma\, \partial W^r_\phi/\partial\Sigma\right]$.  The angular momentum gradient is denoted $U_\phi'$, and $U^0$ is the temporal component of the discs four velocity $U^\mu$.  In common with all analytic models of TDE evolution, our model has necessarily been simplified (see section 4.9 of Mummery \& Balbus 2020a).  If the disc bolometric luminosity remains sub-Eddington, a thin disc model should provide a reasonably accurate description of the disc evolution.   We use the following prescription for the mass source term:
\beq\label{source}
{\cal S}_M(r,t) \propto (t + \Delta t)^{-5/3} \, \delta(r-r_0) 
\eeq
where $\Delta t$ ensures the feeding rate is finite as $t \rightarrow 0$, and was taken equal to one code time step. This model assumes that the matter is fed into the disc at the rate at which disrupted stellar material returns to pericentre (the so-called `fall-back' rate $\dot M_{\rm fb} \sim t^{-5/3}$, Rees 1988). 

As in our earlier work (Mummery \& Balbus 2020a, b) we model the dynamical disc stress with a simple constant profile, $\W = w = $ constant. The reason for this simplification is three-fold. Firstly, we demonstrated in earlier work (Mummery \& Balbus 2019a) that the exact turbulent stress specification only  effects the evolution of the accretion disc in a very weak manner. Secondly, this simplification makes the governing evolution equation linear which greatly reduces the computational complexity of fitting individual TDE X-ray and UV light curves. Thirdly, this model is also physically rather natural: it would emerge, for example, if the turbulent velocity fluctuations follow a Keplerian scaling of the form $\delta v^r \sim \delta v^\phi \sim r^{-1/2}$. Finally, the stress at the ISCO is set to a small but non-zero value, $\gamma = 1$ in the notation of (Mummery \& Balbus 2019b). While TDEs represent a promising observational path to probe the properties of the dynamical disc stress at the inner edge of relativistic accretion discs,  it is not the purpose of this present work to perform a best-fit analysis of the ISCO stress of any particular TDE. Furthermore, the  ISCO stress is very much a second order parameter in the unified TDE model (Papers I, II, III), and so this assumption should not meaningfully effect the results.

\subsection{Intrinsic disc emission}
In previous work (Mummery \& Balbus 2020a, b) we have assumed that the disc was able to fully thermalise the liberated free energy, and that the intrinsic disc emission was therefore purely thermal.  In this work we relax this assumption. We model radiative transfer effects resulting from the incomplete thermalisation of the liberated disc energy with a colour-corrected intrinsic disc spectrum with temperature-dependent colour correction factor $f_{\rm col}(T)$. The intrinsic specific intensity of the disc is given by
\beq
I_\nu(\nu, T) = f_{\rm col}^{-4} \, B_\nu\left(\nu, f_{\rm col} T\right),
\eeq
where every disc annulus will have a temperature $T(r,t)$ resulting from solving the underlying disc equations, and $B_\nu$ is the Planck function 
\beq
 B_\nu(\nu,T) = \frac{2h\nu^3}{c^2} \left[ \exp\left( \frac{h\nu}{k_B T} \right) - 1\right]^{-1} .
\eeq
The colour correction factor $f_{\rm col}$ is a function of temperature, and we use the piece-wise continuous formulation of Done {\it et al}. (2012), explicitly: 
\begin{align}\label{col1}
&f_{\rm col}(T) =  \left(\frac{72\, {\rm keV}}{k_B T}\right)^{1/9}, \quad  T > 1\times10^5 {\rm K}. \\
&f_{\rm col}(T) =  \left(\frac{T}{3\times10^4 {\rm K}} \right)^{0.83}, \, 3\times10^4 {\rm K} < T < 1\times10^5 {\rm K} .\label{col2} \\
&f_{\rm col}(T) = 1, \quad T < 3\times 10^4 {\rm K} . \label{col3}
\end{align}
Once the intrinsic specific intensity of the disc is calculated, the evolving disc light curves are calculated in one of two ways. If the X-ray spectrum of the modelled TDE is observed to be quasi-thermal then we calculate the disc light curves using a relativistic ray tracing calculation (Appendix A, Mummery \& Balbus 2020a). If, on the other hand, the X-ray spectrum is observed to be nonthermal, then the disc light curves are  calculated using the Compton scattering model of Paper II.
 
\subsection{Fitting parameters}
All TDEs modelled in this paper are assumed to be around Schwarzschild black holes, and are further assumed to be observed at an inclination angle of $\theta_{\rm obs} = 30^\circ$.  This obviously represents a simplification, and we reiterate that it is not the intention of this work to provide a definitive study of any individual source. Rather, the purpose of this work is to demonstrate that a wide variety of observed TDE properties can be reproduced by simple disc models with  different disc and black hole parameters. TDEs observed to have a thermal X-ray spectrum are therefore described by 3 fitting parameters: the black hole mass $M$, the total accreted mass  $M_{\rm acc}$ (a normalisation on the source term eq. \ref{source}), and the viscous timescale of the evolving disc $t_{\rm visc}$. We compute $M_{\rm acc}$ in the following manner
\beq
M_{\rm acc} \equiv \int_0^\infty \dot M(r_I, t) \, {\rm d} t,
\eeq 
where $\dot M(r_I,t)$ is the ISCO mass accretion rate. 
Given the simplifications applied in the modelling (the fixed black hole spin, observer inclination angle and ISCO stress) the  best fit parameter values and their associated uncertainties should be treated with some caution, although our previous work found only a moderate change (factor of $\sim 1.5$) in best fit black hole mass over a wide range of black hole spins (Mummery \& Balbus 2020a). 

For those TDEs which are observed to have a non-thermal X-ray spectrum, additional fitting parameters which describe the properties of the electron scattering corona are required. We use the compact electron scattering corona model of Paper II to describe the non-thermal X-ray light curves, to which the reader should refer for more details. The parameters which describe the corona are: $R_{\rm Cor}$, the radial extent of the corona; $f_{SC}$, the fraction of the disc photons emitted at disc radii $R < R_{\rm Cor}$ which are Compton scattered; and $\Gamma$, the photon index of the resulting X-ray spectrum.  The photon index $\Gamma$ is directly observable for a given TDE, and is therefore not treated as a fitting parameter, but is fixed to the observed value. 

In keeping with previous TDE studies (Van Velzen {\it et al}. 2019, Mummery \& Balbus 2020a, b) an additional light curve component is required at early times to model the observed TDE light curves at UV energies, prior to the light curves transitioning to a disc-dominated state. This transition typically occurs at times $t \sim 100-300$ days after the initial disruption.  As in our previous work we model this early time component with an exponentially declining profile 
\beq
F_{\rm exp} = A\, \exp(-t/\tau) ,
\eeq
each UV band with its own $A$ and $\tau$. Empirical correlations between the phenomenological parameters $A$ and $\tau$ and physical disc and black hole parameters will be presented in section \ref{cor_sec}. 

Finally, we determine the best fit system parameters by simultaneously minimising the chi-squared statistic of the different (UV and X-ray, where appropriate) light curves of each TDE. As in our previous work we anticipate to find formally large reduced $\chi^2$ values. These large values result as a consequence of short time-scale fluctuations present in the well-sampled TDE light curves, and is to be expected in any theoretical model using a smooth functional form for the turbulent stress tensor $\W$. This very standard approach implicitly averages over rapid turbulent variations. Short time-scale fluctuations are likely to be highly correlated so that accurately assessing the statistical significance of the fit is not straightforward. We have therefore used $\chi^2$ minimisation as a sensible guide towards finding a best fit, but as $\chi^2$ does not have a Gaussian normal distribution, we have not attempted a quantitative assessment of fit, leaving the plots to speak for themselves. 

In this work we model 6 different TDEs (Table \ref{table1}), two of which (AT2019dsg \& ASASSN-15oi) have bright quasi-thermal X-ray spectra and simultaneous UV observations, two more (ASASSN-14ae \& ASASSN-18pg) were undetected at X-ray energies at early times, but have well-sampled UV light curves, and two more (XMMSL1 J0740 \& XMMSL2 J1446) which have bright non-thermal X-ray light curves and simultaneous UV observations. 

\begin{table}
\renewcommand{\arraystretch}{2}
\centering
\begin{tabular}{|p{2.2cm}|p{2.8cm}|}
\hline
TDE name & X-ray spectral properties  \\ \hline\hline
AT2019dsg & Quasi-thermal  \\ \hline
ASASSN-15oi & Quasi-thermal  \\ \hline
ASASSN-14ae & No detections  \\ \hline
ASASSN-18pg & No detections  \\ \hline
XMMSL1 J0740 & Non-thermal  \\ \hline
XMMSL2 J1446 & Non-thermal  \\ \hline
\end{tabular}
\caption{The  X-ray spectral types of the TDEs studied in this work. The best fitting parameters of all sources are collated in Table \ref{table_appendix}. }
\label{table1}
\end{table}

\section{Bright quasi-thermal X-ray TDEs }
In this section we examine two TDEs which are X-ray bright with observed X-ray spectra well described by quasi-thermal (blackbody) profiles. The two TDEs we have chosen (AT2019dsg and ASASSN-15oi) are both intrinsically interesting in their own right.  Firstly, they represent two temporal extremes of the observed quasi-thermal X-ray TDE population. The X-ray light curve of AT2019dsg  evolves extremely rapidly, dropping in amplitude by over an order of magnitude in a matter of a month (van Velzen {\it et al}. 2020).  Conversely, ASASSN-15oi is potentially the slowest evolving X-ray TDE, reaching the peak of its X-ray light curve $\sim 300$ days after first detection (Gezari {\it et al}. 2017; Holoien {\it et al}. 2018). In addition, AT2019dsg may represent the astrophysical origin of a reported neutrino detection (Stein {\it et al}. 2020). The best fit light curves of both TDEs are presented below. 

\subsection{AT2019dsg}

\begin{figure}
  \includegraphics[width=.5\textwidth]{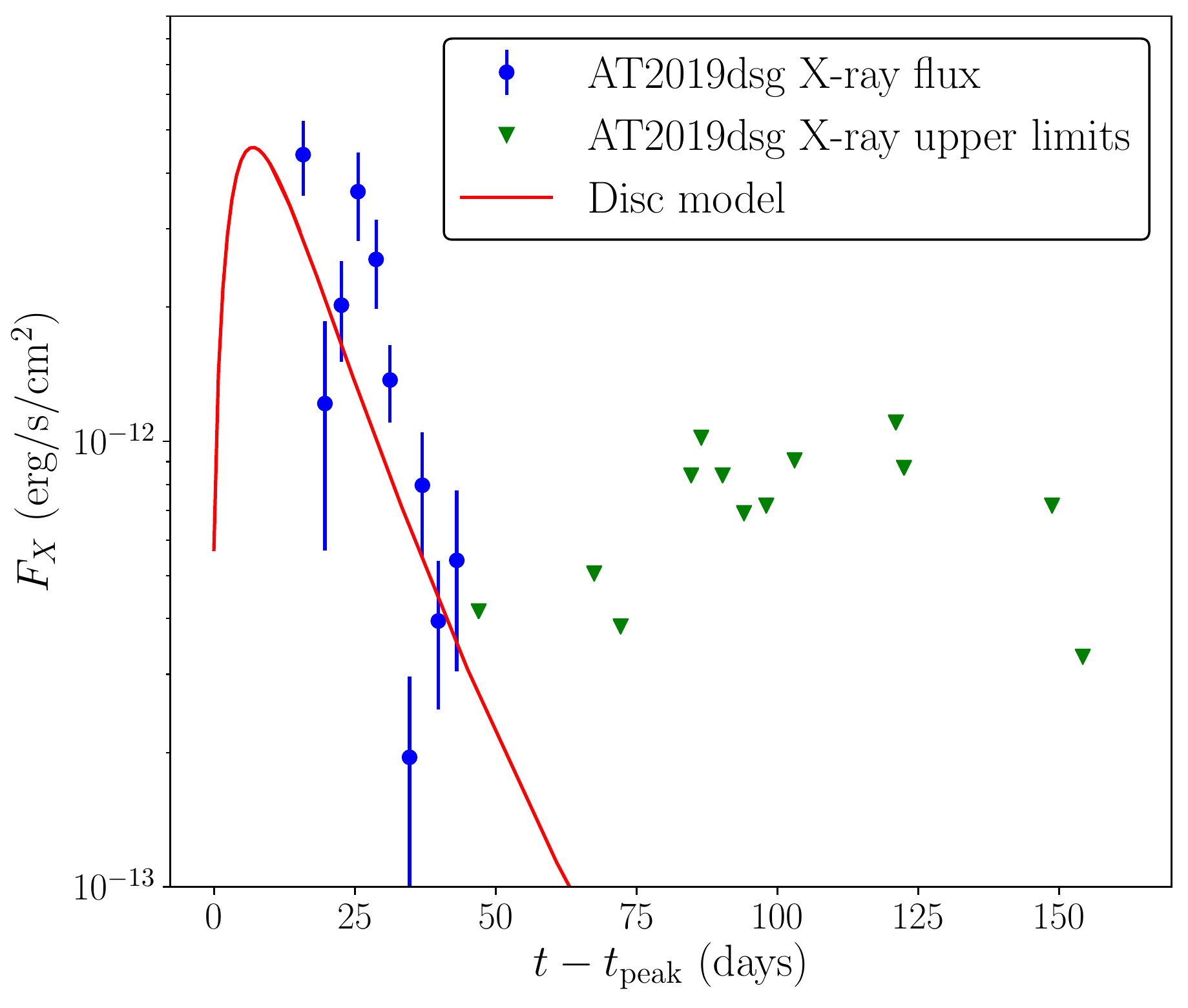} 
 \caption{The best fitting disc model (red curve), fit to the observed X-ray light curve of AT2019dsg (blue points, van Velzen {\it et al}. 2019). X-ray upper detection limits are denoted by green diamonds, and are satisfied by the disc model.   } 
 \label{X19dsg}
\end{figure}

\begin{figure}
  \includegraphics[width=.5\textwidth]{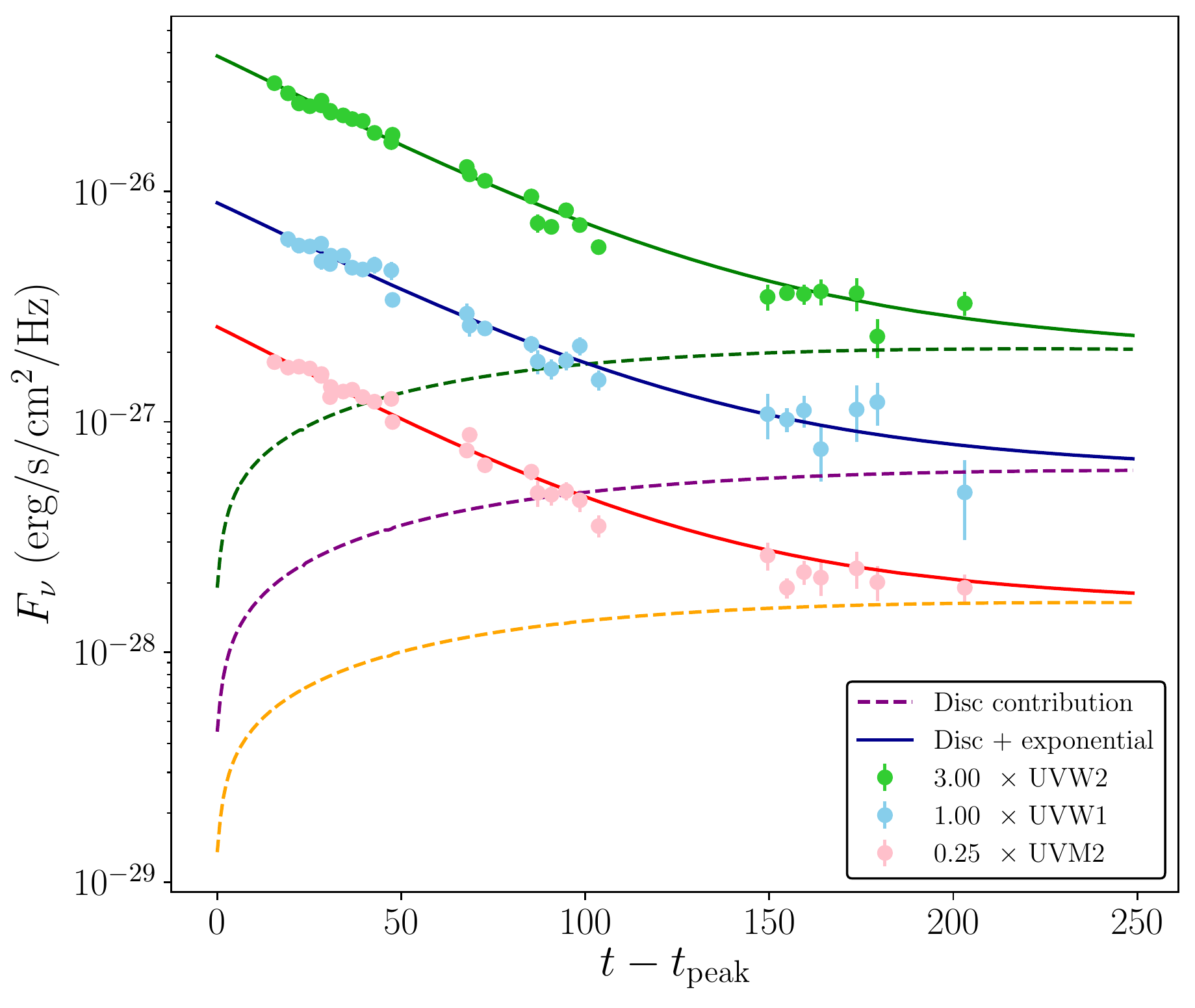} 
 \caption{The best fitting  disc light curves at three different UV frequencies (dashed curves), fit to the observed UV light curves of AT2019dsg (solid points). The total UV model light curve, which includes an exponentially declining component relevant at early times, are denoted by solid curves. The different UV light curves are offset for readability. } 
 \label{UV19dsg}
\end{figure}

The TDE AT2019dsg was first reported by van Velzen {\it et al}. (2020). AT2019dsg has well sampled UV light curves, spanning $\sim 200$ days (Figure \ref{UV19dsg}), and was initially well observed at X-ray energies, before fading to unobservable levels over a 50 day period (van Velzen {\it et al.} 2020, displayed in Figure \ref{X19dsg}). The X-ray spectrum of AT2019dsg is well described by quasi-thermal emission.  As was mentioned above, AT2019dsg may be the astrophysical origin of a reported neutrino detection (Stein {\it et al}. 2020). AT2019dsg is associated with radio emission (Stein {\it et al}. 2020), possibly indicating the presence of a jet. 

AT2019dsg is located at a luminosity distance  $d_L = 226$ Mpc (van Velzen {\it et al}. 2020), and has a mean central black hole mass, inferred from galactic scaling relationships, of $\left\langle M \right\rangle = 2.4^{+2.9}_{-1.5}\times10^7M_\odot$ (Paper I).   Due to the rapid X-ray evolution and the large anticipated black hole mass, we chose a relativistic feeding radius  $r_0 = 10 r_g = 1.67 r_I$.

Table \ref{table19dsg} summarises the best-fitting disc and black hole parameters for the recorded observations of AT2019dsg. The best fit black hole mass, $M = 2.0 \times 10^7 M_\odot$, is consistent with the results of galactic scaling relationships. Fig. \ref{X19dsg} shows the observed AT2019dsg 0.3--10 keV X-ray flux (van Velzen {\it et al}. 2020, blue points), together with the best-fitting evolving X-ray flux from our fiducial disc model (red curve). While the best fit disc model is consistent with all of the observed X-ray upper limits of AT2019dsg, the best fitting disc model was found by fitting to the observed X-ray fluxes only.  Figure \ref{UV19dsg} shows the evolving AT2019dsg UV fluxes (solid points, van Velzen {\it et al}. 2020), the best fitting UV light curves of the disc (dashed curves), and a combined disc and exponentially decaying light curve (solid curves).  The time in Figures \ref{X19dsg} \& \ref{UV19dsg} is measured relative to the time at which the optical emission of this TDE was observed to peak. 

As is clear from Figs. \ref{X19dsg} \& \ref{UV19dsg}, both the evolving X-ray and UV flux of AT2019dsg are well described by an evolving relativistic disc model. The best fit chi-squared value for the X-ray light curve is $\chi^2 = 4.69$, while for the UVW1 light curve: $\chi^2 =1.13$, UVW2: $\chi^2 =1.32$, UVM2: $\chi^2 = 1.67$, and the combined light curves: $\chi^2 =1.80$. The main source of model-data discrepancy results from large-amplitude short-timescale fluctuations in the X-ray data which are not captured by a smoothly decaying model light curve.  

\begin{table}
\renewcommand{\arraystretch}{2}
\centering
\begin{tabular}{|p{2.2cm}|p{2.8cm}|}
\hline
Parameter & Value  \\ \hline\hline
$M/M_\odot$ & $2.0^{+0.3}_{-0.5} \times 10^7$  \\ \hline
 $M_{\rm acc}/M_\odot$ & $0.09^{+0.02}_{-0.03} $  \\ \hline
$t_{\rm visc}$ (days) & $28.1_{-1.4}^{+0.6}$  \\ \hline
\end{tabular}
\caption{Best fit parameters for the X-ray (Fig. \ref{X19dsg}) and UV (Fig. \ref{UV19dsg}) light curves of  AT2019dsg. Reproduced and collated in Table \ref{table_appendix}. }
\label{table19dsg}
\end{table}

\begin{figure}
  \includegraphics[width=.5\textwidth]{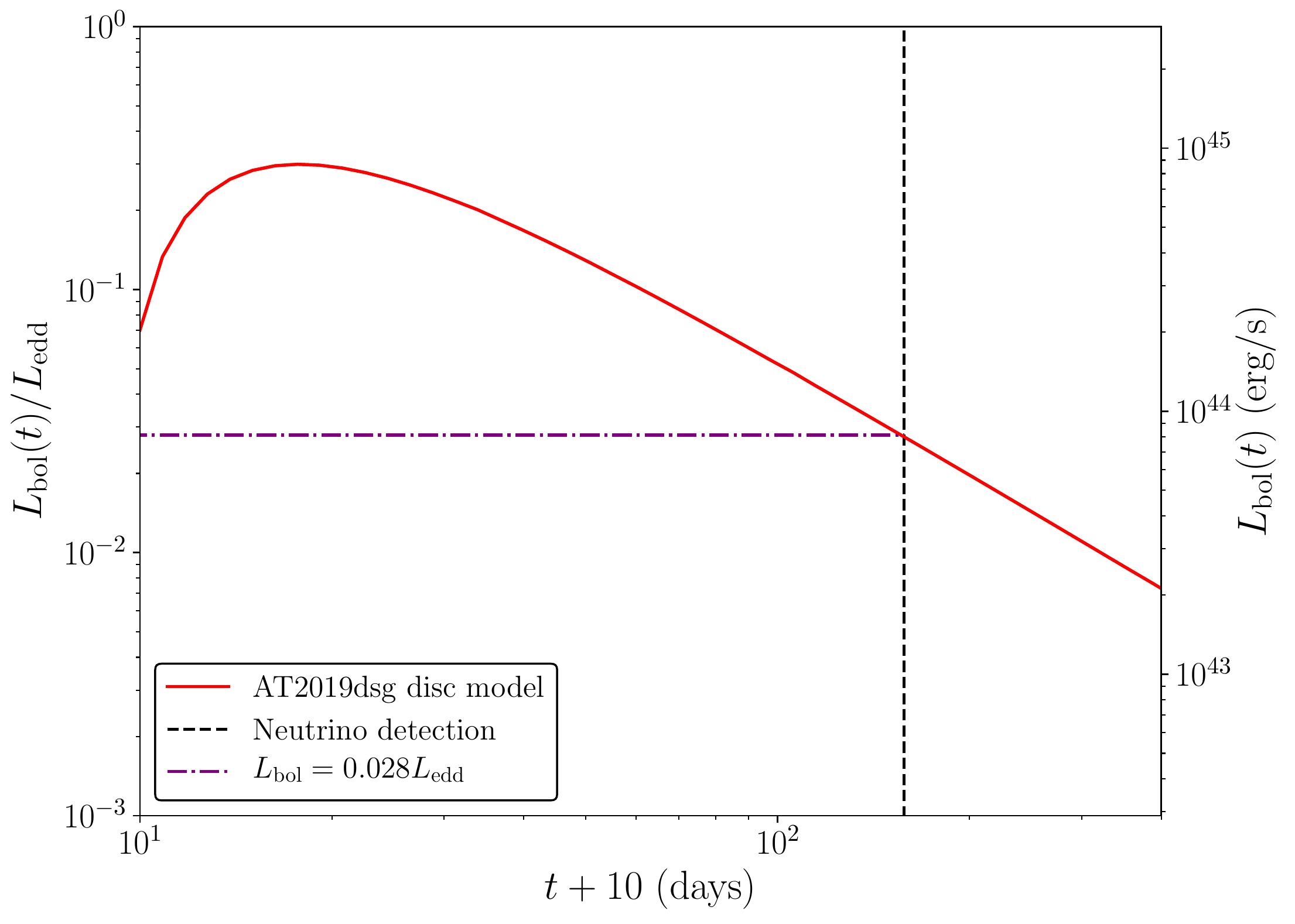} 
 \caption{The evolving bolometric luminosity of the best-fitting AT2019dsg disc model (red curve). Displayed in Eddington units (left scale), and physical units (right scale). The time corresponding to a reported neutrino detection (Stein {\it et al}. 2020) is displayed by a vertical dashed line, at which point the disc luminosity was $L \simeq 0.03 L_{\rm edd}$. } 
 \label{Bol19dsg}
\end{figure}

AT2019dsg is associated with both radio emission and a potential neutrino detection (Stein {\it et al}. 2020), implying the potential presence of a jet. The bolometric luminosity of the best fitting AT2019dsg disc model is displayed in Figure \ref{Bol19dsg} in both Eddington units (left scale), and physical units (right scale).  Our best fit bolometric disc light curve is bright ($L_{\rm bol, peak} \simeq 7 \times 10^{44}$ erg/s), but sub-Eddington $l \equiv L_{\rm bol}/L_{\rm edd} \simeq 0.3$ at peak. Unless the AT2019dsg X-ray emission was significantly brighter prior to the initial X-ray detection (the first X-ray measurement was taken $\sim 14$ days post the optical peak of the TDE, and $\sim 40$ days post the first optical detection), which could potential modify our best fitting parameters, it seems unlikely that the AT2019dsg accretion disc was accreting at super-Eddington rates in the earliest stages of its evolution. What is potentially interesting is that, at the time of the reported neutrino detection, the bolometric luminosity of the best fitting AT2019dsg disc model was at the level typically associated with disc state transitions in X-ray binaries $L_{\rm bol} \simeq 0.03 L_{\rm edd}$ (Maccarone 2003).

It is possible that the detected neutrino is associated with jet launching from the accretion disc as it transitions into a harder accretion state, a result of its low Eddington ratio at this time. This hypothesis could be tested by further  late time X-ray observations of AT2019dsg, where the detection of nonthermal (hard state) X-ray emission could confirm the transition of the AT2019dsg accretion disc into the hard accretion state. It is important to stress that while this late-time transition to the hard accretion state could potentially explain the origin of the neutrino, it is unable to explain the early-time (pre-transition) radio emission. 


\subsection{ASASSN-15oi}

\begin{figure}
  \includegraphics[width=.5\textwidth]{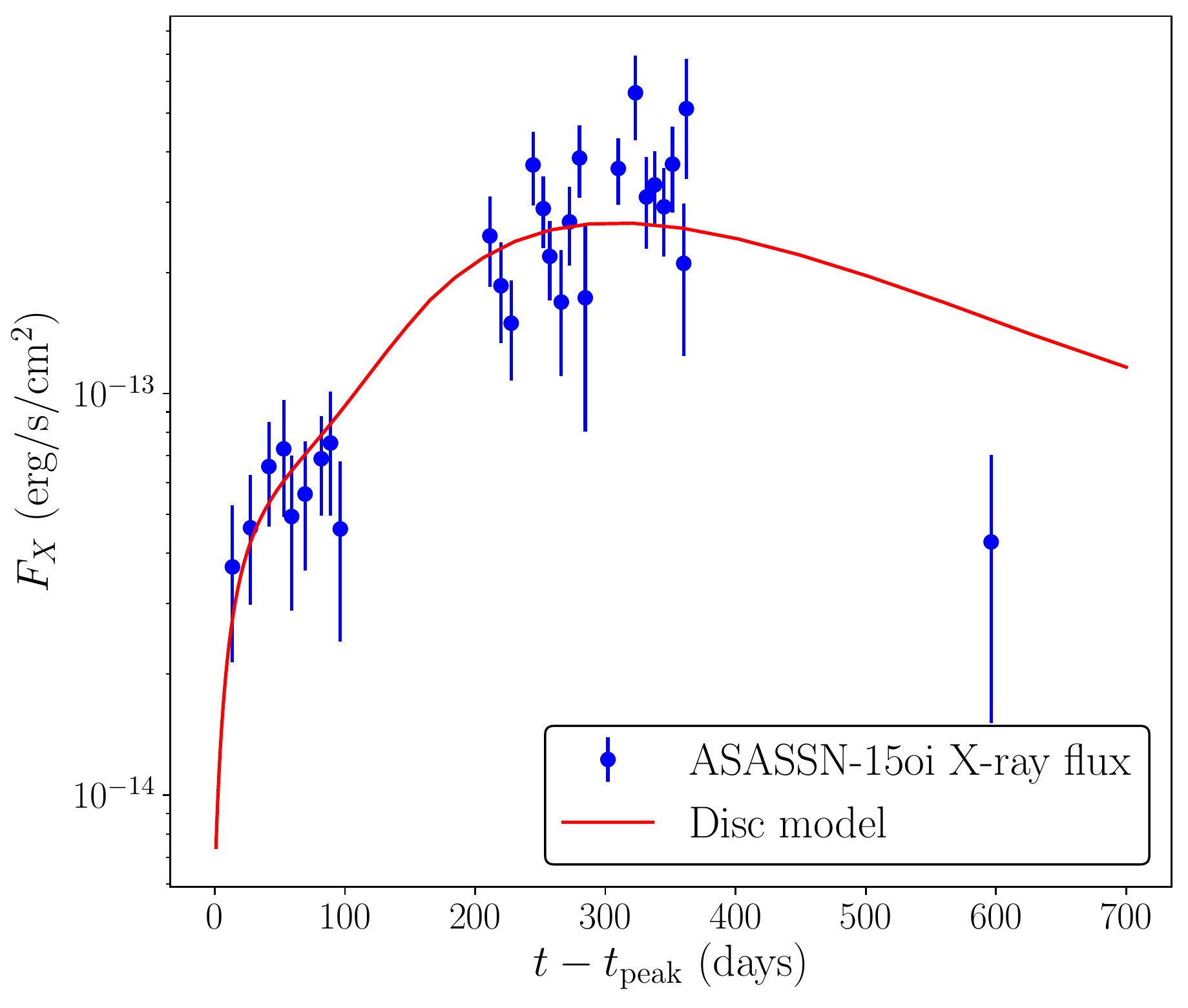} 
 \caption{The best fitting disc model (red curve), fit to the observed X-ray light curve of ASASSN-15oi (blue points, Holoien {\it et al}. 2015, Gezari {\it et al}. 2017).  } 
 \label{X15oi}
\end{figure}

\begin{figure}
  \includegraphics[width=.5\textwidth]{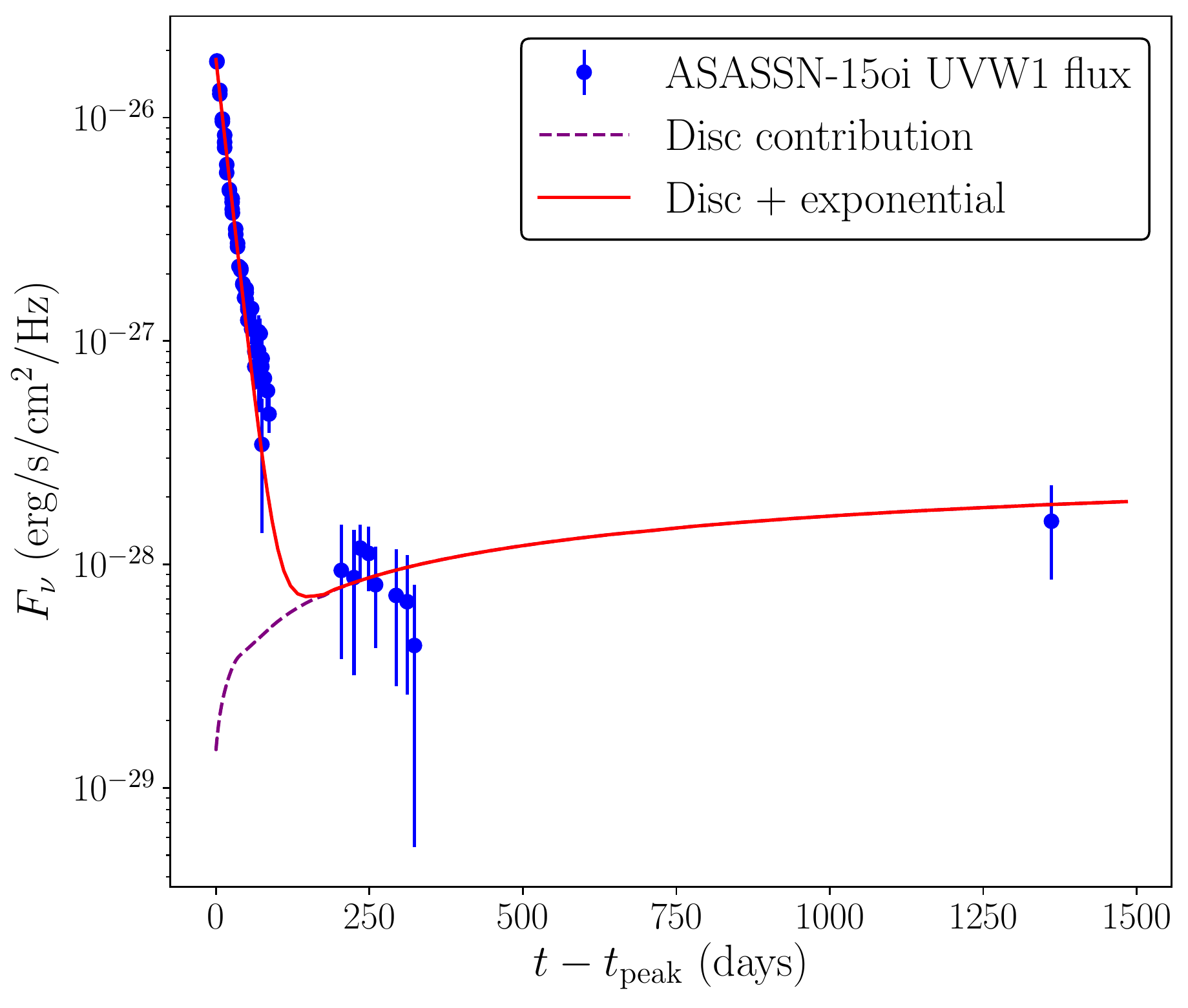} 
 \caption{The best fitting  disc light curve at the UVW1 frequency (dashed curves), fit to the observed UV light curves of ASASSN-15oi (solid points). The total UV model light curve, which includes an exponentially declining component relevant at early times, is denoted by a solid curve. } 
 \label{UV15oi}
\end{figure}

The TDE ASASSN-15oi was first reported by Holoien  {\it et al}. (2016b), with additional late time observations by Gezari {\it et al}. (2017) \& Holoien {\it et al}. (2018). ASASSN-15oi has well sampled UV light curves, spanning $\sim 1300$ days (Fig. \ref{UV15oi}), and has been well observed at X-ray energies (Fig \ref{X15oi}). The X-ray spectrum of ASASSN-15oi is well described by quasi-thermal emission (Holoien {\it et al}. 2016a).  

ASASSN-15oi is located at a luminosity distance $d_L = 216$ Mpc (Holoien {\it et al}. 2016b), and has a mean central black hole mass, inferred from galactic scaling relationships, of $\left\langle M \right\rangle = 8.1^{+7.1}_{-4.1} \times10^6 M_\odot$ (Paper I).

Table \ref{table15oi} summarises the best-fitting disc and black hole parameters for the recorded observations of ASASSN-15oi. Using an identical feeding radius ($10r_g$) as AT2019dsg produced a poor fit to the observations, as it was unable to accurately reproduce the X-ray rise to peak.  We found that  increasing the feeding radius, to $r_0 = 20 r_g$, lead to a significantly improved fit. 
 The best fit black hole mass, $M = 9 \times 10^6 M_\odot$, is consistent with the results of galactic scaling relationships. Fig. \ref{X15oi} shows the observed ASASSN-15oi 0.3--10 keV X-ray flux (blue points, Gezari {\it et al}. 2017), together with the  evolving X-ray flux of the best-fit disc model (red curve). Figure \ref{UV15oi} shows the evolving ASASSN-15oi UVW1 flux (blue points, Holoien {\it et al}. 2018, van Velzen {\it et al}. 2019), the best fitting UV light curve of the disc (purple dashed curve), and a combined disc and exponentially decaying light curve (red solid curve).  The UVW1 light curve was the best sampled of all of the ASASSN-15oi UV light curves. 

As is clear from Figs. \ref{X15oi} \& \ref{UV15oi}, both the evolving X-ray and UV flux of ASASSN-15oi are well described by an evolving relativistic disc model. The best fit chi-squared for the X-ray light curve is: $\chi^2 = 1.76$, for the UVW1 light curve: $\chi^2 = 4.83$, and the combined light curves: $\chi^2 = 3.92$. The main source of model-data discrepancy results from  the UVW1 light curve remaining brighter than predicted at around day $\sim 100$.  This could either result from a UV flare, or the initial UV light curve decaying slightly slower than exponentially. As the early time UV emission does not result from the accretion disc, we do not believe that the formally large UV $\chi^2$ statistic is a cause for concern. The observed UV light curve in the disc dominated regime is clearly well reproduced by our relativistic disc model. 

\begin{table}
\renewcommand{\arraystretch}{2}
\centering
\begin{tabular}{|p{2.2cm}|p{2.8cm}|}
\hline
Parameter & Value  \\ \hline\hline
$M/M_\odot$ & $9.0^{+2.0}_{-2.0} \times 10^6$  \\ \hline
 $M_{\rm acc}/M_\odot$ & $0.31_{-0.09}^{+0.08} $  \\ \hline
$t_{\rm visc}$ (days) & $600^{+95}_{-140}$  \\ \hline
\end{tabular}
\caption{Best fit parameters for the X-ray (Fig. \ref{X15oi}) and UV (Fig. \ref{UV15oi}) light curves of  ASASSN-15oi. Reproduced and collated in Table \ref{table_appendix}.}
\label{table15oi}
\end{table}

ASASSN-15oi has been previously modelled as an evolving disc by Wen {\it et al}. (2020). Wen {\it et al}. put forward an alternative, slim disc, interpretation of ASASSN-15oi. In their analysis Wen {\it et al}. modelled ASASSN-15oi as a disc observed near edge-on, which, as it cooled and `slimmed' (decreasing $H/R$)  with time, revealed more of the innermost disc regions to the observer. The X-ray flux correspondingly increased as more of the inner disc became observable. 

In our modelling  the delayed peak of the ASASSN-15oi X-ray light curve simply results from the large viscous timescale of the best fit disc solution. The peak of the X-ray light curve only occurs once the temperature in the innermost  disc regions reaches its maximum value (Mummery \& Balbus 2020a). Despite the ASASSN-15oi stellar debris  being fed into a marginally relativistic disc radius ($r_0 = 20 r_g$),  the  large best fit viscous timescale ($t_{\rm visc} \simeq 600$ days) results in a substantial period of time elapsing before the majority of the disc material has propagated into the near-ISCO regions of the disc. 

A key difference between our results and the results of Wen {\it et al}. is that our models is able to reproduce the observed late-time UV plateau of ASASSN-15oi.  Wen {\it et al}. found a much lower best fit black hole mass ($M = 4.0^{+2.5}_{-3.1} \times 10^6 M_\odot$), inconsistent with our results. If we were to assume a black hole mass of $M = 4 \times 10^6 M_\odot$ we could, by primarily modifying the best fit accreted mass, reproduce the evolving X-ray light curve of ASASSN-15oi. However, at this low black hole mass we would be unable to reproduce the observed UVW1 plateau flux, as this disc solution would be too dim at UV frequencies. There is strong observational evidence that the late time UV flux of many TDEs stems from an evolving accretion disc (van Velzen {\it et al}. 2019, Mummery \& Balbus 2020a,b), and so the amplitude of the late time UV plateau flux is an extremely important observational constraint on TDE models.

\subsection{Comparison of the two sources}\label{compare}
AT2019dsg and ASASSN-15oi  have very different observed X-ray light curves. This is potentially surprising, particularly as the  best fit black hole and accreted masses of the two sources only differ by a factor $\sim 3$. However, the key difference between the two sources is the magnitude of their respective viscous timescales, or equivalently,  the large differences between their effective $\alpha$-values.

As we described in section \ref{method}, we do not use the canonical $\alpha$-viscosity model when modelling TDE light curves, using a simpler `constant $w$' model. We can however use $\alpha$-type scaling relationships to relate effective $\alpha$ parameters of the two different TDEs. By assuming that the ratio of the orbital to viscous timescales of the two sources are described by the usual $\alpha$-parameter scaling law:
\beq
\left({ t_{\rm visc, 15oi} \over t_{\rm orb, 15oi} }\right) \left({t_{\rm orb, 19dsg} \over t_{\rm visc, 19dsg} }\right) =  \left({\alpha_{\rm 19dsg} \over \alpha_{\rm 15oi}}\right) \left({(H/R)_{\rm 19dsg} \over (H/R)_{\rm 15oi}}\right)^2,
\eeq
we can quantify the difference between the two sources effective $\alpha$-values. The left hand side of the above expression can be evaluated explicitly. Using the orbital timescales evaluated at the feeding radius of both sources, we have: 
\beq
 \left({ t_{\rm visc, 15oi} \over t_{\rm orb, 15oi} }\right) \left({t_{\rm orb, 19dsg} \over t_{\rm visc, 19dsg} }\right)  = 16.6. 
\eeq
If we further assume that the disc aspect ratio scales like the ratio of the  sound and orbital velocities $H/R = c_S/v_K$, with the sound speed scaling proportional to the square root of the disc temperature $c_S \propto \sqrt{T}$, then the usual orbital velocity relationship $v_K \propto \sqrt{GM/r_0}$ leaves
\beq
\left({(H/R)_{\rm 19dsg} \over (H/R)_{\rm 15oi}}\right)^2 = {1 \over 2} \left( { T_{\rm 19dsg} \over T_{\rm 15oi} } \right)  .
\eeq
Here the factor of $1/2$ results from the different feeding radii of the two discs. The temperature scales are measurable quantities. We use the peak temperature reached at the disc feeding radius during the evolution of the two TDEs, which evaluates explicitly to ${ T_{\rm 19dsg} / T_{\rm 15oi} } = 1.38$, to determine that
\beq
{\alpha_{\rm 19dsg} \over \alpha_{\rm 15oi}} = 24.1. 
\eeq
This large ratio of effective $\alpha$-values both explains the very different viscous timescales of the two sources, but also why AT2019dsg was able to be observed at X-ray energies around such a large mass black hole. 

{While the values of the two effective $\alpha$ parameters are themselves not unusual (for $(H/R)_{\rm 15oi}\sim 0.1$, we have $\alpha_{\rm 15oi} \sim 0.01$, $\alpha_{\rm 19dsg} \sim 0.2$), the variance between sources is unexpected, and as of yet unexplained. It is of course important to remember, as discussed at the start of this section, that we chose ASASSN-15oi and AT2019dsg as they represented temporal extremes of the TDE X-ray light curve population. Discovering  parameter values at the extremes of expected values for events chosen for their extreme evolution is perhaps less surprising. 

Furthermore, angular momentum transport within a TDE disc is an extremely complex physical problem, and it may well simply be that the effective $\alpha$ values of a population of TDE sources will vary by $\sim$ an order of magnitude. Crucially, both TDE sources (as well as all six sources modelled in this paper, Table \ref{table_appendix}) have viscosity parameters ${\cal V} \equiv t_{\rm orb}/t_{\rm visc} \sim 10^{-3}-10^{-4}$. Low values of ${\cal V}$ mean that the disc evolution in both cases is well into the diffusive regime, where $\alpha$-type modelling will well describe the disc evolution. }

\section{Dim quasi-thermal X-ray TDEs}

A large population of TDEs are observed to be optically bright, but X-ray dim (see e.g., Wevers {\it et al}. 2019, and references therein). As was argued above, in our TDE unification scheme there are two distinct physical mechanisms which can result in a TDE being only observed at optical and UV frequencies. The first, only expected to be relevant for TDEs around black holes of the lowest masses, results from the obscuration of the innermost X-ray-producing disc regions by a large-mass radiatively-driven outflow. The second, expected to be important for larger black hole masses, occurs when the peak disc temperatures reached in the disc evolution are insufficiently hot to produce observable levels of X-ray emission.

Detailed calculations of the evolving optical light curves of the first class of optical TDEs -- those light curves produced by the reprocessing of the X-ray emission of an obscured disc by a large mass outflow -- lie beyond the scope of the current TDE disc model. However, the evolving UV light curves of those TDEs which reach insufficiently hot temperatures to be observed at X-ray frequencies can be well modelled by our evolving disc solutions.  

In this section we model two TDEs -- ASASSN-14ae and ASASSN-18pg -- which are X-ray dim while being bright UV and optical sources. Crucially, both of these sources exhibit UV light curves which are characteristic of disc-dominated TDEs. After an initial period ($\sim 100-200$ days) of very rapidly decreasing UV flux, both sources transition into a disc-dominated UV plateau, characterised by a near-constant UV flux (Figs \ref{UV18pg}, \ref{UV14ae}). We therefore believe it is likely that these two sources belong to the second class of optical-bright TDEs, those which are insufficiently hot to produce observable X-ray emission, and can be modelled by our evolving disc solutions. The key question is: can the UV plateau flux of these sources be reproduced by relativistic disc models which respect the observed X-ray upper-limits of each source? If so, do the physical parameters of these solutions (disc and black hole mass, peak Eddington ratio) make sense?

There is naturally intrinsic difficulty in obtaining best-fit disc and black hole parameters for TDEs without X-ray detections, as the model parameters are strongly constrained by the simultaneous fitting of both UV and X-ray light curves. This is particularly true for the viscous timescale of the disc, as the UV light curves (the only available light curves in these cases) are effectively time-independent in the disc dominated regime.  In contrast, the black hole mass sets the overarching flux scales of both the X-ray and UV light curves. Although a best fit black hole mass cannot itself be reliably inferred from UV-only observations, a lower black hole mass limit can be inferred. At black hole masses below this limit, the disc mass required to reproduce the UV plateau flux would produce too bright an X-ray flux, exceeding the observed upper X-ray limits. 

Given the intrinsic uncertainties involved, we make two simplifying assumptions. Firstly, we assume that the X-ray emission of each TDE peaks exactly at its reported X-ray upper limit. This then allows us to determine the lower bound on acceptable black hole masses. Secondly, as the viscous timescale of the disc is extremely hard to constrain, we fix its value so that it satisfies $t_{\rm visc} = 10^{4} t_{\rm orb} $. Here $t_{\rm orb}$ is the orbital timescale at the discs feeding radius. This means that the solutions are safely in the diffusive regime required for the $\alpha$-type modelling we have performed. Given the difficulty in properly assessing the significance of our model fits to UV only observations, we do not quote uncertainties on these fitted parameters, noting that a wide range of  disc and black hole parameters can reproduce the observed UV fluxes.

Finally, there is further difficulty in modelling TDEs undetected at X-ray energies due to the lack of available spectral information. We assume here that both TDEs have unobservable quasi-thermal X-ray spectra, i.e., are evolving in the soft state.

\subsection{ASASSN-18pg}
\begin{table}
\renewcommand{\arraystretch}{2}
\centering
\begin{tabular}{|p{2.2cm}|p{2.8cm}|}
\hline
Parameter & Value  \\ \hline\hline
$M/M_\odot$ & $3.1  \times 10^7$  \\ \hline
 $M_{\rm acc}/M_\odot$ & $0.22 $  \\ \hline
$t_{\rm visc}$ (days) & $560$  \\ \hline
\end{tabular}
\caption{System parameters used to reproduce the four ASASSN-18pg UV light curves (Figure \ref{UV18pg}) while satisfying the X-ray flux upper detection limits reported in Holoien {\it et al}. (2020). Reproduced and collated in Table \ref{table_appendix}.}
\label{table18pg}
\end{table}
The TDE ASASSN-18pg is a recently reported TDE (Leloudas {\it et al.} (2019), Holoien  {\it et al}. (2020)). ASASSN-18pg has well sampled UV light curves, spanning $\sim 450$ days, but has only two marginal (2$\sigma$ above background) detections at X-ray energies (Holoien {\it et al}. 2020), with many other X-ray observations resulting in only upper-limits (Leloudas {\it et al}. 2019, Miller \& Cenko 2018). According to Holoien {\it et al}. (2020), given the large number (54 epochs) of observations, they would expect 1.23 spurious 2$\sigma$ detections, with a roughly 35\% chance of detecting two 2$\sigma$ detections. In this work we therefore assume that the two marginal  detections reported by Holoien {\it et al}. (2020) are not real detections of the ASASSN-18pg accretion disc. Numerous upper X-ray detection limits of ASASSN-18pg have been reported, with the most stringent being $F_X = 2.6 \times 10^{-14}$ erg/s/cm${}^2$ (Holoien {\it et al}. 2020). This upper limit corresponds to a stacked upper limit, combining all 54 epochs of X-ray observations. While a stacked upper limit may underestimate the true maximum flux, particularly if the X-ray flux of ASASSN-18pg is rapidly evolving, we use this most stringent upper limit in our modelling.

\begin{figure}
  \includegraphics[width=.5\textwidth]{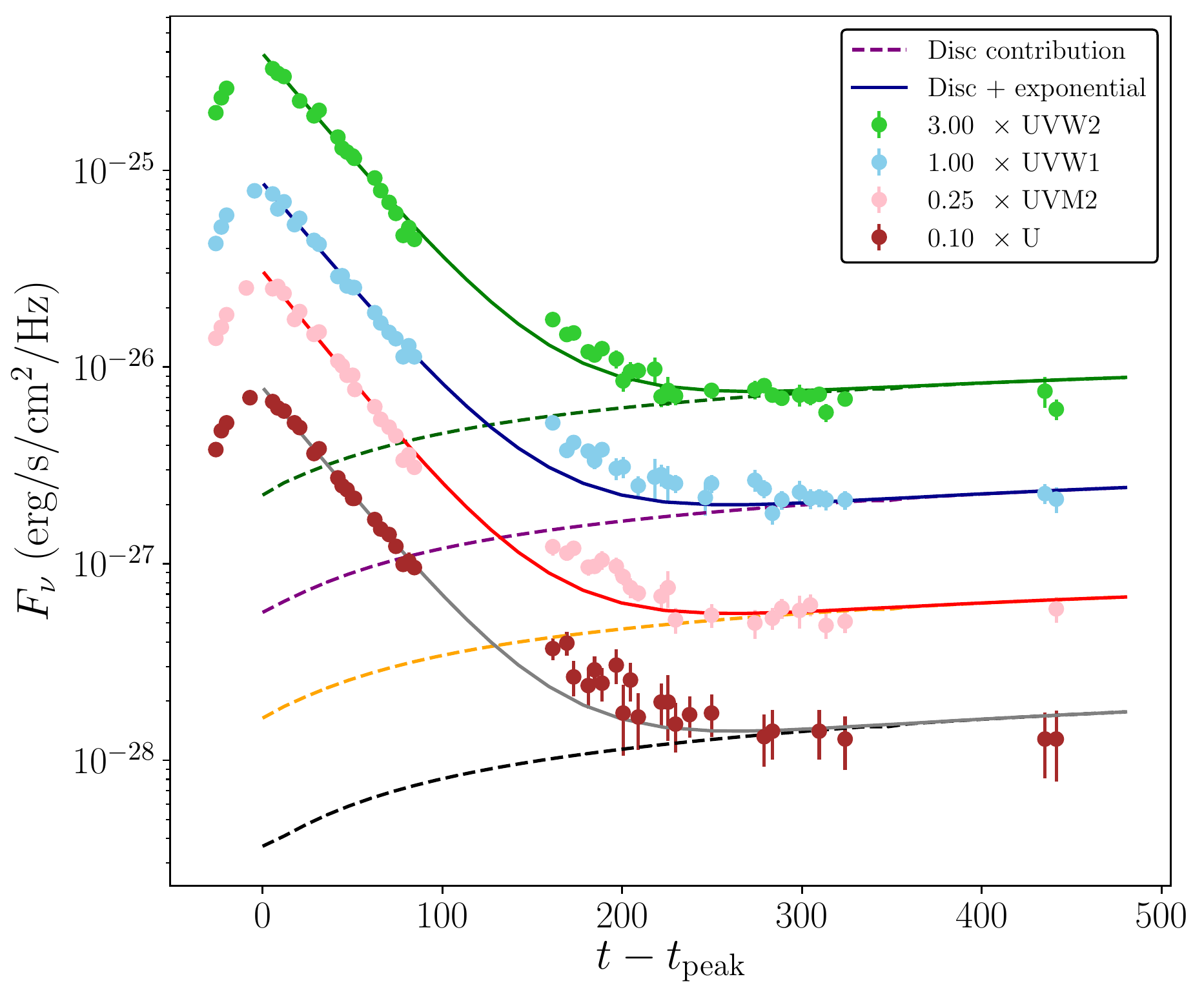} 
 \caption{The best fitting  disc light curves at four different UV frequencies (dashed curves), fit to the observed UV light curves of ASASSN-18pg (solid points). The total UV model light curve, which includes an exponentially declining component relevant at early times, are denoted by solid curves. The different UV light curves are offset for readability. This best-fitting disc model simultaneously satisfies an upper X-ray limit of ASASSN-18pg.  } 
 \label{UV18pg}
\end{figure}

ASASSN-18pg is a nearby TDE, located at a luminosity distance of $d_L = 78.6$ Mpc (Holoien {\it et al}. 2020). The central black hole mass has been estimated from galactic scaling relationships, with a measurement of the galactic bulge mass implying $M \sim 10^{7 \pm 0.4} M_\odot$, (Holoien {\it et al}. 2020), while the $M:\sigma$ relationship implies a lower mass $M = 3^{+5}_{-2}\times10^6 M_\odot$ Leloudas {\it et al.} (2019).  The UV light curves of ASASSN-18pg have been modelled using the MOSFiT software (Guillochon {\it et al}. 2018) by two groups, with conflicting best fit black hole masses: $M = 1.6^{+ 1.1}_{- 0.7} \times 10^7 M_\odot$ (Holoein {\it et al}. 2020), and $M = 4^{+5}_{-2}\times 10^6 M_\odot$ Leloudas {\it et al.} (2019).

 Figure \ref{UV18pg} shows the evolving ASASSN-18pg UV flux (solid points, Holoien {\it et al}. 2020), the UV light curves of the disc (dashed curves), and a combined disc and exponentially decaying light curve (solid curves).  For this set of system parameters the X-ray flux peaked at $F_{X, {\rm max}} = 2.6 \times 10^{-14}$, just below the stacked upper limit.   

Table \ref{table18pg} summarises the disc and black hole parameters used to reproduce the recorded observations of ASASSN-18pg (figure \ref{UV18pg}). The inferred black hole mass, $M = 3.1 \times 10^7 M_\odot$, clearly favours the two larger estimates of the black hole mass (Holoien {\it et al}. 2020), but is actually just above the uncertainties of the black hole masses inferred from both the galactic bulge mass and the Holoien {\it et al}. (2020) MOSFiT analysis. As we argued in the introduction to this section, this black hole mass value can be thought of as a lower black hole mass bound inferred from the ASASSN-18pg observations. The reason for this is that a combination of a larger black hole mass and smaller disc mass can reproduce the observed UV plateau, with an even lower peak X-ray flux.

If we force the black hole mass to be consistent with the upper limit from the Holoien {\it et al} (2020) MOSFiT analysis, $M = 2.7\times10^7M_\odot$ then the disc model produces an X-ray flux larger than the stacked X-ray upper-limit. Interestingly, the peak amplitude of the resulting disc X-ray emission lies exactly at the level of the two marginal detections reported in Holoien {\it et al}. (2020),  $F_{X, {\rm max}} = 1 \times 10^{-13}$ erg/s.

As is clear from Figure \ref{UV18pg},  the evolving UV flux of ASASSN-18pg can be well described by a relativistic disc model, which respects the upper limits of the X-ray observations. The best fit chi-squared of the 4 UV light curves is: 3.38. The main source of model-data discrepancy in the UV light curves results from the 4 UV light curves remaining brighter than predicted at around day $\sim 180$, just before the transition into the disc-dominated plateau.  This could either result from a UV flare, or the initial UV light curve decaying slightly slower than exponentially.


\subsection{ASASSN-14ae}

Another optically-bright X-ray-dim TDE, ASASSN-14ae, was first reported by Holoien {\it et al}.  (2014), with subsequent late time UV observations in Brown {\it et al}. (2016). ASASSN-14ae has well sampled UV light curves, spanning $\sim 700$ days, and  was not observed at X-ray energies at early times (Holoien {\it et al}. 2014).  ASASSN-14ae has been re-observed at X-ray energies at very late times (Jonker {\it et al}. 2020). In these extremely late-time observations, ASASSN-14ae was shown to now have observable levels of X-ray flux. This suggests that at early-times  ASASSN-14ae was in the X-ray dim soft state, producing undetectable levels of quasi-thermal X-ray emission, but at later times the disc underwent a state transition into the low-hard state, thereafter producing detectable nonthermal X-ray emission. This hypothesis can be tested with our model, as we are able to produce  disc bolometric light curves in addition to  X-ray and UV light curves. If TDE accretion discs are similar to the accretion discs in X-ray binaries, then we would expect the transition between accretion states to occur at an Eddington ratio of approximately $l \equiv L_{\rm bol}/L_{\rm edd} \simeq 10^{-2}$.

\begin{table}
\renewcommand{\arraystretch}{2}
\centering
\begin{tabular}{|p{2.2cm}|p{2.8cm}|}
\hline
Parameter & Value  \\ \hline\hline
$M/M_\odot$ & $1.3 \times 10^7$  \\ \hline
 $M_{\rm acc}/M_\odot$ & $0.05  $  \\ \hline
$t_{\rm visc}$ (days) & $250 $  \\ \hline
\end{tabular}
\caption{System parameters used to reproduce the ASASSN-14ae UVW1 light curve (Figure \ref{UV14ae}) while satisfying the X-ray flux upper detection limits reported in Holoien {\it et al}. (2014), Figure \ref{X14ae}. Reproduced and collated in Table \ref{table_appendix}.}
\label{table14ae}
\end{table}

ASASSN-14ae is located at a luminosity distance of $d_L = 193$ Mpc (Holoien {\it et al}. 2014), and has two existing estimates for the central black hole mass, both inferred from galactic scaling relationships. From the $M:\sigma$ relationship Holoien {\it et al}. (2014) inferred  $M \lesssim 10^{6.9} M_\odot$, whereas from the galactic bulge mass $M \sim 10^{6.8 \pm 0.4} M_\odot$ (Holoein {\it et al}. 2014). As we mentioned earlier, ASASSN-14ae was undetected at X-ray frequencies early in its evolution, with a flux upper limit of $F_X = 3 \times 10^{-14}$ erg/s/cm${}^2$ (Holoien {\it et al}. 2014).

Table \ref{table14ae} summarises the disc and black hole parameters used to reproduce the recorded observations of  ASASSN-14ae. The lower bound black hole mass, $M = 1.3 \times 10^7 M_\odot$, while consistent with the galactic bulge measurement of Holoien {\it et al}. (2014), is in tension with the black hole mass inferred from a measurement of the galactic velocity dispersion (Holoien {\it et al}. 2014). 

\begin{figure}
  \includegraphics[width=.5\textwidth]{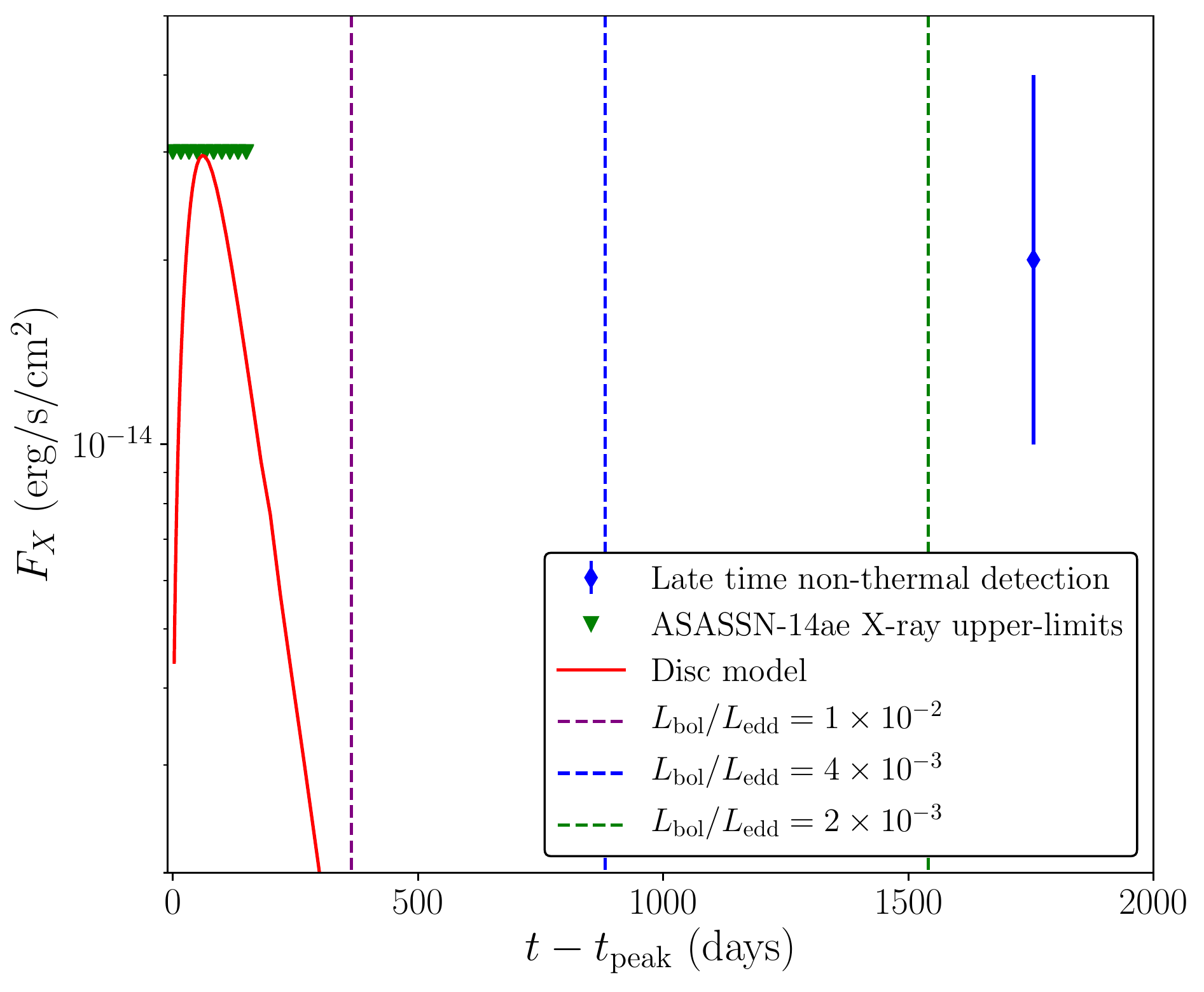} 
 \caption{The evolving X-ray flux of the disc model (red curve), fit to the observed UV light curve of ASASSN-14ae (Figure. \ref{UV14ae}). This disc model satisfies early-time upper X-ray limits. The Eddington ratio of the best-fit disc solutions at three late times in the disc evolution are denoted by vertical dashed lines, demonstrating that the disc has likely transitioned into a harder accretion state by the time of the late nonthermal X-ray detection.   } 
 \label{X14ae}
\end{figure}

\begin{figure}
  \includegraphics[width=.5\textwidth]{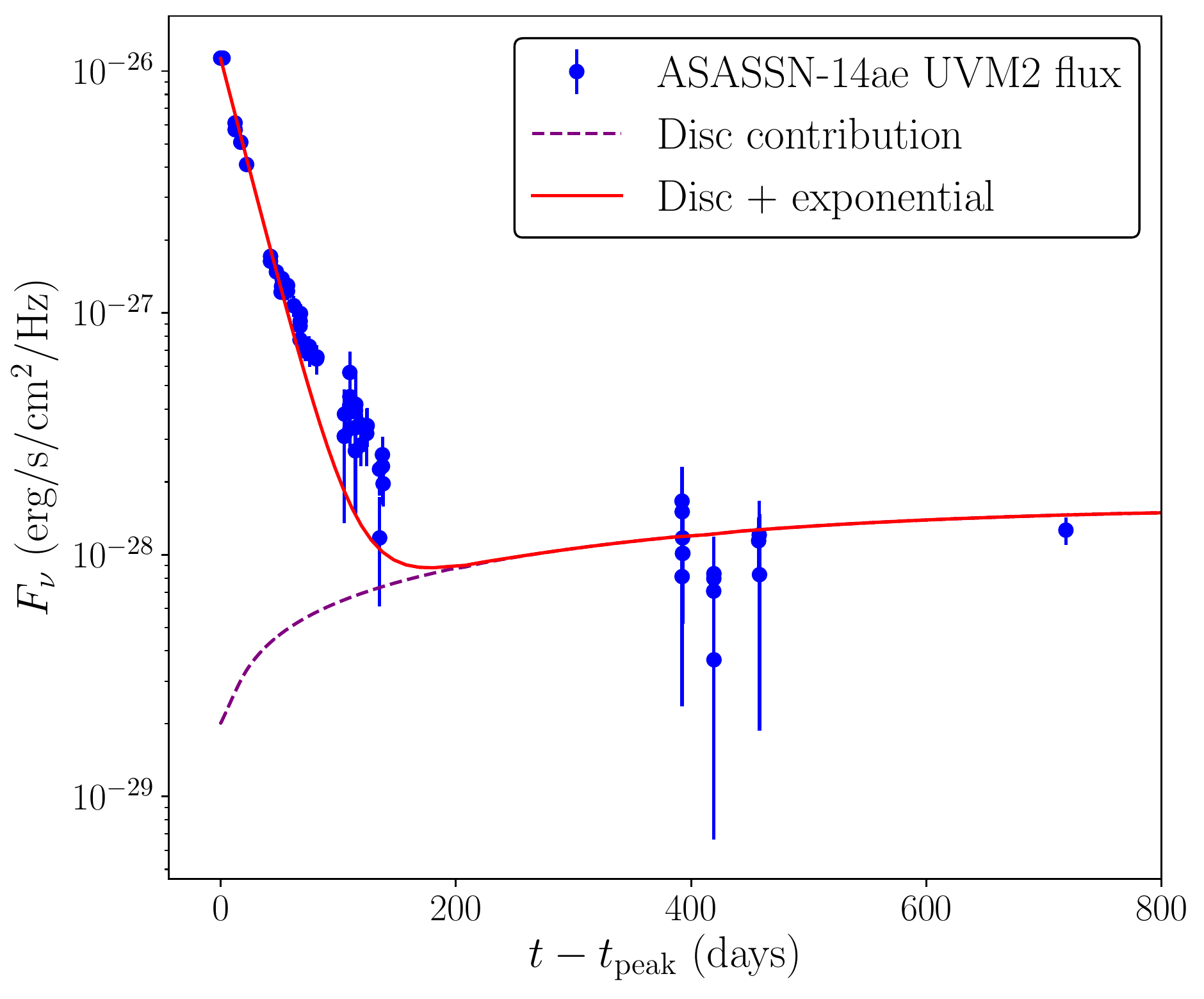} 
 \caption{The best fitting  disc light curve at the UVM2 frequency (dashed curves), fit to the observed UV light curves of ASASSN-14ae (solid points). The total UV model light curve, which includes an exponentially declining component relevant at early times, is denoted by a solid curve.} 
 \label{UV14ae}
\end{figure}

Fig. \ref{X14ae} shows the observed ASASSN-14ae 0.3--10 keV X-ray upper limits (Holoien {\it et al}. 2014), together with the evolving X-ray flux from our fiducial disc model. The system parameters in Table \ref{table14ae} have been chosen so that the disc X-ray light curve peak just below the observed X-ray upper limit. Figure \ref{UV14ae} shows the evolving ASASSN-14ae UVM2 flux (the best sampled of all the UV light curves, blue points, Brown {\it et al}. 2016), the best fitting UVM2 light curve of the disc (purple dashed curve), and a combined disc and exponentially decaying light curve (red solid curve).  

As is clear from Figs. \ref{X14ae} \& \ref{UV14ae},  the evolving UVM2 flux of ASASSN-14ae can be well described by a  relativistic disc model, while still respecting the upper limits of the X-ray observations. The best fit chi-squared for the UVM2 light curve is: $\chi^2 = 6.74$. The main source of model-data discrepancy in the UV light curve stems from, in common with ASASSN-18pg, a slower than exponential initial decline (or a potential flare) in the UVM2 band around day $\sim 150$.  

Over-plotted in Figure \ref{X14ae} are three vertical lines denoting the disc Eddington ratio at different times during its evolution.  As is clear from Figure \ref{X14ae}, by the time at which the late-time X-ray emission was detected (Jonker {\it et al}. 2020) the disc Eddington ratio is comfortably lower than the values at which state transitions are observed in X-ray binaries (e.g. Macarone 2003). It is therefore likely that ASASSN-14ae has transitioned into a harder accretion state, a result of its bolometric luminosity decreasing with time. Using our compact corona model (Paper II) we are able to reproduce the amplitude of the late time X-ray flux with a wide range of fitted parameters. With just a single late-time X-ray observation, we do not attempt to analyse the properties of ASASSN-14ae in this harder accretion state.

\section{Bright non-thermal X-ray TDEs}
While ASASSN-14ae appears to have transitioned into a harder accretion state at a late time in its evolution, a population of TDEs have been observed to have non-thermal (hard-state) X-ray spectra right from their first observation. In Paper II we developed a model whereby this non-thermal X-ray flux is produced by Compton up-scattering of thermal disc photons by a compact electron corona covering the innermost regions of the accretion disc. As we summarised in section \ref{UM}, we expect to find TDEs with non-thermal X-ray spectra  predominantly around TDEs with the largest black hole masses (a result of these TDEs forming discs with the lowest Eddington ratios). 

We assume that, though do not model in detail,  a Comptonizing corona forms and is active when TDE discs form in (or transition into) the hard accretion state.
This corona, parameterised by a radial size $R_{\rm Cor}$, then Compton up-scatters a fraction $f_{\rm SC}$ of the thermal disc photons emitted from disc radii $R < R_{\rm Cor}$ into a power-law spectrum characterised by a photon index $\Gamma$. The detailed disc spectrum is calculated using photon-number conservation, and is described in detail in Paper II. 

In this section we examine two TDEs from the XMM slew catalogue XMMSL1 J0740 \& XMMSL2 J1446. While both TDEs have X-ray spectra which are dominated by a non-thermal power-law component, XMMSL1 J0740 also has a prominent `thermal excess', which will be examined in detail in section 6.3.

While the disc-corona model  technically has three free parameters describing the electron-scattering corona, one, the X-ray spectrum photon index $\Gamma$, is directly observable and we fix this parameter to the observed value of each source. The final two free parameters, $R_{\rm Cor}$ and $f_{SC}$, are somewhat degenerate. A larger corona (larger $R_{\rm Cor}$) which scatters a smaller fraction $f_{ SC}$ of the thermal disc photons will produce similar X-ray fluxes to a smaller corona which scatters a larger fraction of photons.   The X-ray light curves of both XMMSL1 J0740 \& XMMSL2 J1446 have relatively large amplitude temporal fluctuations, and it is not possible with the current data to break this intrinsic parameter degeneracy. We therefore fix the coronal radius to $R_{\rm Cor} = 8 r_g$, and treat the photon scattering fraction $f_{SC}$ as the only free parameter of the corona.

\subsection{XMMSL2 J1446}
\begin{table}
\renewcommand{\arraystretch}{2}
\centering
\begin{tabular}{|p{2.2cm}|p{2.8cm}|}
\hline
Parameter & Value  \\ \hline\hline
$M/M_\odot$ & $2.7^{+0.3}_{-0.2} \times 10^7$  \\ \hline
 $M_{\rm acc}/M_\odot$ & $2.0^{+0.6}_{-0.2} \times 10^{-2} $  \\ \hline
$t_{\rm visc}$ (days) & $109^{+25}_{-4} $  \\ \hline
$f_{ SC}$ & $0.12^{+0.16}_{-0.02}$ \\ \hline
\end{tabular}
\caption{Best fit parameters for the X-ray (Fig. \ref{Xj1446}) and UV (Fig. \ref{UVj1446}) light curves of  XMMSL2 J1446. Reproduced and collated in Table \ref{table_appendix}.}
\label{tablej1446}
\end{table}

\begin{figure}
  \includegraphics[width=.5\textwidth]{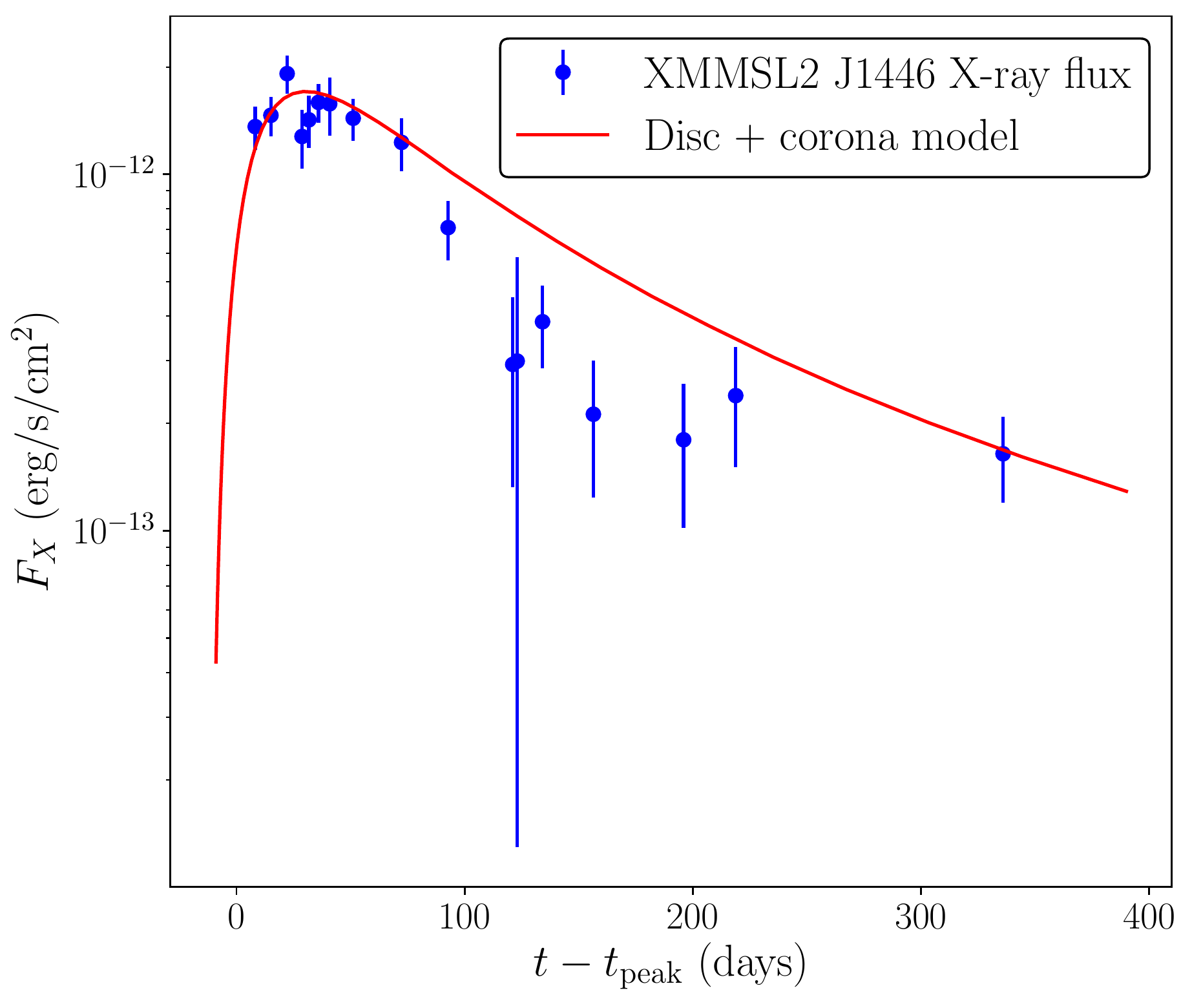} 
 \caption{The best fitting disc and corona model (red curve), fit to the observed X-ray light curve of XMMSL2 J1446 (blue points, Saxton {\it et al}. 2019). } 
 \label{Xj1446}
\end{figure}

\begin{figure}
  \includegraphics[width=.5\textwidth]{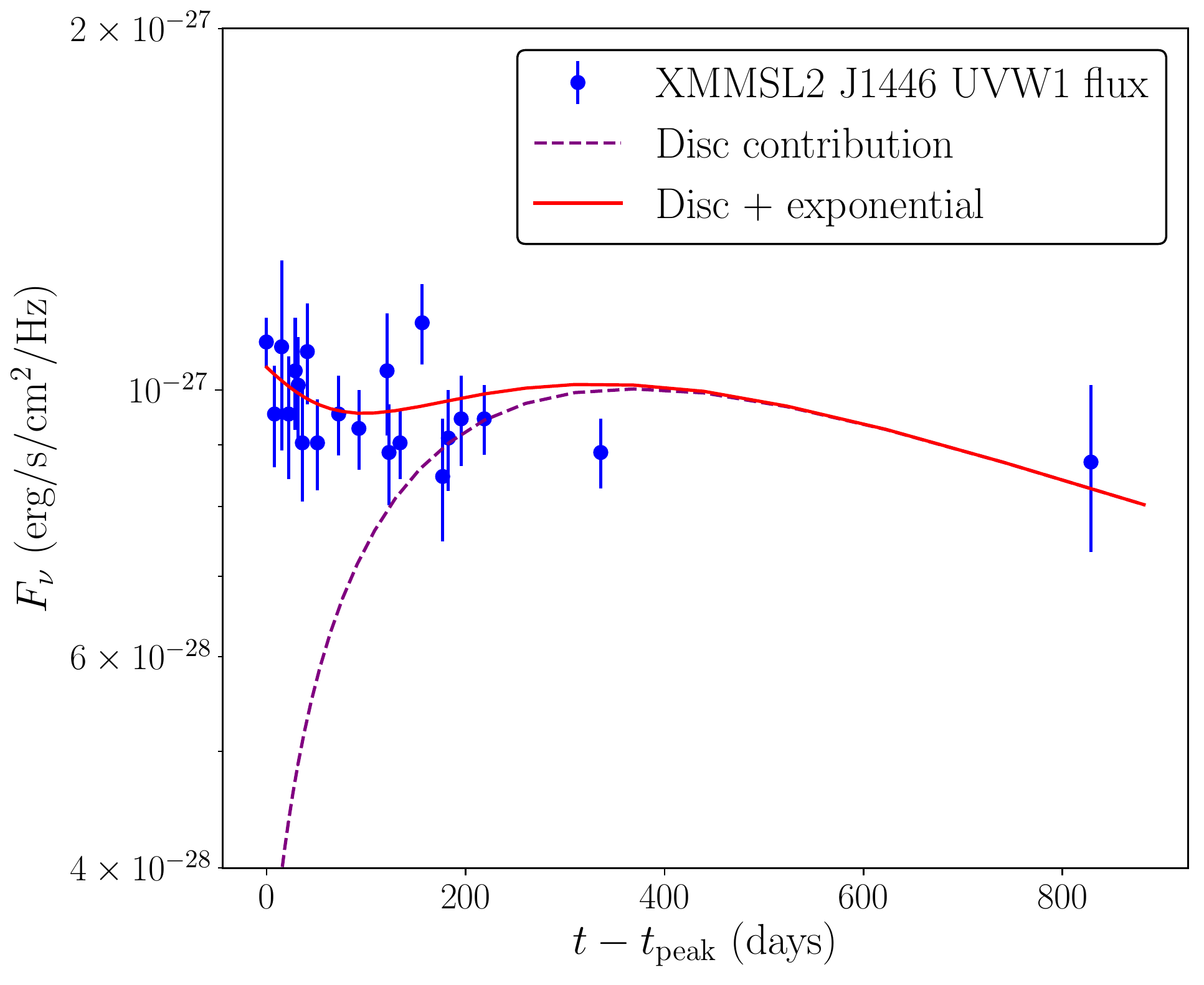} 
 \caption{The best fitting disc and corona model (dashed curve), fit to the observed UV light curve of XMMSL2 J1446 (blue points, Saxton {\it et al}. 2019). The total UV model light curve, which includes an exponentially declining component relevant at early times, is denoted by a solid curve. } 
 \label{UVj1446}
\end{figure}

The TDE XMMSL2 J1446 was first reported by Saxton {\it et al}. (2019). XMMSL2 J1446 has well sampled UV light curves, spanning $\sim 800$ days, and has been well observed at X-ray energies (Saxton {\it et al}. 2019), with $\sim 400$ days of available X-ray data. The X-ray spectrum of XMMSL2 J1446 is well described by pure power-law emission, with a photon index of  $\Gamma = 2.58$ (Wevers 2020, Saxton {\it et al}. 2019).

XMMSL2 J1446 is located at a luminosity distance of $d_L = 127$ Mpc (Saxton {\it et al}. 2019), and has a mean central black hole mass, inferred from galactic scaling relationships, of  $\left\langle M \right\rangle = 4.1^{+4.7}_{-2.8} \times 10^7 M_\odot$ (Paper II).  

Table \ref{tablej1446} summarises the best-fitting disc, corona and black hole parameters for the recorded observations of XMMSL2 J1446. The best fit black hole mass, $M = 2.7\times 10^7 M_\odot$, is consistent with the results of galactic scaling relationships. If we treat the observed light curves in isolation, without considering additional spectral information, then discs around lower black hole masses can produce acceptable fits. However, for these lower black hole masses the disc X-ray spectrum would be dominated by thermal emission at low photon energies, in contrast to what is observed. Fig. \ref{Xj1446} shows the observed XMMSL2 J1446 0.3--10 keV X-ray flux (Saxton {\it et al}. 2019), together with the best-fitting evolving X-ray flux from our fiducial disc-corona model. Figure \ref{UVj1446} shows the evolving XMMSL2 J1446 UVW1 flux (blue points, Saxton {\it et al}. 2019), the best fitting UV light curve of the disc (purple dashed curve), and a combined disc and exponentially decaying light curve (red solid curve).  

An interesting result clear from Figure \ref{UVj1446} is that, unlike the four quasi-thermal TDEs considered earlier, the exponentially decaying UV component in XMMSL2 J1446 is similarly bright to the disc-dominated plateau flux. In fact, if the initial disruption substantially preceded the first observation the entire  UV light curve of XMMSL2 J1446 could be described entirely by an evolving relativistic accretion disc.

As is clear from Figs.  \ref{Xj1446} \&  \ref{UVj1446}, both the evolving X-ray and UV flux of XMMSL2 J1446 are well described by an evolving relativistic disc model. The best fit chi-squared for the X-ray light curve is: $\chi^2 = 1.21$, for the UVW1 light curve: $\chi^2 = 1.17$, and the combined light curves: $\chi^2 = 1.05$. The main source of model-data discrepancy  results from short timescale fluctuations in the X-ray observations, not captured by a smoothly decaying disc model.

\subsection{XMMSL1 J0740}
\begin{table}
\renewcommand{\arraystretch}{2}
\centering
\begin{tabular}{|p{2.2cm}|p{2.8cm}|}
\hline
Parameter & Value  \\ \hline\hline
$M/M_\odot$ & $3.1_{-0.4}^{+0.9} \times 10^7$  \\ \hline
 $M_{\rm acc}/M_\odot$ & $1.8^{+2.4}_{-1.0} \times10^{-2} $  \\ \hline
$t_{\rm visc}$ (days) & $84.3_{-10.6}^{+7.1} $  \\ \hline
$f_{ SC}$ & $0.032_{-0.019}^{+0.038}$ \\ \hline
\end{tabular}
\caption{Best fit parameters for the X-ray (Fig. \ref{Xj0740}) and UV (Fig. \ref{UVj0740}) light curves of  XMMSL1 J0740. Reproduced and collated in Table \ref{table_appendix}.}
\label{tablej0740}
\end{table}

\begin{figure}
  \includegraphics[width=.5\textwidth]{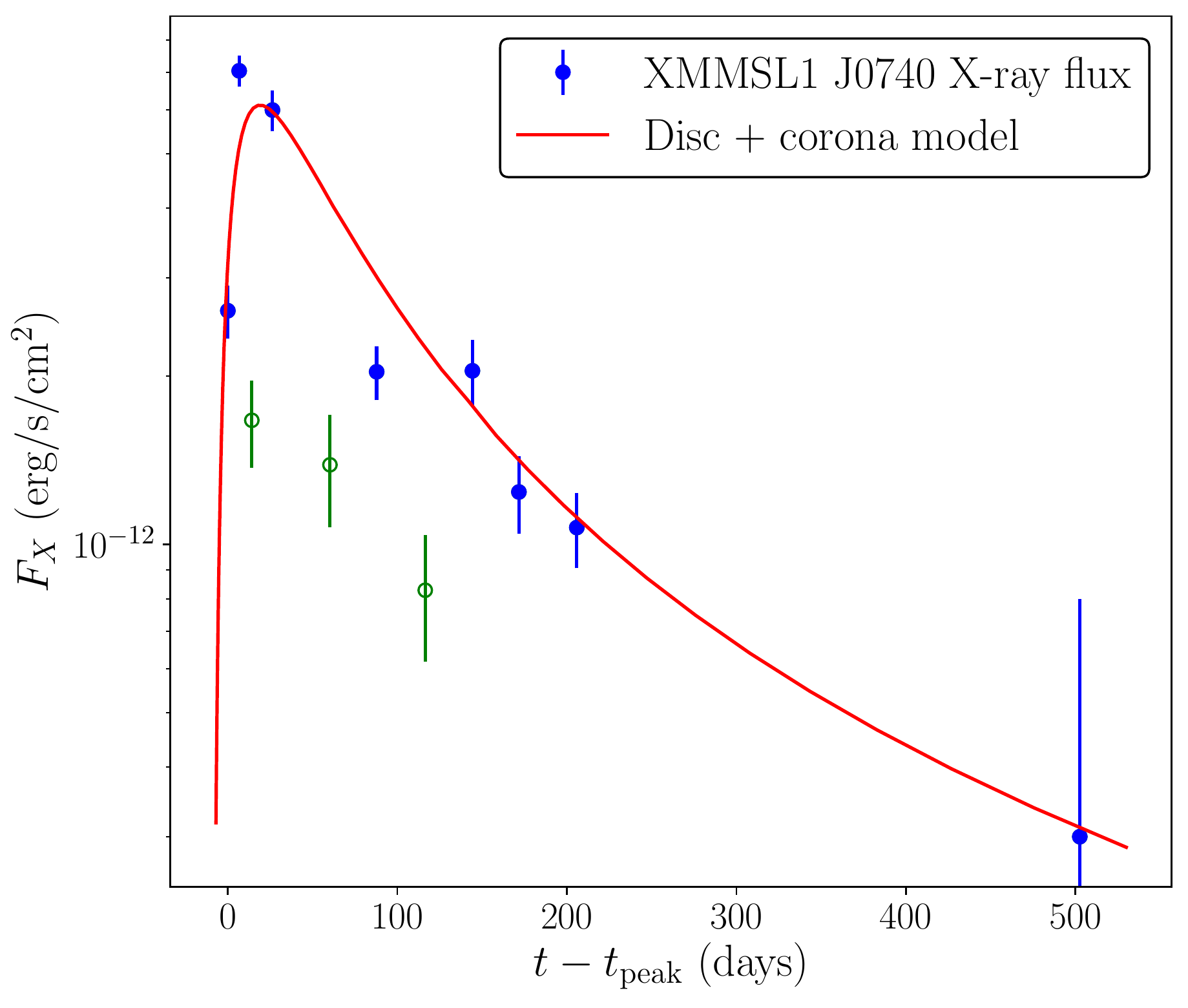} 
 \caption{The best fitting disc and corona model (red curve), fit to the observed X-ray light curve of XMMSL1 J0740 (blue points, Saxton {\it et al}. 2017). Points shown in unfilled green were not included in the fitting process (see text).  } 
 \label{Xj0740}
\end{figure}

\begin{figure}
  \includegraphics[width=.5\textwidth]{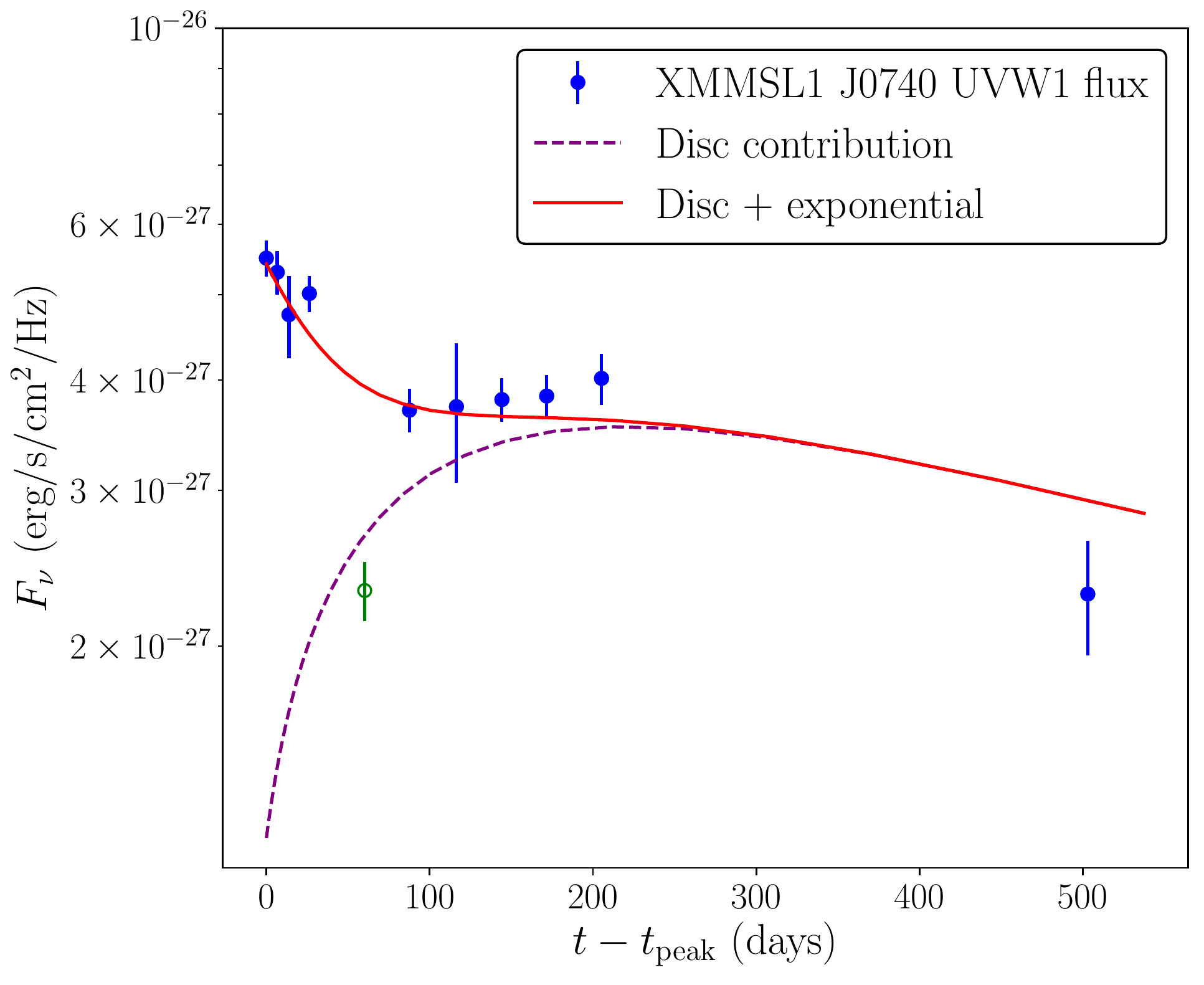} 
 \caption{The best fitting disc and corona model (dashed curve), fit to the observed UV light curve of XMMSL2 J1446 (blue points, Saxton {\it et al}. 2019).  Points shown in unfilled green were not included in the fitting process (see text). The total UV model light curve, which includes an exponentially declining component relevant at early times, is denoted by a solid curve.} 
 \label{UVj0740}
\end{figure}

The TDE XMMSL1 J0740 was first reported by Saxton {\it et al}. (2017). XMMSL1 J0740 has been observed at UV frequencies for $\sim 500$ days, and has been  observed for $\sim 500$ days at X-ray energies (Saxton {\it et al}. 2017). The X-ray spectrum of XMMSL1 J0740 is well described by non-thermal emission, with a photon index of  $\Gamma = 1.95$ (Wevers 2020, Saxton {\it et al}. 2017), with a `soft excess' at low energies. This additional spectral component is well modelled as a  blackbody with temperature $kT \sim 100$ eV, and will be discussed further in the following subsection.  

XMMSL1 J0740 is located at a luminosity distance of $d_L = 75$ Mpc (Saxton {\it et al}. 2017), and has an estimated central black hole mass, inferred from the $M:\sigma$ galactic scaling relationship, of $M = 7.9^{+4.4}_{-2.9} \times 10^6 M_\odot$ (Paper II, Saxton {\it et al}. 2017).

The X-ray and UV light curves of XMMSL1 J0740 are both relatively poorly sampled, and both have large-amplitude short-timescale variations, which makes finding best fitting parameters difficult. Fitting to the entirety of the observed data results in extremely poor fits ($\chi^2 > 10$),  primarily a result of large-amplitude dips in both the UV and X-ray light curves at early times ($t \lesssim 100$ days). We therefore decided to neglect three observations at X-ray energies (denoted by green unfilled points in Fig. \ref{Xj0740}), and one observation in the UVW1 band (denoted by green unfilled points in Fig. \ref{UVj0740}). To the remaining observations we are able to find reasonable best-fit parameters. The best-fitting disc X-ray light curve has a chi-square statistic $\chi^2 = 5.35$, and the UVW1 light curve $\chi^2 = 1.25$. The combined UVW1 and X-ray light curves are fit with total $\chi^2$ statistic $\chi^2 = 3.07$.

Table \ref{tablej0740} summarises the best-fitting disc and black hole parameters for the recorded observations of XMMSL1 J0740. The best fit black hole mass, $M = 3.1 \times 10^7 M_\odot$, is not consistent with the mass inferred from the $M:\sigma$ relationship. Fig. \ref{Xj0740} shows the observed XMMSL1 J0740 0.3--10 keV X-ray flux (Saxton {\it et al}. 2017), together with the best-fitting evolving X-ray flux from our fiducial disc model. Figure \ref{UVj0740} shows the evolving XMMSL1 J0740 UV flux (blue points, Saxton {\it et al}. 2017), the best fitting UV light curve of the disc (purple dashed curve), and a combined disc and exponentially decaying light curve (red solid curve).  Similarly to XMMSL2 J1446, the early-time UV component of XMMSL1 J0740 is much less bright than those observed around quasi-thermal TDEs.

\subsection{Additional spectral information}

The primary  reason we model XMMSL1 J0740, despite its poor quality light curves, is that it represents a distinct sub-class of nonthermal X-ray TDEs from XMMSL2 J1446. The reason for this is that its X-ray spectrum shows the presence of a `thermal excess' at low photon energies. Explicitly, XMMSL1 J0740 shows the presence of a secondary blackbody component with temperature $kT \sim 100$ eV, while the X-ray spectrum of XMMSL2 J1446 is well described with a power-law profile. If our combined disc and corona model is accurate, we should expect it to reproduce the observed thermal excess at low photon energies for XMMSL1 J0740, but not for XMMSL2 J1446.

\begin{figure}
  \includegraphics[width=.5\textwidth]{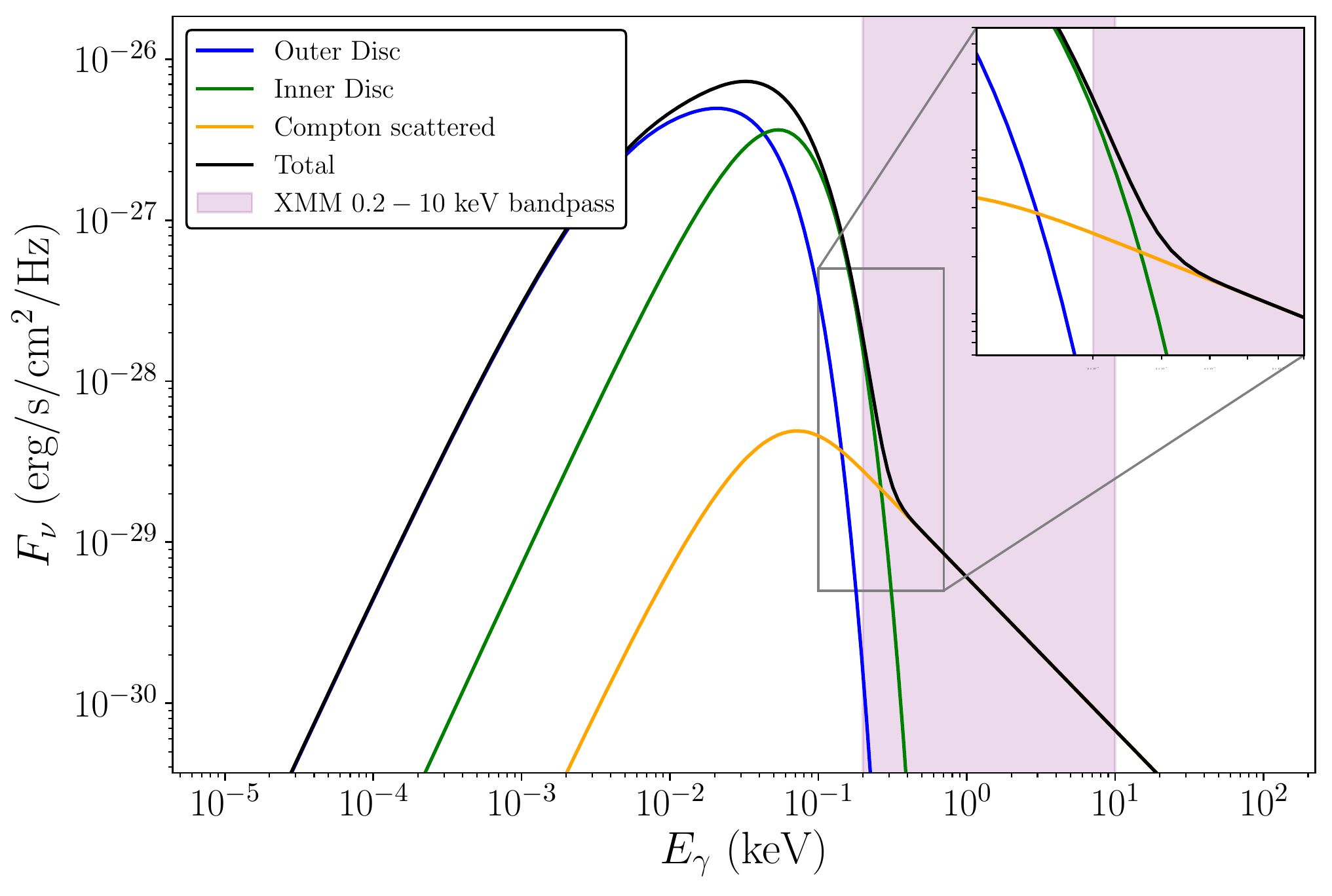} 
 \caption{A snapshot of the disc spectrum of XMMSL1 J0740 after one viscous timescale. Different spectral components are displayed by different colours (see text). As can be seen in the inset, the X-ray spectrum as observed between $0.2 - 10$ keV would be well described by a power-law profile, with an additional blackbody component with temperature $kT \sim 100$ eV. This is exactly as was found by Saxton {\it et al}. (2017). } 
 \label{specj0740}
\end{figure}

To further test our model we therefore look in detail at the disc spectra of both XMMSL1 J0740 and XMMSL2 J1446. In figure  \ref{specj0740} we plot the disc spectrum of the best-fitting XMMSL1 J0740 disc model at a time equal to one viscous timescale into its evolution. The three coloured curves represent the contributions to the total disc spectrum from the outer $R > R_{\rm Cor}$ disc (blue curve), the contribution from the un-scattered photons from the inner $R < R_{\rm Cor}$ disc (green curve), and the contribution from the Compton-scattered photons (orange curve). The total disc spectrum is shown by a black curve. 

In the inset of Figure \ref{specj0740} we zoom into the  photon energy region around  $E_\gamma \simeq 0.3$ keV. What is clear from Figure \ref{specj0740} is that at low photon energies, but still within the XMM $0.2 - 10$ keV observing bandpass, there is a clear additional component resulting from the contribution from un-scattered photons emitted in the inner disc regions. This is exactly as was observed by Saxton {\it et al}. (2017). When observed exclusively between $0.2-10$ keV, this additional component would likely be well modelled by a single temperature blackbody of temperature $kT \sim 100$ eV.

\begin{figure}
  \includegraphics[width=.5\textwidth]{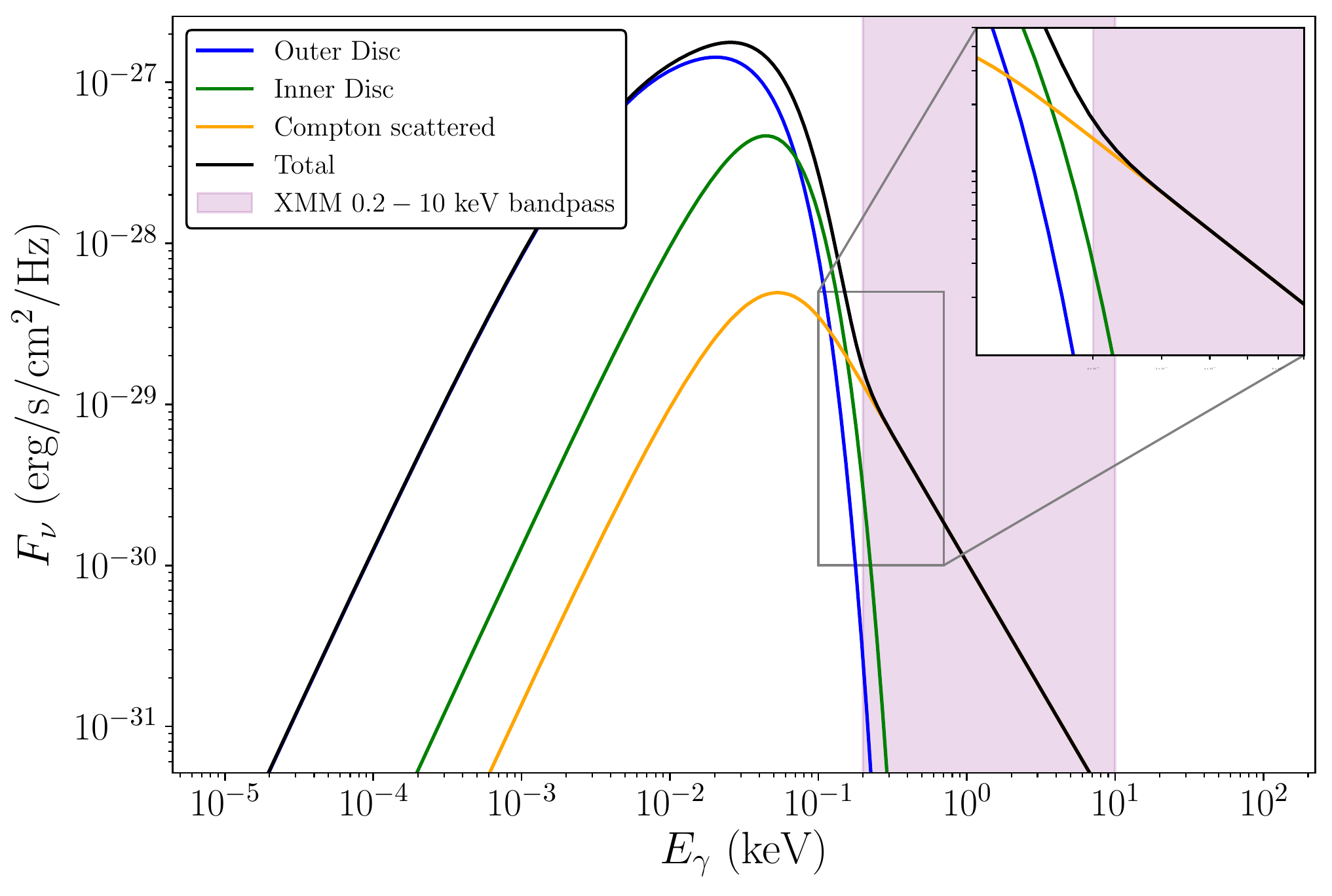} 
 \caption{A snapshot of the disc spectrum of XMMSL2 J1446 after one viscous timescale. Different spectral components are displayed by different colours (see text). As can be seen in the inset, the X-ray spectrum as observed between $0.2 - 10$ keV is well described by a pure power-law profile, as found by Saxton {\it et al}. (2019).  } 
 \label{specj1446}
\end{figure}

\begin{figure}
  \includegraphics[width=.5\textwidth]{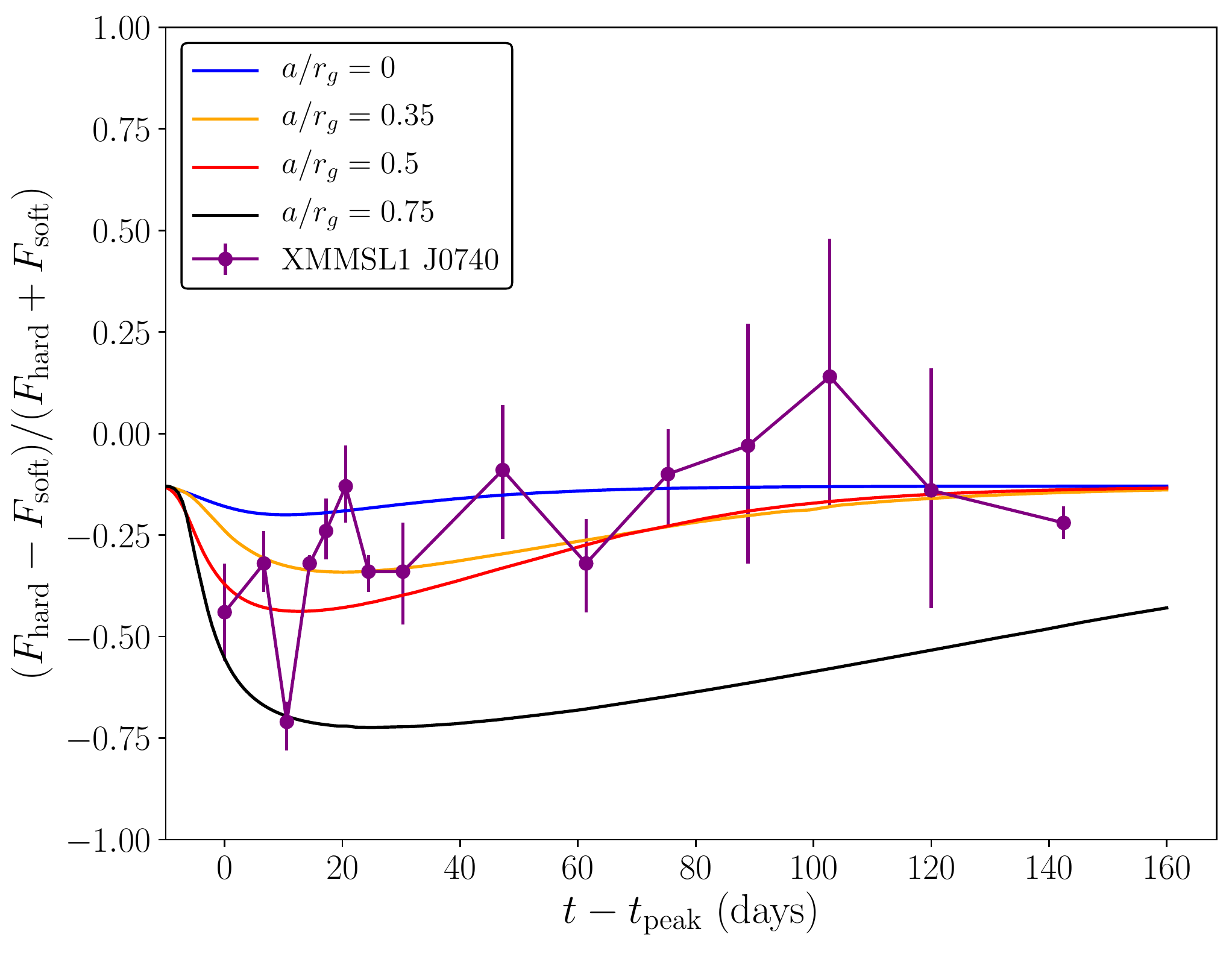} 
 \caption{The evolving hardness ratio (defined in text) of the TDE XMMSL1 J0740 (solid points, Saxton {\it et al}. 2017). Also plotted are the evolving hardness ratios of best-fitting disc and corona models for discs around black holes of different spins (solid curves). The best fitting disc models are quantitatively consistent with the observations. Simultaneously modelling the evolving light curves and hardness ratios of future, better observed, TDEs will provide strong constraints on black hole spin. }
 \label{HRt}
\end{figure}

Similarly, figure \ref{specj1446} shows the disc spectrum of XMMSL2 J1446 one viscous timescale into its evolution. Unlike XMMSL1 J0740,  XMMSL2 J1446 shows no evidence of a thermal component at low photon energies. When observed exclusively between $0.2-10$ keV the X-ray spectrum of XMMSL2 J1446 would be best modelled with a pure power-law, exactly as was found by Saxton {\it et al}. (2019). 

In combination these two results provide compelling evidence for the disc and corona model developed in Paper II and used here. The best fit parameters of both disc models were found from the evolving amplitudes of the X-ray and UVW1 fluxes. Despite this, the spectral properties of both best fitting disc solutions are in good accord with observations.

We can perform a more quantitive analysis of the soft excess present in XMMSL1 J0740 by computing the so-called `hardness ratio' as a function of time. The hardness ratio ${\rm HR}$ is defined as 
\beq
{\rm HR} \equiv {F_{\rm hard} - F_{\rm soft} \over F_{\rm hard} + F_{\rm soft}} ,
\eeq
where $F_{\rm hard}$ is the X-ray flux observed between $2-10$ keV, and $F_{\rm soft}$ is the X-ray flux observed between $0.2-2$ keV. The evolving hardness ratio of XMMSL1 J0740 was recorded by Saxton {\it et al}. (2017), and is plotted in Figure \ref{HRt}. In Figure \ref{HRt} we also plot the evolving hardness ratio of the best-fitting XMMSL1 J0740 disc models for a number of black hole spins. Each of the best-fitting disc profiles fit the observed X-ray and UV light curves of XMMSL1 J0740 similarly well. It is important to note that the best-fitting hardness ratios of disc solutions around black holes of different spins correspond closely with the observed hardness ratios of XMMSL1 J0740 (Fig. \ref{HRt}). The evolving best-fit disc spectra are therefore quantitively, as well qualitatively, consistent with the observations of XMMSL1 J0740. 

What is clear from Figure \ref{HRt} is that  the evolving hardness ratio of TDEs encodes additional information about the best fitting properties of that TDE disc system. While the disc models for different black hole spins produced very similar X-ray and UV light curves, the X-ray spectrums of the different models are clearly quantitively different. This differences is quantified in the evolving hardness ratios. Neither the evolving X-ray and UV light curves, nor the evolving hardness ratio of XMMSL1 J0740 are constraining enough to provide a definitive measurement of the best-fitting black hole spin of the XMMSL1 J0740 system. However, we can tentatively rule out large $a/r_g \gtrsim 0.75$ black hole spins, which produce evolving disc spectra which remain too `soft' at large times. The evolving hardness ratio of X-ray TDEs may be able, in future, better observed sources, to constrain the black hole spin of supermassive black holes.

\section{Population analysis}
We now have a small sample of TDEs modelled in a near-identical manner (best fitting parameters summarised in Table \ref{table_appendix}). In this section we analyse the properties of this  population of TDEs. We begin by examining whether their properties fit with what is expected from the unified model of disc-dominated TDEs (summarised in section \ref{UM}). 

The unified model of disc-dominated TDEs predicts that the three sub-populations of observed TDEs examined here: bright X-ray TDEs with quasi-thermal X-ray spectra; X-ray-dim, UV-bright TDEs which undergo disc-dominated plateaus; and bright nonthermal X-ray TDEs should lie in regions of distinct peak Eddington ratios. Explicitly, we would expect the three sub-populations to be separated by decreasing Eddington ratio, with the bright quasi-thermal TDEs having peak Eddington ratios which satisfy $0.1 < l_{\rm max} < 1$, the X-ray-dim UV-bright TDEs: $0.01 < l_{\rm max} < 0.1$, and the bright nonthermal TDEs $l_{\rm max} < 0.01$. We remind the reader that there is expected to be a second population of UV-bright X-ray-dim TDEs, for those TDEs which form with initially extremely super-Eddington luminosities. The population of UV-bright TDEs we are discussing in this section are X-ray-dim as a result of insufficiently high disc temperatures  reached during their evolution. 

The different sub-populations are primarily expected to correspond to sources of different black hole masses, due to the strong dependence of the disc Eddington ratio on black hole mass (eq. \ref{edrat}). The black hole mass dependence of the different TDE sub-populations examined in this work are in qualitative agreement with the predictions of the unified model. As predicted, the largest inferred black hole mass was for the bright nonthermal X-ray TDE XMMSL1 J0740, while the smallest black hole mass was for the bright quasi-thermal X-ray TDE ASASSN-15oi. There is however, as is to be expected for sources with different disc masses and effective $\alpha$-values, some overlap between the black hole masses of the different sub-populations at intermediate $M \sim 10^7 M_\odot$ black hole masses. The principal outlier of the TDEs examined in this work is AT2019dsg, which has an extremely large black hole mass for a quasi-thermal X-ray TDE. However, as we demonstrated in section \ref{compare}, AT2019dsg was observable around a  black hole with large mass  because of its large effective $\alpha$-value, which also caused its rapid X-ray evolution.

\begin{figure}
  \includegraphics[width=.5\textwidth]{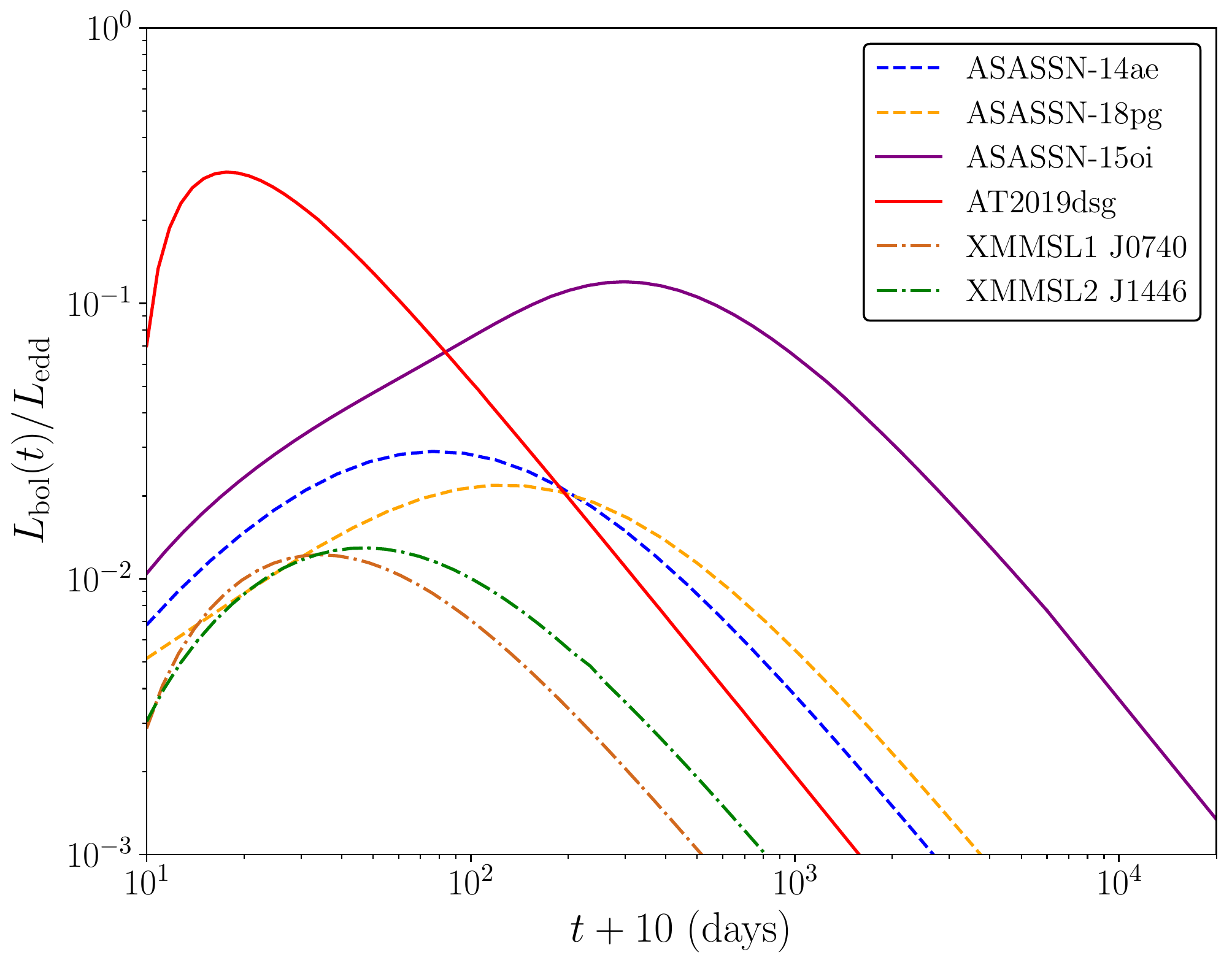} 
 \caption{The evolving bolometric luminosities, expressed in Eddington units, of the six TDEs studied in this work. The different spectral types of TDEs lie in different regions of peak Eddington ratio space, exactly as predicted by the unified model of disc-dominated TDEs.  } 
 \label{BOL}
\end{figure}

Furthermore, and more importantly, the Eddington-ratio-dependence of the different TDE sub-populations examined in this work are in complete quantitative agreement with the predictions of the unified model of disc-dominated TDEs. This is demonstrated in Figure \ref{BOL}, perhaps the most important figure in this paper. Figure \ref{BOL} shows the evolving bolometric luminosities (expressed in Eddington units) of the six TDEs examined in this paper. As is clear from figure \ref{BOL}, all six TDEs have sub-Eddington luminosities at all times. The different sub-populations of TDEs clearly occupy distinct regions of luminosity space, with the bright quasi-thermal X-ray TDEs reaching Eddington ratios $l \gtrsim 0.1$, and the nonthermal X-ray TDEs evolving with $l \lesssim 0.01$ at all times. The UV-bright and X-ray-dim TDEs occupy the intermediate region of bolometric luminosity space. 

This result is further strengthened by incorporating the two TDEs we have studied in previous works, ASASSN-14li and ASASSN-15lh. ASASSN-14li is a bright quasi-thermal X-ray TDE, and was found to have a peak Eddington ratio of $l_{\rm max} = 0.85$ (Mummery \& Balbus 2020a), while ASASSN-15lh was a UV bright X-ray dim TDE around a black hole of extremely large mass (Mummery \& Balbus 2020b), and was found to have a peak Eddington ratio of $l_{\rm max} = 0.02$. While these two TDEs were modelled in a slightly different manner to the present work, and so are not directly comparable, their behaviour further strengthens the evidence for the unified TDE disc model. 

In the remainder of the section we examine some of the other properties of this population of TDEs. The `bolometric' properties of the six TDEs analysed in this paper are collated in Table \ref{tablebol}.

\begin{table}
\renewcommand{\arraystretch}{2}
\centering
\begin{tabular}{|p{1.9cm}|p{2.cm}|p{1.2cm}|p{1.5cm}|}
\hline
TDE  & $L_{\rm bol, max}$ (erg/s) & $l_{\rm max}$ & $E_{\rm rad}$ (erg) \\ \hline\hline
AT2019dsg & $6.9 \times 10^{44}$ & $0.27$  & $4.0 \times 10^{51}$ \\ \hline
ASASSN-15oi & $1.4 \times 10^{44}$ & $0.12$ & $1.9 \times 10^{52}$ \\ \hline
ASASSN-14ae &  $4.8 \times 10^{43}$  & $0.03$ & $2.4 \times 10^{51}$ \\ \hline
ASASSN-18pg & $9.1 \times 10^{43}$  & $0.02$ & $1.0 \times 10^{52}$ \\ \hline
XMMSL1 J0740 &  $4.8 \times 10^{43}$ & $0.01$ & $8.6 \times 10^{50}$  \\ \hline
XMMSL2 J1446 & $4.1 \times 10^{43}$ & $0.01$ & $9.1 \times 10^{50}$ \\ \hline
\end{tabular}
\caption{ The bolometric properties of the best-fit disc models for the TDEs studied in this work. The quantity $l_{\rm max}$ is defined   $l_{\rm max} \equiv L_{\rm bol, max} / L_{\rm edd}$. } 
\label{tablebol}
\end{table}

\subsection{The resolution of the missing energy problem}
The so-called `missing energy problem' can be stated in the following manner: if the typical mass of a star which undergoes a complete tidal destruction is of order $M_\star \sim 0.1 - 1 M_\odot$, then why does the integrated energy observed in typical TDE light curves only reach $E_{\rm rad, obs} \sim 10^{50}$ erg/s (e.g. Holoien {\it et al}. 2014b)? This observed radiated energy, if we assume a nominal radiative efficiency $\eta \simeq 0.1$, corresponds to a mass of $M_{\rm rad} \sim 10^{-4} - 10^{-3} M_\odot$. In effect: where has all the mass gone?

This solution to the missing energy problem is the following: the vast majority of the `missing' energy is simply radiated in EUV frequencies, which are unobservable from earth. This can be clearly seen in the snapshots of the disc spectrum presented in Figs. \ref{specj0740} \& \ref{specj1446}, where the vast majority of the disc energy is being radiated at EUV energies $E_\gamma \simeq 0.01-0.1$ keV.

In total we have modelled 8 TDEs, with a wide range of observed properties. Every TDE we have modelled has a best fitting disc mass $M_{\rm acc} > 0.01 M_\odot$, ranging up to $M_{\rm acc} \sim 0.3 M_\odot$.  The average disc mass of the six TDEs we have modelled in an identical manner in this work is $\left \langle M_{\rm acc} \right \rangle \simeq 0.12 M_\odot$. Not all of the stellar debris will settle into an evolving accretion disc in the aftermath of a TDE, with the fraction of material accreted expected to be $f < 1/2$ (Rees 1988). We can therefore state that the average stellar mass of the six TDEs we have modelled is $\left \langle M_\star \right \rangle > 0.24 M_\odot$, in line with what is expected from typical stellar populations. It is worth stressing that this mass value does not include any of the mass which powers the early time UV and optical emission, which can be extremely bright, nor any radio emission, and so is an underestimate.

\begin{figure}
  \includegraphics[width=.5\textwidth]{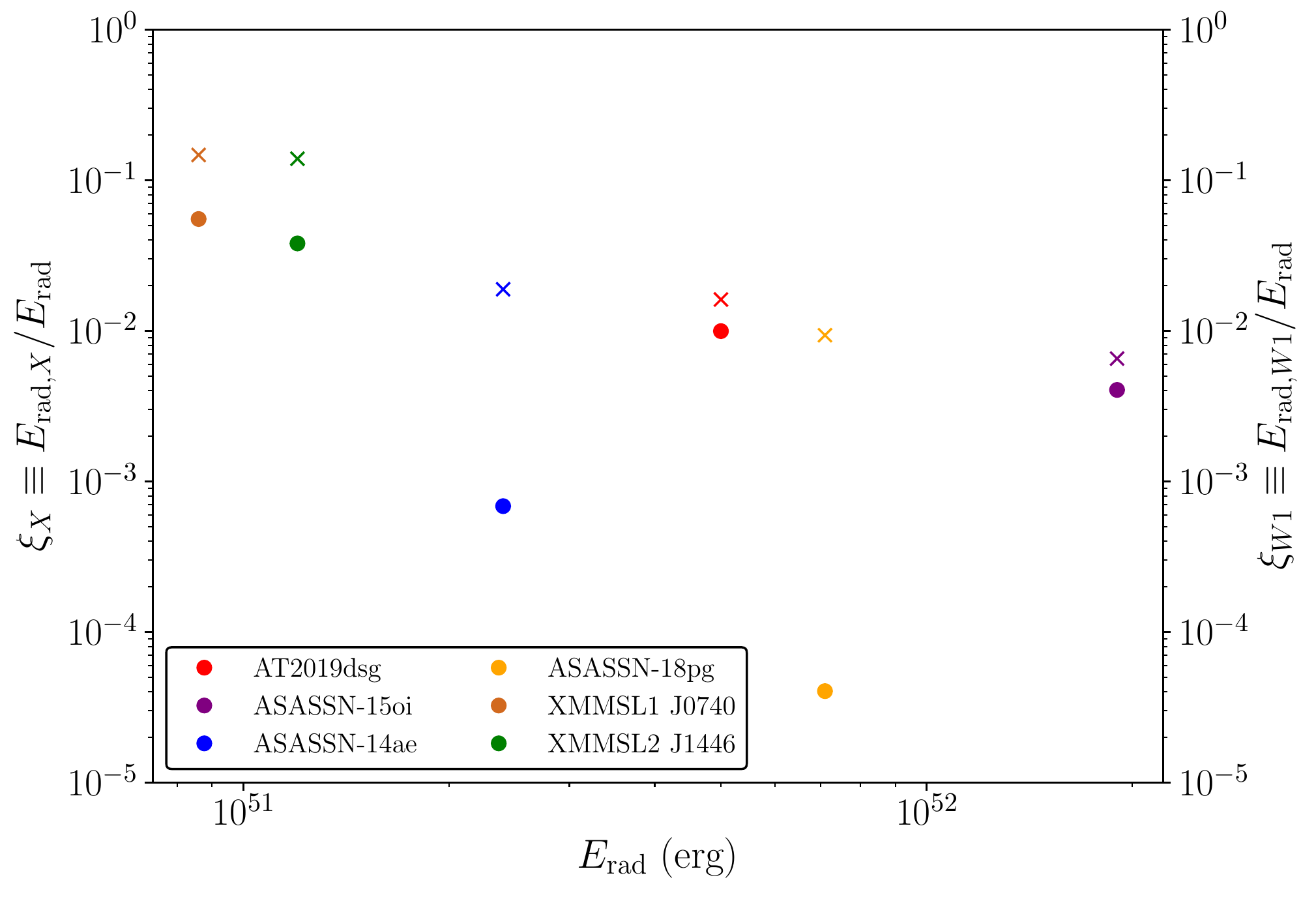} 
 \caption{The observed UV and X-ray energy deficits of the six TDEs studied in this work. We define $\xi_X$ and $\xi_{W1}$ (respectively) as the ratio of the energy observed in the X-ray and UVW1 bands to the total radiated disc energy. Both $\xi_X$ (dots) and $\xi_{W1}$ (crosses) are typically of order $\sim 0.01$, meaning that observed TDE light curves only capture $\sim 1 \%$ of the total energy emitted from TDE discs.   } 
 \label{MEP}
\end{figure}

\begin{figure}
  \includegraphics[width=.5\textwidth]{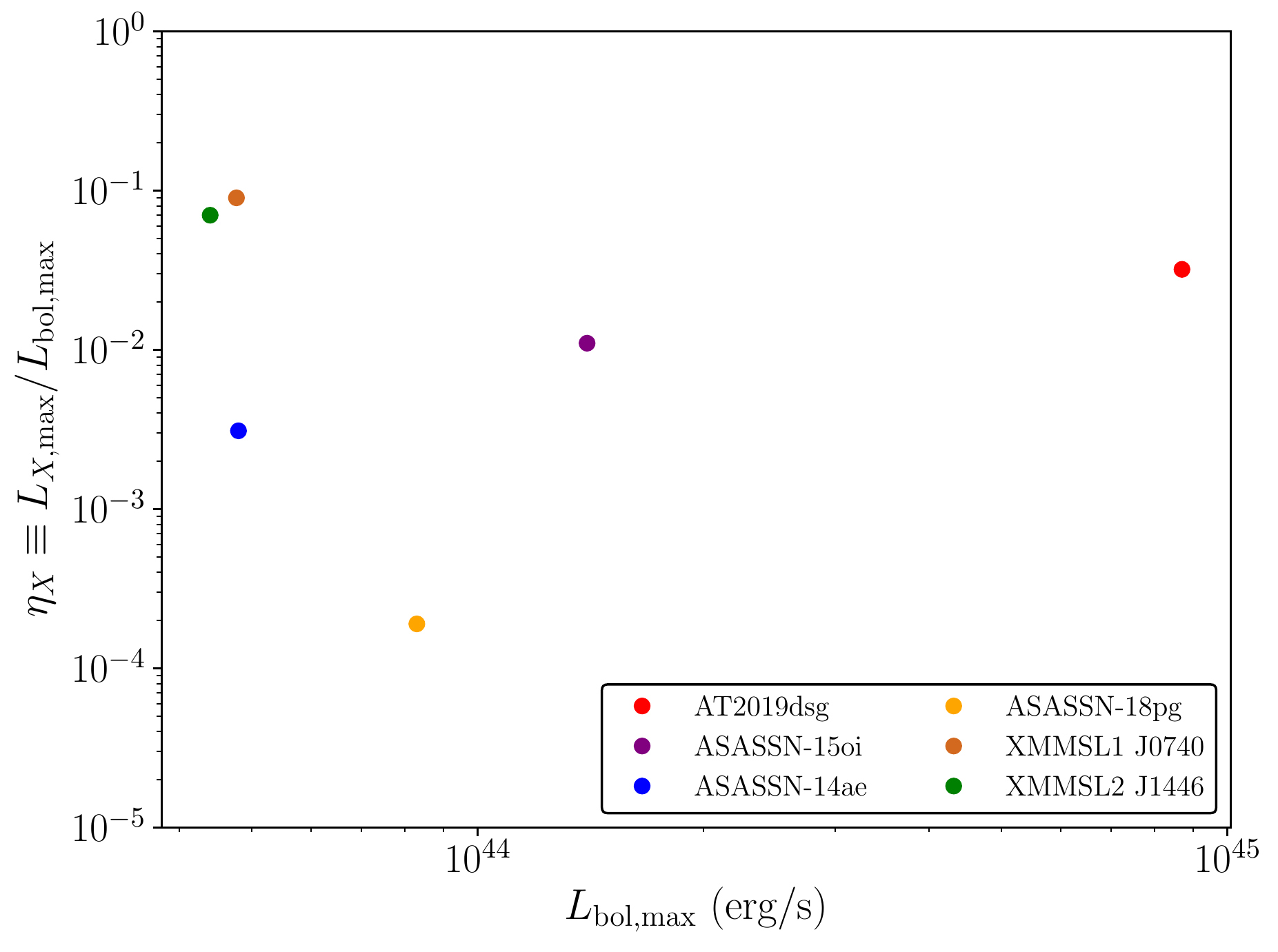} 
 \caption{The ratio of the peak X-ray and bolometric luminosities of the six TDEs considered in this study. X-ray bright TDEs have bolometric luminosities typically $\sim 10 - 100$ times brighter than the peak of their observed X-ray light curves.   } 
 \label{ETAX}
\end{figure}

To be more quantitative, we can define the total radiated energy:
\beq
E_{\rm rad} \equiv \int_0^\infty L_{\rm bol}(t) \,{\rm d}t,
\eeq
the total observed X-ray energy:
\beq
E_{{\rm rad}, X} \equiv 4\pi d_L^2 \int_0^\infty F_X(t) \,{\rm d}t,
\eeq
and the total observed UV energy:
\beq
E_{{\rm rad}, W1} \equiv 4\pi d_L^2 \int_0^\infty \nu_{W1} F_\nu(\nu_{W1}, t) \,{\rm d}t,
\eeq
where we take the UVW1 band as a representative example. {The three light curves are: $L_{\rm bol}(t)$, the models bolometric luminosity}, $F_X(t)$ the models X-ray light curve, and $F_\nu(\nu_{W1}, t)$ the evolving UVW1 flux from the accretion disc. All of these quantities can be determined numerically for our best fitting disc parameters. To resolve the `missing energy problem' we are in effect interested in the amplitude of two key dimensionless parameters
\beq
\xi_X \equiv E_{{\rm rad}, X}/E_{\rm rad} , \quad \xi_{W1} \equiv E_{{\rm rad}, W1}/E_{\rm rad} .
\eeq
These two parameters quantify the degree to which the observed X-ray and UV energies underestimate the total radiated energy. Both $\xi_X$ (dots) and $\xi_{W1}$ (crosses) are plotted against the total radiated energy of each TDE (summarised in Table \ref{tablebol}) in Figure \ref{MEP}. As can be clearly seen in Figure \ref{MEP}, the observed X-ray and UV energies underestimate the total radiated energy of TDE disc systems by a very large factor. Nonthermal X-ray TDEs typically radiated a fraction more of their energy at X-ray frequencies, but even these systems only radiated $\sim 5\%$ of their energy as X-rays. Quasi-thermal X-ray TDEs radiated even less of their energy at X-ray frequencies, typically at the $1\%$ level. The same is true for all TDEs at UV energies (Fig. \ref{MEP}). The nonthermal X-ray TDEs again radiated the largest fraction of their energy at observable frequencies ($\sim 10\%$), but the quasi-thermal X-ray bright, and UV-bright X-ray-dim TDEs only radiated $\sim 1\%$ of their energy at UV energies.

A further interesting parameter which quantifies the `missing' energy is 
\beq
\eta_X \equiv L_{\rm X, peak}/ L_{\rm bol, peak} .
\eeq
This parameter can be thought of as a `bolometric' correction from the disc X-ray to bolometric luminosity. As is clear from Fig. \ref{ETAX}, the `bolometric correction' of typical TDEs is extremely large. Nonthermal X-ray TDEs have bolometric luminosities typically a factor $\sim 10$ brighter than their observed X-ray luminosity, while quasi-thermal X-ray TDEs are closer to a factor $\sim 100$ brighter.

\subsection{Correlations of observed UV components with best-fit system parameters}\label{cor_sec}
The unified model of disc-dominated TDEs is not currently able to qualitatively explain the properties of a rapidly-decaying early-time component observed across UV and optical frequencies. This  early-time UV emission  is modelled with a simple exponentially decaying profile 
\beq
F_{\rm exp} = A \, \exp(-t/\tau) .
\eeq
We now have best-fitting physical parameters, and best fitting amplitudes $A$ and timescales $\tau$,  for six TDEs modelled in an identical manner. In this section we examine whether there is any correlation between physical system parameters ($M, E_{\rm rad}$ etc.) and the phenomenological parameters $A$ and $\tau$. 

We stress here  that our results should be considered preliminary, due to the small sample size we are working with. However, if correlations between the phenomenological  and physical parameters of our TDE solutions do exist, they may elucidate the physical nature of the early-time UV component. Furthermore, if the correlations are strong enough, observations of the early-time UV emission of TDEs may be used to estimate physical parameters of the TDE system. 

We begin by defining a distance independent measure of the luminosity of the early time UV emission
\beq
L_{W1, {\rm peak}} \equiv 4 \pi d_L^2 \nu_{W1} A_{W1} ,
\eeq
where we take the UVW1 band as representative of the UV flux. The values of $L_{W1, {\rm peak}}$ and $\tau_{W1}$ for the six TDEs considered in this paper are displayed in Table \ref{early_params}. We also include the TDEs ASASSN-14li and ASASSN-15lh in this analysis. These two TDEs were modelled using different techniques to the six TDEs in this paper, with ASASSN-14li having a larger ISCO stress value and higher black hole spin (Mummery \& Balbus 2020a), and ASASSN-15lh having a near maximal black hole spin $a/r_g = 0.99$ (Mummery \& Balbus 2020b). There were further methodological differences in the treatment of the disc initial condition and radiative transfer physics for these two sources. 

It is possible that the inferred parameter correlations will be affected by the different modelling techniques used for the different sources, and so these results should be treated with caution. The reason we include ASASSN-14li and ASASSN-15lh in our analysis is that they inhabit markedly different regions of the potential TDE parameter space when compared to the six TDEs examined in this paper. We believe that the increased size of the parameter space spanned by incorporating these two TDEs is more important than any possible complications resulting from methodological differences in the source modelling. 

\begin{table}
\renewcommand{\arraystretch}{2}
\centering
\begin{tabular}{|p{1.9cm}|p{2.2cm}|p{1.5cm}|}
\hline
TDE  & $L_{W1, {\rm peak}}$ (erg/s) & $\tau_{W1}$ (days) \\ \hline\hline
AT2019dsg & $6.8^{+0.2}_{-0.2} \times 10^{43}$ & $55.10^{+2.6}_{-2.6}$   \\ \hline
ASASSN-15oi & $1.3^{+0.1}_{-0.1} \times 10^{44}$ & $17.7^{+0.4}_{-0.4}$  \\ \hline
ASASSN-14ae &  $5.8^{+0.1}_{-0.1} \times 10^{43}$  & $25.2^{+0.5}_{-0.5}$ \\ \hline
ASASSN-18pg & $8.1^{+0.1}_{-0.1} \times 10^{43}$  & $40.1^{+1}_{-1}$  \\ \hline
XMMSL1 J0740 &  $3.6^{+0.3}_{-0.2} \times 10^{42}$ & $50^{+10}_{-5}$   \\ \hline
XMMSL2 J1446 & $1.9_{-0.2}^{+0.2} \times 10^{42}$ & $65_{-15}^{+15}$  \\ \hline
\end{tabular}
\caption{ The  properties of the early-time UVW1 emission of the six TDEs studied in this work. } 
\label{early_params}
\end{table}

\begin{figure}
  \includegraphics[width=.5\textwidth]{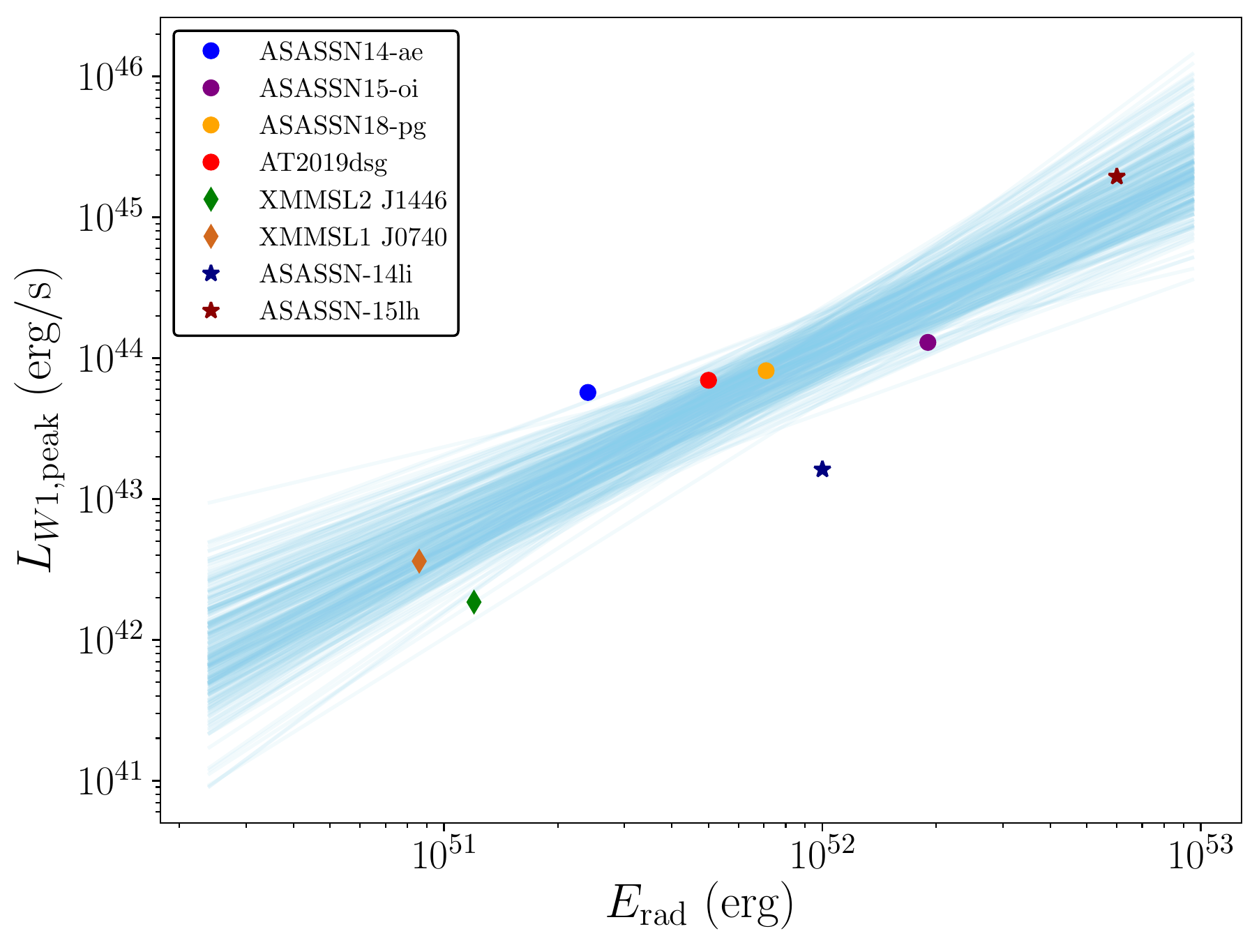} 
 \caption{A positive correlation between the peak UVW1 luminosity (defined in text), and the total radiated energy of the best-fitting TDE disc systems. Blue lines are samples drawn from the posterior distributions of the best fit regression line. Uncertainties may be smaller than the marker size.  } 
 \label{peakw1_erad_cor}
\end{figure}

We are able to determine that the peak of the early-time UVW1 light curve positively correlates with the total radiated energy of the best-fitting TDE accretion disc (Figure \ref{peakw1_erad_cor}), with  correlation $L_{W1, {\rm peak}} \sim E_{\rm rad}^{1.26 \pm 0.20}$. 
This is an extremely interesting result, as it suggests that the amplitude of the initial UV flare is correlated with the amount of material in the TDE accretion disc. Furthermore, there does not appear to be an accretion-state dependence of the amplitude of the initial UV flare (Figure \ref{peakw1_erad_cor}). As we discussed earlier, the two TDEs observed to be in the hard accretion state, characterised by nonthermal X-ray emission, appear to have strongly suppressed initial UV flares when compared to the quasi-thermal population of TDEs. However, when examined in combination with the other TDE sources, it appears that this is a result of their discs having lower matter content, rather than an intrinsic accretion-state dependence of the initial UV flare.

The magnitude of the initial UV flare was found to not correlate with the best-fit black hole mass of the TDEs, nor the magnitude of the discs bolometric luminosity (in physical or Eddington units).

\begin{figure}
  \includegraphics[width=.5\textwidth]{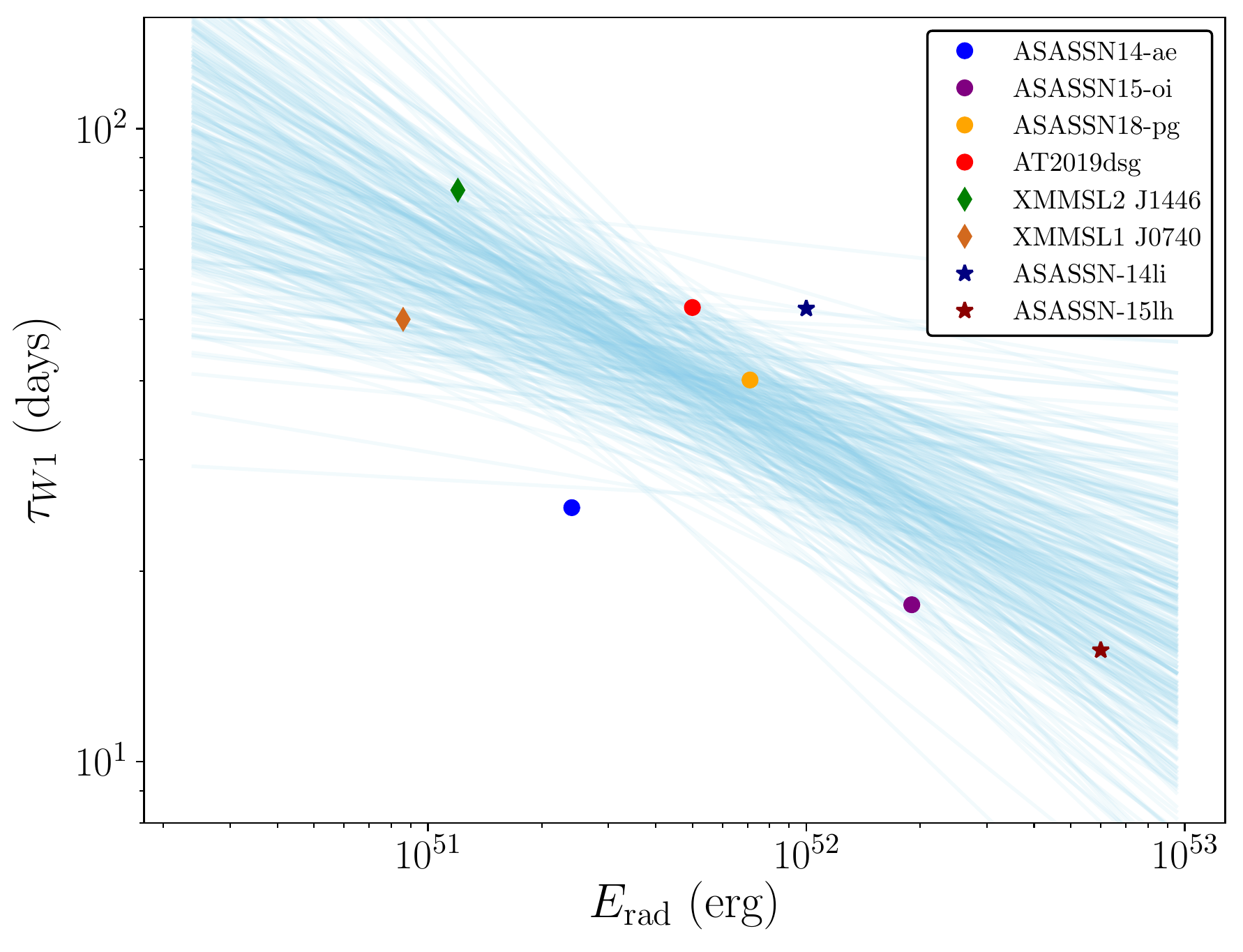} 
 \caption{A negative correlation between the UVW1 luminosity  decay timescale, and the total radiated energy of the best-fitting TDE disc systems. Blue lines are samples drawn from the posterior distributions of the best fit regression line. Uncertainties may be smaller than the marker size.   } 
 \label{tw1_cor}
\end{figure}

The UV flare decay timescale $\tau_{W1}$ appears to negatively correlate with the total radiated energy of the best fit TDE disc solutions (Figure \ref{tw1_cor}), with correlation $\tau_{W1} \sim E_{\rm rad}^{-0.26 \pm 0.13}$. If we were to exclude ASASSN-14li and ASASSN-15lh, then a positive correlation between the decay timescale $\tau_{W1}$ and black hole mass $M$ of the six TDEs studied in this paper is found. This is similar to what was found in previous works (e.g. van Velzen {\it et al}. 2020b). However, neither ASASSN-14li or ASASSN-15lh lie near this correlation, with ASASSN-15lh a particularly stark outlier $M\simeq 10^9M_\odot$, $\tau \sim 20$ days. It will be interesting to see whether a larger population of modelled TDEs reproduces a decay-time black hole mass correlation, which has theoretical support (Stone {\it et al}. 2013). 

\begin{figure}
  \includegraphics[width=.5\textwidth]{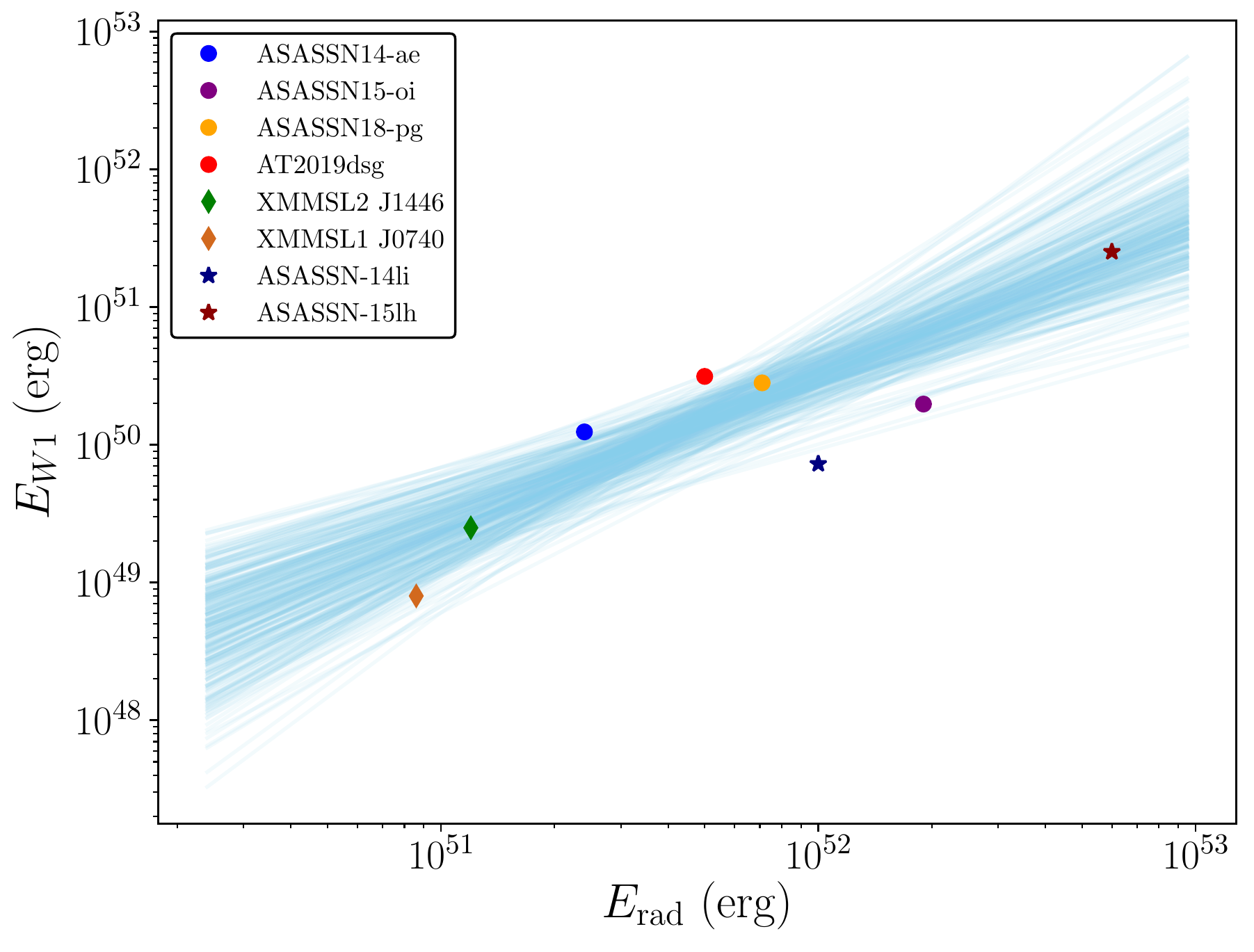} 
 \caption{A positive correlation between the early-time UVW1 energy (defined in text), and the total radiated energy of the best-fitting TDE disc systems. Blue lines are samples drawn from the posterior distributions of the best fit regression line. Uncertainties may be smaller than the marker size.  } 
 \label{EW1_cor}
\end{figure}

An interesting result of the dual correlations of the UVW1 luminosity amplitude and decay timescale with the total radiated energy is that the combination $E_{W1} \equiv L_{W1, {\rm peak}} \, \tau_{W1} $, a proxy for the total radiated energy in the early-time UV component, scales approximately linearly proportional to the total radiated energy of the disc $E_{W1} \sim E_{\rm rad}^{1.1 \pm 0.2} $ (Figure \ref{EW1_cor}).  The early-time UVW1 energy, in common with the total observed disc UVW1 and X-ray  energies, only corresponds to $\sim 1 \%$ of the total energy radiated from the disc.

\begin{figure}
  \includegraphics[width=.5\textwidth]{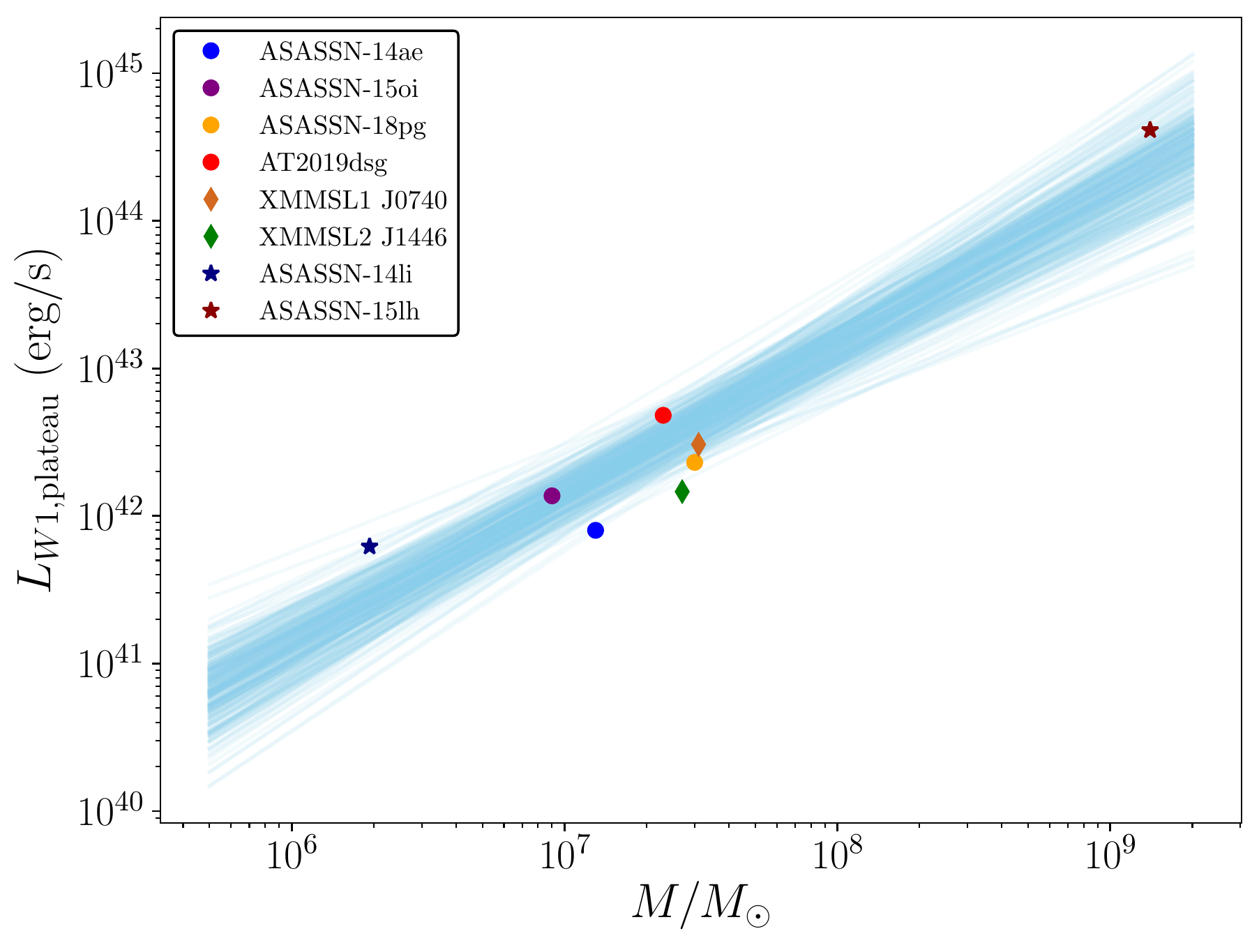} 
 \caption{A positive correlation between the late-time UVW1 plateau luminosity (defined in text), and the black hole mass of the best-fitting TDE disc systems. Blue lines are samples drawn from the posterior distributions of the best fit regression line. Uncertainties may be smaller than the marker size.  } 
 \label{plat_m_cor}
\end{figure}

Unlike the radiated disc energy, the central black hole mass $M$ does not correlate with the amplitude  of  the early-time UV light curve. However, the amplitude of the UV flux once it transitions into the disc-dominated state, the ``plateau flux'', does in fact correlate with black hole mass. To be more quantitative we define $L_{W1, {\rm plateau}} $ as 
\beq
L_{W1, {\rm plateau}} \equiv 4 \pi d_L^2 \nu_{W1} \, \max\left[ F_\nu(\nu_{W1}, t) \right] ,
\eeq
where $ F_\nu(\nu_{W1}, t)$ is the evolving disc UVW1 flux. Figure \ref{plat_m_cor} shows the UVW1 plateau luminosity plotted against best fitting black hole mass for the eight TDEs considered in this section. As is clear from Figure \ref{plat_m_cor}, there appears to be a positive correlation between the UV plateau flux and the central black hole masses of the TDEs considered in this work, $L_{W1, {\rm plateau}} \sim M^{1.02 \pm 0.1}$.   

There are therefore four correlations which we we suspect may represent true correlations in the observed TDE population. We use simple MCMC fits (Foreman-Mackey {\it et al}. 2013) of the functional form $y = a x^b$, to the eight available TDE data points to determine each explicit scaling relationship. The best fitting parameters with 1$\sigma$ uncertainties are summarised below. Firstly, the amplitude of the early-time UV luminosity appears to positively correlate with the total radiated energy of the best fitting TDE disc solutions: 
\beq
L_{W1, {\rm peak}} \simeq 1.33^{+0.49}_{-0.40} \times 10^{44} \left( { E_{\rm rad} \over 10^{52} \, {\rm erg } } \right)^{1.25^{+0.19}_{-0.15} } \, {\rm erg/s} .
\eeq
In contrast, the decay timescale of the early-time UV emission appears to negatively correlate with the total radiated disc energy:
\beq
\tau_{W1} \simeq 32^{+6}_{-5} \left( { E_{\rm rad} \over 10^{52} \, {\rm erg} }\right)^{-0.26^{+0.13}_{-0.13}}  \, {\rm days} .
\eeq
The combination $E_{W1} \equiv L_{W1, {\rm peak}} \, \tau_{W1}$ scales approximately linearly with the total radiated energy of the TDE disc solutions 
\beq\label{EEcor}
E_{W1}  \simeq 3.3^{+1.10}_{-0.88} \times 10^{50}  \left( { E_{\rm rad} \over 10^{52} \, {\rm erg } } \right)^{1.10^{+0.23}_{-0.19}} \, {\rm erg}. 
\eeq
Finally, the UV plateau flux of TDEs appears to correlate with the central black hole mass of the TDE
\beq
L_{W1, {\rm plateau}} \simeq 1.37^{+0.42}_{-0.34} \times 10^{42} \left( { M \over 10^7M_\odot } \right)^{1.02^{+0.10}_{-0.09}} \, {\rm erg/s} .
\eeq

\subsection{Physical interpretation of empirical correlations}
{In this paper we have therefore discovered two notable correlations between the UV light curves of TDEs and the physical properties of their best fitting discs and black hole systems. In this subsection we discuss possible physical interpretations of these results. 

A possible interpretation of the linear (within the error-bars) correlation between the early-time UV energy and the disc radiated energy is the following. Suppose the early-time UV energy ultimately results from the liberation of the gravitational energy of some mass content $M_{UV}$, with efficiency $\eta_{UV}$. This mass content will be a fraction $f_{UV}$ of the initial stellar mass. The disc mass will be a different fraction $f_d$ of the initial stellar mass, and the disc radiated energy will relate to the disc mass by an efficiency $\eta_d$, which is a function of black hole spin but is expected to be around $\eta_d \sim 0.1$. A linear relationship between $E_{\rm rad}$ and $E_{W1}$ then implies $f_{UV}\eta_{UV}/f_{d}\eta_{d} \sim 0.03$, constant across all TDEs (eq. \ref{EEcor}). {If, as expected, $f_d \sim 1/2$ and $\eta_d \sim 0.1$, then we can ascertain some properties of this UV bright component, namely: $f_{UV}\eta_{UV} \sim 10^{-3}$. The trivial constraint  $f_{\rm  UV} < 1$ implies that the early-time UV component is relatively efficient $\eta_{UV} \gtrsim 10^{-3}$.  } 

{Perhaps the most natural explanation for the constancy of the ratio between the disc and early-time-UV energies is that the early time UV flux is generated by some mechanism involved in the disc formation process. Gravitationally bound debris streams launched from the initial tidal disruption of the star must shed some angular momentum before they can form into a disc.  The behaviour of these debris streams, specifically how they are eventually circularised into a disc,  is an extremely complex, and still unsolved, problem. The solution is believed to involve the debris streams shocking, either on themselves or on the forming disc (Bonnerot \& Stone 2021). These shocks, in addition to circularising the disc material, will themselves liberate energy from the debris, and produce bright emission (Bonnerot {\it et al}. 2021).  Given that the amount of material which shocks in the early phases of the TDE evolution will be directly related to the amount of material which eventually forms into a disc ($f_{UV} \simeq f_d$), a linear relationship between the two energies would then simply result from the differing efficiencies of the two processes $\eta_{UV}/\eta_d \sim 0.03$. If we take the canonical $\eta_d \sim 0.1$, then the efficiency of the shocking/disc forming process can be inferred to be $\eta_{UV} \sim 3\times10^{-3}$. 

In addition to naturally explaining a linear-type relationship between the two energies, UV emission resulting from the disc formation process would aid in explaining why the early time  UV emission is so prompt. Unlike the X-ray light curves of TDEs (which result from disc emission), the UV light curves of TDEs are rarely,   if ever, observed during the rising phase. If the early UV emission results from a disc formation process then the UV light curves of TDEs must peak prior to their X-ray light curves, which can only reach peak brightness  once the disc is fully formed.    }

Irrespective of physical cause, the linear relationship between energies suggests that the physics of TDEs are universal: every TDE resulting in the same fractions of matter powering early-time UV emission and forming into a disc, with each of these processes then having their own characteristic efficiencies.  

The UV plateau-luminosity black hole mass correlation is more simple to qualitatively explain. TDE discs around black holes of larger masses are cooler (eq. \ref{temp}). The typical photon energy at which a disc radiates the majority of its energy is related to the disc temperature by $\nu_p \sim k_B T_p/h$, meaning that TDEs around large mass black holes radiate a larger fraction of their total energy at lower photon frequencies.  As typical TDE disc temperatures are higher than the UV scale, this disc cooling increases the fraction of energy radiated at UV energies. Furthermore, as the majority of UV flux results from the outer regions of TDE accretion discs, the increased emitting disc area ($\sim M^2$) of larger mass black holes  acts to increase the observed UV disc flux.

We reiterate that these correlations are built upon a small number of data points and should be viewed with caution. There is the possibility that `hidden variables' not considered in this analysis (e.g. the black hole spin) may play an important role in determining the properties of the early-time UV flux. With future observations and modelling of a much wider population of TDEs we hope to examine these correlations in more detail in future studies. 
}

\subsection{Black hole masses: comparing the masses inferred from TDE light curves and host galaxy properties} 
An important property of the TDE disc model is that physical properties of the central black hole  can be constrained from the fitting of multi-wavelength observations.  Each of the TDEs studied in this paper have black hole mass estimates inferred from galactic scaling relationships, these measurements represent an important consistency test for the TDE disc model.  

The black hole masses of TDEs are generally inferred from measurements of the galactic velocity dispersion $\sigma$,  galactic bulge mass $M_B$, or galactic bulge luminosity $L_B$. These different measurements offer independent estimates of the central black hole mass. Unfortunately, all of these scaling relationships have large intrinsic scatter (typically $\sim 0.3-0.5$ dex), and different scaling relationships often imply mutually incompatible black hole masses for a single TDE. This makes quantitatively comparing our best-fit black hole masses to those inferred from galactic scaling relationships inherently difficult. 

We find that the black hole masses inferred from light curve modelling (with the exception of ASASSN-14li and XMMSL2 J1446) are generally larger than the black hole masses inferred from the $M:\sigma$ relationship. However, the black hole masses of the majority (six out of eight) of the TDES we have analysed are in agreement with either the masses inferred from the galactic bulge mass/luminosity, or with the mean black hole mass found from combining multiple scaling relationships. 

In total we have modelled three TDEs with quasi-thermal X-ray spectra and UV light curves which transition into a disc-dominated plateau (ASASSN-14li, Mummery \& Balbus 2020a, AT2019dsg \& ASASSN-15oi, this work). These TDEs should be best described by the TDE disc model, and it is therefore extremely encouraging that the black hole masses of all three are consistent (i.e., lie within 1$\sigma$ error bars) with the mean black hole mass estimates obtained from galactic scaling relationships.  

The three TDEs which we have modelled which are only observed at UV energies (ASASSN-15lh, Mummery \& Balbus 2020b, ASASSN-14ae \& ASASSN-18pg, this work), have been generally consistent with, but on the large-mass end of,  the black hole masses inferred from the galactic bulge mass scaling relationship. As was discussed above, the inferred black hole masses for these sources are generally more massive than the black hole masses inferred from the $M:\sigma$ relationship. Of the two TDEs observed to have nonthermal X-ray spectra, one (XMMSL2 J1446) has a best-fit black hole mass in good agreement with the mean black hole mass inferred from galactic scaling relationships. The other, XMMSL1 J0740, only has a mass estimate inferred from the $M:\sigma$ relationship, this mass is  smaller than the black hole mass inferred from our light curve modelling. 

We have modelled eight individual TDEs, of which six have best-fitting black hole masses consistent with either an individual black hole mass inferred from galactic scaling relationships, or with the mean black hole mass inferred from multiple galactic scaling relationships. In general, we believe that the good agreement between the black hole mass estimates resulting from galactic scaling relationships (in particular the galactic bulge mass) and the black hole masses inferred from our detailed multi-wavelength light curve modelling strengthens the case for a disc origin of the observed emission. 

\section{TDE Radio emission and incorporating jets into the unified model}

All black holes accreting at relatively high rates produce, at times, relativistic jets which are predominantly (although not exclusively) revealed by their radio emission. The global coupling between the accretion `state' (temperature, geometry, optical depth and variability spectrum of the accretion 
flow) and the power and variability of the jet is well known, empirically, for stellar-mass black holes in X-ray binary systems (Fender {\it et al}. 2004). This pattern of coupling may well apply also to supermassive black holes in AGN (e.g. Koerding, Jester \& Fender 2006), and it makes sense to apply it also to TDEs. 

X-ray binaries (XRBs) reveal two main accretion states (hard and soft). The radio properties of XRBs show a clear accretion-state dependence. Our unified TDE model makes predictions about the X-ray properties of TDEs on the population level by assuming that TDEs have the same Eddington-ratio-dependent accretion states as XRBs. By making further analogy with XRBs we can therefore make predictions about the observed distribution of radio-bright TDEs.

An XRBs hard state, which can be observed over a range of Eddington-ratioed luminosities from $10^{-9} - 0.1$ (although there may be smoothly evolving changes in the properties over this range), is always associated with persistent, flat-spectrum radio emission (Fender 2001). This radio emission is correlated with the X-ray luminosity in a non-linear way (Gallo, Fender \& Pooley 2003; Corbel {\it et al}. 2003). This correlation has been extended to include supermassive black holes with the inclusion of a mass term (Merloni, Heinz \& di Matteo 2003; Falcke, Koerding \& Markoff 2003; Plotkin {\it et al}. 2012). There is good evidence that this flat-spectrum component is associated with a jet which carries a large fraction of the available accretion power, and may even be the dominant power output channel at the lowest luminosities. 

The analogous TDE accretion state to the XRB hard state occurs around black holes of the largest masses, $M \gtrsim 2 \times 10^7 M_\odot$ (eq. \ref{MHS}). One (XMMSL1 J0740 Saxton {\it et al}. 2017) of the two ``hard accretion state'' TDEs studied in this paper   has been detected at radio frequencies. Detailed radio follow up (Alexander {\it et al}. 2017) showed that this emission could result from a weak initially relativistic but decelerated jet with an energy of $\sim 2 \times10^{50}$ erg.  Alexander {\it et al}. argued that the emission could also result from a mildly relativistic outflow, as these two scenarios produce indistinguishable radio emission at the time of the first radio observation of XMMSL1 J0740. However, the other `hard-state' TDE XMMSL2 J1446 (Saxton {\it et al}. 2019) was not detected at radio frequencies, with a deep $5\sigma$ upper limit of $L_R \sim 10^{37}$ erg/s. While both the detection of XMMSL1 J0740 and the non-detection of XMMSL2 J1446 do lie within the (large) scatter of the fundamental radio-X-ray plane of activity (Merloni, Heinz \& di Matteo 2003), the low radio luminosity of XMMSL2 J1446 is somewhat surprising. We anticipate that TDEs which form around black holes with the largest masses should be the best targets for observing radio emission.

X-ray binaries with super-Eddington luminosities are also expected to produce jetted radio emission. In the disc-dominated TDE unification scheme these sources would correspond to TDEs with the lowest ($M \lesssim {\rm few} \times  10^6 M_\odot$) black hole masses.  There are hints that XRBs accreting at super-Eddington rates, in very dense environments, produce slower, more mass-loaded jets. The classic example of this is SS433, in which slowly-precessing mass loaded jets propagate away from the central, enshrouded, binary at $0.26$c (with some small variation). This is much lower than the speeds observed in the transient jets from low-mass XRBs in much less dense environments, most of which are observed to launch with initial Lorentz factors $\geq 2$. Interestingly, in the 2015 outburst of the black hole V404 Cyg, which appeared to also occur while the source was heavily affected by local absorption, precessing and relatively slow $\leq 0.5$c jets were observed on small scales (Miller-Jones {\it et al}. 2019). 

The potential for differences in the observed properties of TDE jets launched from super-Eddington and hard-state discs may lead to future avenues for testing the TDE unification scheme. We would expect any observed slower-moving mass-loaded jets to be around black holes of the lowest masses, containing super-Eddington discs. 

 Of the eight TDEs we have modelled so far,  ASASSN-14li was found to have the lowest black hole mass $M \simeq 2\times10^6 M_\odot$, and the largest initial Eddington ratio $l \simeq 0.85$.  ASASSN-14li has become the canonical `weak' radio TDE. The radio emission of ASASSN-14li has been associated both with a mildly-relativistic outflow (Alexander {\it et al}. 2016a), and also with a jet (van Velzen {\it et al}. 2016; Pasham \& van Velzen 2018).

In XRBs the jet appears to disappear in the soft X-ray state, which is usually observed at Eddington ratios $\geq 1$\% where the flat spectrum radio component appears to drop in luminosity by a factor of at least $10^3$ (Russell {\it et al}. 2011, Bright {\it et al}. 2020). The soft state corresponds roughly to Eddington ratios $1 > l > 0.01$, and black hole masses in the TDE unification scheme of $3 \times 10^6 \lesssim M/M_\odot \lesssim 3 \times 10^7$ (exact values are of course dependent on the disc mass, $\alpha$ parameter and black hole spin, Papers I, II). If TDE discs are similarly radio-quiet in the soft accretion state, then a simple prediction of the TDE unification scheme is that there will exist a mass-gap in the radio-loud TDE distribution. Explicitly, we would expect radio-loud TDEs to only occur around the largest and smallest mass black holes. The observed properties of TDE jets will provide future avenues for testing our TDE unification scheme.

Finally, we consider the properties of the famous jetted TDE {\it Swift} J1644. {\it Swift} J1644 is an extremely luminous ($L_X \sim 10^{48}$ erg/s, $L_R \sim 10^{41}$ erg/s) TDE with X-ray and radio light curves dominated by jetted emission. There are a number of existing black hole mass estimates of {\it Swift} J1644, with an upper-bound from the mass of the galactic bulge of $M < 2 \times 10^7 M_\odot$ (Burrows {\it et al}. 2011). This upper bound mass is at the level of the hard-state transitional mass, and we therefore suggest {\it Swift} J1644 was likely a TDE which formed in the super-Eddington accretion state. This is further supported by simultaneous X-ray and radio measurements (Miller {\it et al}. 2011), which imply a black hole mass of $\log(M/M_\odot) \simeq 5.5 \pm 1.1$, and a QPO measurement in the {\it Swift} J1644 X-ray spectrum (Reis {\it et al}. 2012), which implies a black hole mass $M \simeq 5 \times 10^5 - 5\times 10^6 M_\odot$. {\it Swift} J1644 has also been suggested to be the result of a white dwarf being tidally disrupted by an intermediate mass black hole $M < 2 \times 10^5 M_\odot$ (Krolik {\it et al.} 2011). These low inferred black hole masses point towards {\it Swift} J1644 being consistent, as expected within the unified TDE framework, with a jet launched from a super-Eddington accretion flow.

\section{Conclusions}
In this paper we have developed a unification scheme which explains the varied properties of TDE light curves in terms of simple disc physics. Our model, built upon detailed analysis of the X-ray properties of TDE discs performed in three separate papers, makes a number of predictions about the properties of TDEs on the population level. We demonstrate that the peak Eddington ratio of the disc which forms in the aftermath of a TDE is the principal parameter which controls the observed properties of an individual TDE. This peak Eddington ratio is primarily controlled by the central black hole mass of a TDE host, and we therefore make quantitative predictions about the mass distributions of different sub-populations of observed TDEs. These predictions are in good agreement with the observed mass distributions of the current observed TDE population (Fig. \ref{UM1}). In addition, we predict a black hole mass-independent maximum X-ray luminosity scale of bright TDEs, at the level $L_M \simeq 10^{44}$ erg/s. This is also reproduced by the observed TDE population (Figs. \ref{UM2}, \ref{UM3}). 

To test this unification scheme on the individual source level we have analysed six TDEs from across the different observed spectral TDE sub-populations (X-ray dim, X-ray bright with thermal emission, and X-ray bright with nonthermal emission). We demonstrate that the UV and X-ray light curves of all six sources can be reproduced by evolving relativistic disc models, with disc and black hole parameters consistent with expectations. These best fitting disc solutions are in strong agreement with the predictions of the TDE unification scheme (Fig \ref{BOL}). 

In addition, we resolve the the so-called ``missing energy problem'' by demonstrating that only $\sim1\%$ of the radiated accretion disc energy is observed at X-ray and UV photon frequencies. Similarly, early time non-disc emission observed at optical and UV frequencies only represent $\sim 1\%$ of the total radiated disc energy. The physical origin of the early-time UV and optical emission observed in TDEs is currently poorly understood. We present empirical scaling relationships between the amplitude, decay timescale, and total radiated energy of the early-time UV luminosity with the total radiated energy of the accretion disc (Figs. \ref{peakw1_erad_cor} -- \ref{plat_m_cor}). We hope that these scaling relationships will elucidate the physical origin of this early-time luminosity, and provide a useful tool for estimating physical system parameters from observations.  We further demonstrate that the late-time UV ``plateau'' luminosity positively correlates with the central black hole mass of a TDE.

In the final parts of this paper we discuss how radio observations can be incorporated into the unification scheme by making analogy with disc-jet coupling observed in X-ray binary systems. The observed properties of TDE jets will provide future avenues for testing our TDE unification scheme. 

We believe that the unification scheme put forward in this paper will provide a useful, physically motivated and testable framework with which future observations of TDEs can be interpreted.   

\section*{Acknowledgments} 
The author is extremely grateful for detailed discussions with Rob Fender on the properties of X-ray binaries and disc-jet coupling. 

\section*{Data availability}
No new data was used in this paper. All data used to make the figures which appear in this paper were taken from previous works cited at relevant points.

\appendix{}
\section{TDE population black hole masses}\label{masses}
\begin{table}
\renewcommand{\arraystretch}{2}
\centering
\begin{tabular}{|p{2.2cm}|p{2.cm}|}\hline
TDE name  & $\left\langle M_{\rm BH}\right\rangle/10^6M_\odot$  \\ \hline\hline
ASASSN-18jd & $ 145^{+105}_{-116} $  \\ \hline
AT2018fyk & $ 37.7^{+53.4}_{-12.0} $ \\ \hline
XMMSL1 J0740 & $7.9^{+4.4}_{-2.9}  $ \\ \hline
XMMSL2 J1446 & $41.2^{+47.1}_{-27.9}  $\\ \hline
SDSS J1323 & $14.7^{+20.7}_{-10.2}  $ \\ \hline
PTF-10iya & $34.9^{+46.7}_{-22.6}  $ \\ \hline
XMMSL1 J0619 &$33.3^{+19.6}_{-19.6}  $ \\ \hline
\end{tabular}
\caption{The  mean black hole mass of the 7 nonthermal (hard state)  X-ray TDEs from the literature.  }
\end{table}

\begin{table}
\renewcommand{\arraystretch}{2}
\centering
\begin{tabular}{|p{2.2cm}|p{2.cm}|p{2cm} |}\hline
TDE name  & $\left\langle M_{\rm BH}\right\rangle/10^6M_\odot$ \\ \hline\hline
ASASSN-14li & $2.9^{+2.9}_{-1.6} $  \\ \hline
ASASSN-15oi & $8.1^{+7.1}_{-4.3}$ \\ \hline
AT2018hyz & $4.3^{+6.9}_{-3.3} $ \\ \hline
AT2019dsg & $20.4^{+28.3}_{-14.7} $ \\ \hline
AT2019azh & $4.5^{+8.0}_{-3.5} $ \\ \hline
AT2019ehz &$6.6^{+8.0}_{-3.8}$ \\ \hline
AT2018zr & $11.0^{+14}_{-6.7}  $ \\ \hline
SDSS J1311 & $5.2^{+8.9}_{-3.3} $ \\ \hline
XMMSL1 J1404 & $2.8^{+1.4}_{-1.0} $ \\ \hline
OGLE 16aaa & $26.0^{+35}_{-16} $ \\ \hline
3XMM J1521 & $5.4^{+5.1}_{-3.0}$ \\ \hline
\end{tabular}
\caption{The mean black hole mass of the 11 Thermal (soft state) X-ray TDEs from the literature.  }
\end{table}
{In Papers I \& II we used well-established galactic scaling relationships between the black hole mass and (i) the galactic bulge mass $M : M_{\rm bulge}$, (ii) the galactic velocity dispersion $M : \sigma$, and (iii) the bulge V-band luminosity $M : L_V$. All of the scaling relationships are taken from McConnell \& Ma (2013).  Where available, values of $ M_{\rm bulge}$, $\sigma$ and $L_V$ were taken from the literature for each TDE, and the mean black hole mass for each TDE was computed. The mean masses are reproduced in the two tables of this Appendix, see Papers I (thermal) or II (nonthermal) for a summary of the galactic scaling measurements which make up each mean mass. For the optical/UV-only TDEs black hole masses are calculated from the velocity dispersion ($\sigma$) measurements of Wevers {\it et al}. (2019a), their table A1. We include 13 of the 15 optical TDEs catalogued by Wevers {\it et al}., but do not include ASASSN-14li and ASASSN-15oi as they were also observed at X-ray energies. }

\begin{table}
\renewcommand{\arraystretch}{2}
\centering
\begin{tabular}{|p{2.2cm}|p{2.cm}|p{3cm} |}\hline
TDE name  & $\sigma$ (km/s) & $\log_{10}\left(M_{\rm BH,\sigma}/M_\odot\right)$ \\ \hline\hline
GALEX D1-9 & $65\pm 6$ & $5.85^{+0.54}_{-0.53}$  \\ \hline
GALEX D23-H1 & $84\pm 4$ & $6.39^{+0.44}_{-0.44}$  \\ \hline
GALEX D3-13 & $133\pm 6$ & $7.36^{+0.43}_{-0.44}$  \\ \hline
ASASSN-14ae & $53\pm 2$ & $5.42^{+0.46}_{-0.46}$  \\ \hline
ASASSN-15lh & $210\pm 7$ & $8.32^{+0.41}_{-0.41}$  \\ \hline
PTF--09ge & $82 \pm 2$ & $6.31_{-0.39}^{+0.48}$ \\ \hline 
PTF--09axc & $60 \pm 4$ &  $5.68_{-0.49}^{ +0.56}$\\ \hline 
PTF--09djl &  $64 \pm 7$ & $ 5.82_{-0.58}^{ +0.38}$ \\ \hline
iPTF--15af &  $106 \pm 2$ & $ 6.88_{-0.38}^{ +0.42}$ \\ \hline 
iPTF--16axa &  $82 \pm 3$ & $6.34_{-0.42}^{ +0.42}$ \\ \hline 
iPTF--16fnl &  $55 \pm 2$ &  $  5.50_{-0.42}^{ +0.44}$\\ \hline
PS1--10jh  & $65 \pm 3$ & $ 5.85_{-0.44}^{ +0.45}$ \\ \hline 
SDSS TDE1 & $126 \pm 7$ & $ 7.25_{-0.46}^{+0.45}$ \\ \hline 

\end{tabular}
\caption{The black hole masses of the optical-only TDEs inferred from the $M:\sigma$ relationship. Reproduced from Wevers {\it et al}. 2019, their Table A1.  }
\end{table}

\section{TDE disc model parameters}
We reproduce the best-fitting model parameters for the six TDEs studied in this paper in Table \ref{table_appendix}. 

\begin{table}
\renewcommand{\arraystretch}{2}
\centering
\begin{tabular}{|p{2.2cm}|p{2.cm}| p{2. cm} | p{1.5 cm} | p{2. cm} | p{2. cm} | p{2. cm} | p{1.2 cm} |}
\hline
TDE name  & $M_{\rm BH}/M_\odot$  &  $M_{\rm acc}/M_\odot$ & $t_{\rm visc}$ (days)  &  ${\cal V} = t_{\rm orb}/t_{\rm visc}$ & $f_{SC}$  &  $\Gamma$  \\ \hline\hline
ASASSN-14ae & $1.3  \times 10^7$ & $0.05 $ &  $250$ &$1.0\times10^{-4}$ &  --- & ---  \\  \hline
ASASSN-15oi & $9.0^{+2.0}_{-2.0}  \times 10^6$ & $0.31^{+0.08}_{-0.09} $ &  $600^{+95}_{-140}$ &$7.8\times 10^{-5}$ &  --- & ---  \\  \hline
ASASSN-18pg  & $3.1  \times 10^7$ & $0.22 $ &  $560$ &$1.0\times10^{-4}$ &  --- & ---  \\  \hline
AT2019dsg & $2.0^{+0.3}_{-0.5}  \times 10^7$ & $0.09^{+0.02}_{-0.03} $ &  $28.1^{+0.6}_{-1.4}$ &$1.3\times 10^{-3}$ &  --- & ---  \\  \hline
XMMSL1 J0740 & $3.1^{+0.9}_{-0.4}  \times 10^7$ & $1.8^{+2.4}_{-1.0}\times10^{-2} $ &  $84.3^{+7.1}_{-10.6}$ &$6.4\times 10^{-4}$ &  $0.032^{+0.038}_{-0.019}$ & 1.95  \\  \hline
XMMSL2 J1446 & $2.7^{+0.3}_{-0.2}  \times 10^7$ & $2.0^{+0.6}_{-0.2}\times10^{-2} $ &  $109^{+25}_{-4}$ &$4.5\times 10^{-4}$ &  $0.12^{+0.16}_{-0.2}$ & 2.58  \\  \hline

\hline
\end{tabular}
\caption{The best-fitting properties of the 6 TDEs modelled in this work. Other than ASASSN-15oi, all TDEs had a disc feeding radius $r_0 = 10r_g$ ($r_{0, {\rm 15oi}} = 20r_g$). All TDEs were around Schwarzschild black holes $a=0$. The disc viscosity parameter ${\cal V}$ varies by roughly one order of magnitude across the population of modelled TDEs. }
\label{table_appendix}
\end{table}

\label{lastpage}
\end{document}